\documentclass{aa}  
\makeatletter
\def\@to{to}
\def\as     {\ifmmode {\rlap.}$\,$''$\,$\! \else ${\rlap.}$\,$''$\,$\!$\fi}
     \def\decsec  {\ifmmode {\rlap.}$\,$^{\rm s}$\,$\! \else ${\rlap.}$\,$^{\rm s}$\,$\!$\fi}\def\decs  {\ifmmode {\rlap.}$\,$^{\rm s}$\,$\! \else ${\rlap.}$\,$^{\rm s}$\,$\!$\fi}
\makeatother
\usepackage{graphicx}
\usepackage{color}
\usepackage{xcolor}
\usepackage[flushleft]{threeparttable}
\usepackage{python}
\usepackage{float}
\usepackage{graphics,graphicx}
\usepackage{epstopdf}
\usepackage{amssymb, amsmath}
\usepackage{graphicx}%
\usepackage[normalem]{ulem}
\usepackage[draft]{hyperref}
\usepackage{soul,xcolor}

\usepackage{booktabs}
\usepackage{txfonts}
\begin{document} 
\setstcolor{red}


   \title{Fragmentation and filaments at the onset of star and cluster formation }
    \subtitle{{\emph{SABOCA}} 350 $\mu$m view of ATLASGAL selected massive clumps}

   \author{Y. Lin
          \inst{1}\thanks{Member of the International Max-Planck Research School (IMPRS)
for Astronomy and Astrophysics at the Universities of Bonn and
Cologne.} \and T. Csengeri \inst{1, 2} \and F. Wyrowski \inst{1} \and J. S. Urquhart \inst{3} \and F. Schuller \inst{1} \and A. Weiss \inst{1} \and K. M. Menten \inst{1}}

   \institute{Max-Planck Institute f\"{u}r Radio Astronomy, Auf dem H\"{u}gel 69, 53121 Bonn\\
              \email{ylin@mpifr-bonn.mpg.de}
         \and
        OASU/LAB-UMR5804, CNRS, Universit\'e Bordeaux, all\'ee Geoffroy Saint-Hilaire, 33615 Pessac, France
        \and
        Centre for Astrophysics and Planetary Science, University of Kent,
Canterbury CT2 7NH, UK
       }


 
  \abstract
  {The structure formation of the dense interstellar material and the fragmentation of clumps into cores is a fundamental step to understand how stars and stellar clusters form.}
   {We aim to establish a statistical view of clump fragmentation  
   at sub-parsec scales based on a large sample of massive clumps selected from the ATLASGAL survey. } 
   {We used the APEX/SABOCA camera at 350\,$\mu$m to image clumps at a resolution of $8\as5$, corresponding to physical scales of $<$0.2\,pc at a distance $<$5\,kpc. The majority of the sample consists of massive clumps that are weak or in absorption at 24\,$\mu$m. We resolve spherical and filamentary structures and identify the population of compact sources. Complemented with
   archival {\it{Herschel}} data, we derive the physical properties, such as dust temperature, mass and bolometric luminosity of clumps and cores. We use association with mid-infrared 22-24\,$\mu$m and 70\,$\mu$m point sources to pin down the star formation activity of the cores. We then statistically assess their physical properties, and the fragmentation characteristics of massive clumps.}  
   {We detect emission at 350\,$\mu$m towards all targets and find that it typically exhibits a filamentary(-like) morphology and hosts a population of compact sources. Using {\it{Gaussclumps}} we identify 1120 compact sources and derive the physical parameters and star formation activity for 971 of these,  874 of which are associated with 444 clumps.  
   {We find a moderate correlation between the clump fragmentation levels with the clump gas density and the predicted number of fragments with pure Jeans fragmentation scenario.}
  We find a strong correlation between the mass of the most massive fragment and the total clump mass, suggesting that the self-gravity may play an important role in the clumps' small scale structure formation. 
  Finally, due to the improved angular resolution compared to ATLASGAL, we are able to identify 27 massive quiescent cores with $M_{\rm core}>100$\,M$_{\odot}$ within 5\,kpc; these are massive enough to be self-gravitating but do not yet show any sign of star-formation. This sample comprises, therefore, promising candidates of massive pre-stellar cores, or deeply embedded high-mass protostars. }
 {The submillimeter observations of the massive clumps that are weak or completely dark at 24\,$\mu$m reveal rich filamentary structures and an embedded population of compact cores. The maximum core mass is likely determined by the self-gravity of the clump. The rarity of massive pre-stellar core candidates implies short collapse time-scales for dense structures.}


   \maketitle
%

\section{Introduction}\label{sec:intro}

%

Despite the significant influence of massive stars on their natal environment and the Galaxy as a whole, their formation process remains poorly constrained. Massive stars (M$\gtrsim$10\,M$_{\odot}$)  are considerably rarer than solar-mass stars. They constitute less than $\sim$10$\%$ of the stellar initial mass function (IMF) in mass. The duration of their formation time is short, hence the deeply embedded early stages are difficult to identify. Adding to the complexity of the high-mass star-forming scenario is that massive stars form typically in clusters  
(\citealt{Stahler00}; \citealt{Lada03}), 
yet, the mass assembly and the fragmentation of massive clumps into cores is still not well understood (for recent reviews see \citealt{Tan14}, \citealt{Motte18}).

In the last decade, Galactic plane surveys in the far-infrared and submillimeter regime such as the APEX Telescope Large Area Survey of the Galaxy (ATLASGAL, \citealt{Schuller09}; \citealt{Csengeri14}), the Bolocam Galactic Plane Survey (BGPS, \citealt{Aguirre11}), the Herschel infrared Galactic Plane Survey ({\it{Hi-Gal}}, \citealt{Molinari10}) and the JCMT Plane Survey (JPS, \citealt{Moore15}; \citealt{Eden17}) have proven to be 
excellent finding charts for 
identifying large samples of massive clumps that exhibit typical physical characteristics of $\rm >10^{3}-10^{4}\,M_{\odot}$, an extent 
up to $\sim$1\,pc and a temperature of $\sim$20\,K  typically associated with giant molecular cloud complexes (e.g. \citealt{Csengeri16a}; \citealt{Svoboda16}; \citealt{Konig17}; \citealt{Elia17}; \citealt{Urquhart18}). Massive clumps fragment into cores where star formation takes place (\citealt{Motte18}). Investigating how the dense gas is structured within massive clumps, and how they fragment and form filaments and cores on a 0.1-0.3\,pc scale (e.g. \citealt{Motte07}; \citealt{Andre14}) is, therefore, an important step to understand the mass assembly process to form stars and clusters.

Due to the typically large, several kilo-parsec distance of  high-mass star-forming regions, investigating their structure and fragmentation requires high-angular resolution interferometric studies  (e.g. \citealt{Bontemps10}; {\citealp{Palau13,Palau15}}; \citealt{Csengeri17b}; \citealt{Beuther18}). Reaching physical scales of a few thousands to tens of thousands of au is necessary to resolve the typical Jeans length in these objects, which is 0.1-0.3\,pc, assuming a temperature of 20 K and a volume density of 10$^{4}$-10$^{5}$\,cm$^{-3}$, characteristic of massive cores or clumps (e.g. \citealt{Williams00}; \citealt{Motte07}; \citealt{Wienen12}; \citealt{Urquhart18}).
Mapping of large areas is not feasible with current interferometers in the (sub)millimetre wavelength range. However, ground-based telescopes with sensitive, large field-of-view bolometer arrays provide an efficient way to map large regions at a moderate resolution at short wavelengths (350 and 450\,$\mu$m, i.e. $\sim$10$\arcsec$ corresponding to a physical scale of 0.1 pc at $d\sim$2\,kpc). This allows us to conduct a statistical study of fragmentation from clump to core scales (e.\,g.\,\citealt{Motte03}; \citealt{Minier09}; \citealt{Ragan13}; \citealt{Merello15}; \citealt{Andre16}; \citealt{Rayner17}, \citealt{Heyer18}). 
We have, therefore, carried out observations towards more than 200 ATLASGAL selected sources at 350\,$\mu$m with the Submillimetre APEX Bolometer array camera (SABOCA, \citealt{Siringo10}) at the Atacama Pathfinder Experiment 12-meter telescope (APEX, \citealt{Guesten06}). These observations achieve an angular resolution of $8\as5$, a factor of $>2$ improvement in terms of angular resolution compared to the ATLASGAL survey, allowing us to identify a population of cores with sizes of 0.1-0.2\,pc embedded in clumps with distances $d\lesssim$ 5\,kpc.  The angular resolution achieved corresponds to an intermediate physical scale to dissect structures of massive star-forming clumps, which is a missing gap to link with typical interferometric studies. The sample is twice that of  \citet{Merello15} who targeted a sample of BGPS sources with the SHARC-II camera on the Caltech Submillimeter Observatory (CSO). Combining our results with ancillary submillimeter and infrared data, we estimate the cores' physical parameters and compare them with the properties of clumps in order to constrain their fragmentation and structure formation. The higher angular resolution compared to ATLASGAL allows us to distinguish between quiescent and star-forming cores and deliver a sample of starless or pre-stellar massive cores.

This paper is organised as follows: in Section\,2 we present the observations, data reduction, and the complementary data used. In Section\,3 we present our results and the data analysis, including source extraction, and determination of physical properties of clumps and SABOCA compact sources. In Section\,4 we discuss the statistics of the sample in terms of the fragmentation properties and the relation between clump and core properties. Our main results and conclusions are summarised in Section\,5.

\section{Observations and data reduction}\label{sec:obs}
\subsection{Source selection}

\begin{figure}[h]
\includegraphics[width=\hsize]{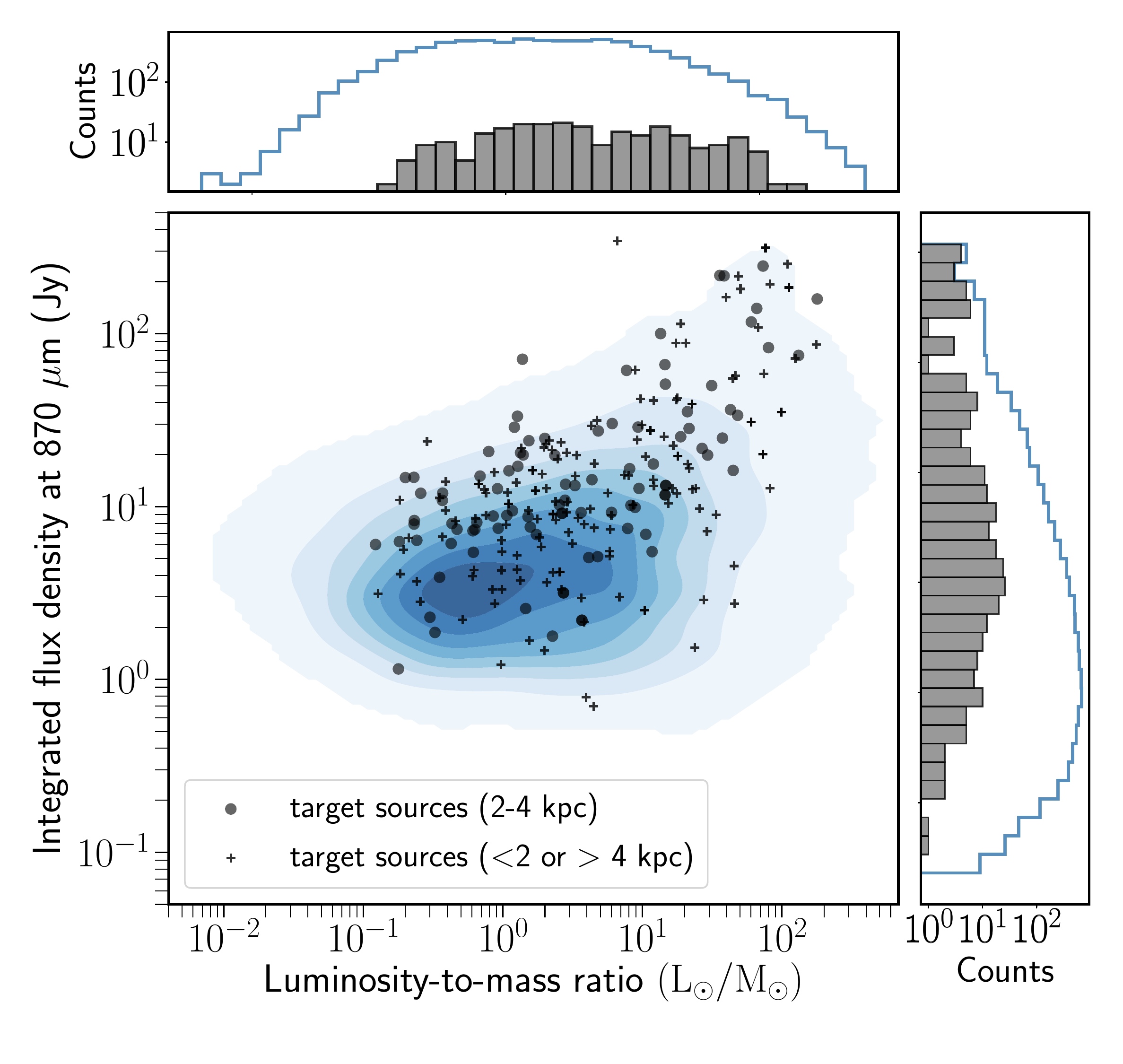}
\caption{Distribution of the integrated flux density and bolometric luminosity-to-mass ratios (L/M) of all ATLASGAL sources (in blue filled contours, the darker region corresponds to higher density) (\citealt{Urquhart18}), and the target sources observed with SABOCA (gray pluses and dots) at different distance ranges. The histograms in the right side and top panel show the 1-dimensional distributions of integrated flux density and L/M of all ATLASGAL sources (blue) and all target sources (gray), respectively.} 
\label{fig:intflux}
\end{figure}

Over 10\,000 compact sources have been identified in the ATLASGAL survey (\citealt{Csengeri14}; \citealt{Urquhart14}), and the majority of them  (8007)  have distance estimates allowing their physical parameters to be determined (\citealt{Urquhart18}). In this study, we use 350\,$\mu$m SABOCA observations to investigate small scale structure of the clumps. In comparison with the extensive follow-up observations of the brightest submillimetre clumps selected from the ATLASGAL survey using both single dish (\citealp{Giannetti14, Giannetti17}; \citealt{Csengeri16a}; \citealp{Kim17, Kim18}), and interferometric observations \citep{Csengeri17b, Csengeri18} this study focuses more on a population of massive, but lower surface density clumps (with typically lower submillimeter flux density). Our targets have been selected to be weak or in absorption at 24\,$\mu$m based on the Spitzer MIPS GALactic plane survey (MIPSGAL; \citealt{Carey09}). This initial sample of 95 clumps has been complemented with sources at different evolutionary stages based on their mid-infrared color selection at 24 and 70 $\mu$m mostly having a bolometric luminosity to mass ratio ($L_{bol}/M$) less than 10. In addition, we include the data for the brightest 25 sources from ATLASGAL survey. 


Altogether we targeted 204 ATLASGAL clumps, which is the largest number of clumps targeted with SABOCA. Since we obtained maps of the targets, there are 371 other ATLASGAL sources covered in the fields, on average we have 2-3 ATLASGAL clumps per field. The total number of observed clumps from the \emph{GaussClump} compact source catalog of \citet{Csengeri14} is, therefore, 575. In Fig.\,\ref{fig:intflux}, we compare the integrated flux density and $L_{bol}/M$ of our sample with the same quantities of the whole ATLASGAL sample where available (\citealt{Urquhart18}). This shows that our sample is representative of the entire population of sub-millimetre sources,  having a $L_{bol}/M$ range of 0.1 to 100\,L$_{\odot}$/M$_{\odot}$. Many of these sources are good candidates to host objects in early stages of high-mass star and cluster formation (see for example \citealt{Csengeri17b}). The sample also comprises, however, massive clumps in more advanced stages of high-mass star formation that are associated with bright mid-infrared sources and ultra-compact H\textsc{ii} regions exhibiting $L/M>$ 100\,L$_{\odot}$/M$_{\odot}$. 
In the following, we perform all the analysis on the entire sample homogeneously, and clarify in the discussion when only focusing on the mid-infrared dark sources.

\begin{figure}[h]
\includegraphics[scale=0.42]{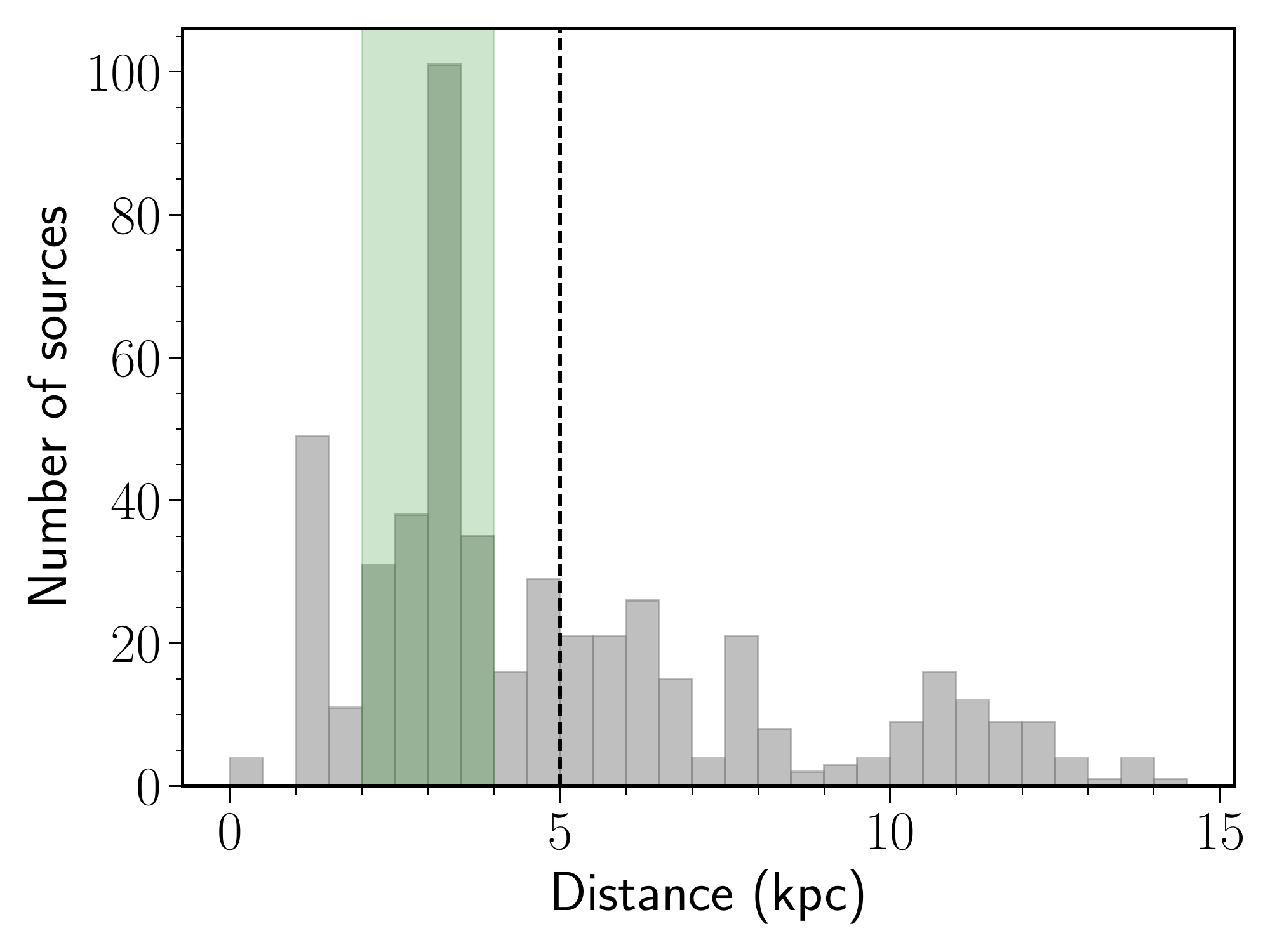}
\caption{Distance distribution of the observed  ATLASGAL clumps. The vertical dashed line indicates the 5\,kpc distance. The filled green region indicates 2-4 kpc range.}\label{fig:dist_distri}
\end{figure}

We show the distance distribution of all sources in Fig.\,\ref{fig:dist_distri} for which a distance has been determined by \citet{Urquhart18}; these are primarily kinematic distances. Altogether, this provides us with a distance to 507 clumps, which corresponds to 88\% of our sample.

The entire sample covers a distance range from 1 to 14.5\,kpc (Fig.\,\ref{fig:dist_distri}), corresponding to a physical resolution between 0.04 and 0.6\,pc. The large range of linear scales ($\sim 10$) has an impact the observed morphologies and derived physical properties. To mitigate any possible observational biases we restrict our statistical analysis to a distance limited sample ($d\lesssim$2-4\,kpc); this corresponds to physical sizes of 0.07 and 0.15\,pc, and thus, reducing the range of linear scales to a factor of 2. The majority of the clumps are located within 5\,kpc (314 out of the 507 clumps corresponding to 62$\%$ of the sample), 205 of which are located in 2-4 kpc. The more distant sources are mostly luminous OB clusters that tend to be associated with very active star forming regions, which belong to the 25 strongest submillimeter sources with ATLASGAL survey, such as the W51, W43 complexes (e.g. \citealt{Ginsburg15}; \citealt{Motte14}).

\subsection{SABOCA observations and data reduction}\label{subset:sabocaobs}

\begin{figure}[h]
\includegraphics[width=\hsize]{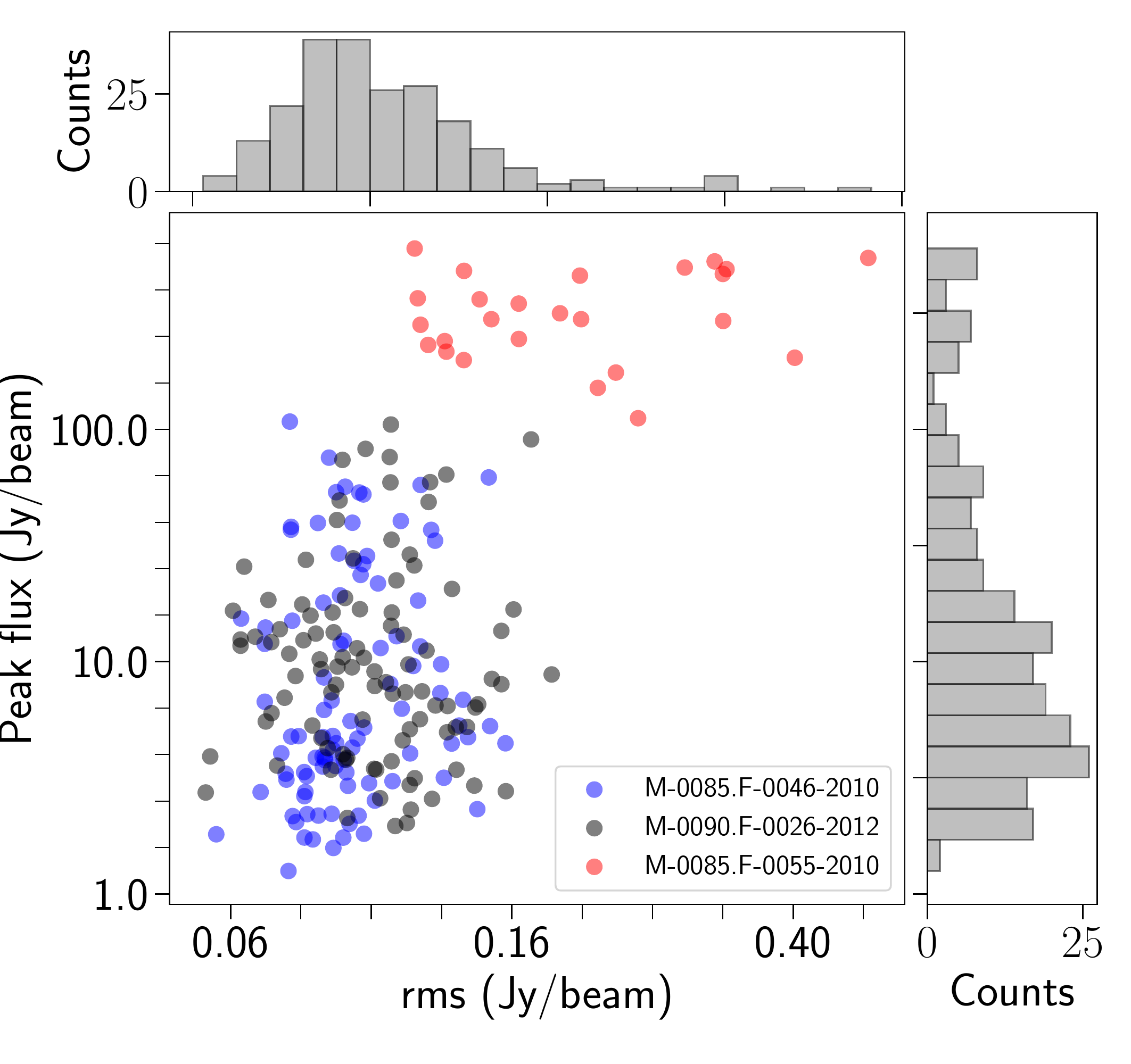}
\caption{350\,$\mu$m peak flux density versus the 1$\sigma$ rms noise level observed with SABOCA towards all fields. The histogram on the right side, and the top panel show the distribution of peak flux densities and the noise levels, respectively.}
\label{fig:peak_rms}
\end{figure}

We targeted the selected ATLASGAL sources using the SABOCA instrument installed at the APEX telescope which offers $\sim$8$''$ angular resolution at 350\,$\mu$m. We used a raster spiral mapping strategy for each source that mapped a  4$'$ region centered on the source position. The observations were conducted in 2010 and 2012 with a median precipitable water vapor (PWV) of 0.3~mm and all less than 0.5 mm. The typical on-source integration time is $\sim$5 minutes, which gives a better than $\sim$160\,mJy/beam noise level for most of the sources (Fig.\,\ref{fig:peak_rms}).  This corresponds to a mass sensitivity of 1.84\,M$_{\odot}$ at 3\,kpc, assuming a dust temperature of 20\,K, $\kappa_{350\mu {\rm m}}=0.076$\,cm$^{2}$\,g$^{-1}$, and a gas-to-dust ratio of 100 (Eqs.\,1 and 2). 

We used the BoA software (\citealt{Schuller12}) with standard procedures and iterative masking for the data reduction.
The atmospheric zenith opacity determined from the radiometer in each science scan was used to correct for atmospheric opacity. These values were compared to the results of the \texttt{skydip} measurement and found to be consistent. Mars and standard calibrators, such as IRAS 13134-6264, were regularly observed and used as primary flux calibrators. Typical calibration uncertainties are within 20$\%$. Pointing and focus were updated every 1-2 hours. The telescope pointing accuracy is within 2-3\arcsec. 

We used the standard, iterative source-masking reduction procedures optimised for the faint and the bright sources, respectively. In the first step of the iterative processing, an initial map is obtained after opacity correction, and then the correlated noise and bad channels are removed. The resulting map is then smoothed by 2/3 of the beam size (yielding a resolution of 9.5$''$) to achieve a better signal-to-noise ratio. This is then used as an input model for the following iterations to represent the real source structure. The residual flux is modeled by subtracting the model image from the initial map, with a cut-off level at a signal-to-noise ratio of 3. After adding the model back to the image, the process is repeated until the peak flux level remains unchanged. For the final maps{\color{purple},} we applied 3\arcsec\ smoothing which gives a resolution of 8\rlap{.}{\arcsec}5 (original beam size $\sim$8$''$). We checked the convergence of the data reduction process by examining the relative change of the peak and integrated flux density around the brightest pixels during the iterative process. The pixel size is 1$\as$85 in the final map.

In order to have a uniform and consistent noise level estimation for all fields, we use a wavelet-based estimator implemented in the \texttt{skimage} python package; this assumes a Gaussian noise distribution and calculates the standard deviation in each map (\citealt{Donoho94}). Before this estimation, we trimmed the edge of the maps, because of the increased noise level due to the non-uniform sampling at the map edges caused by the scanning strategy. We selected several sources to check the derived values and found them to be consistent with measurements made of an emission-free region of the map. The achieved rms noise distribution of all the sources is shown in Fig.\,\ref{fig:peak_rms}. The mean noise level for the relatively weak sources (from projects M-0085.F-0046-2010, M-0090.F-0026-2012) is 0.12\,Jy/beam with a standard deviation of 0.07\,Jy/beam, and for the brightest sources (M-0085.F-0055-2010) it is 0.21\,Jy/beam with a standard deviation of 0.10\,Jy/beam.

We find that the brightest sources, above 100\,Jy/beam peak flux density, have up to a factor of 5 higher rms noise compared to the weaker sources (Fig.\,\ref{fig:peak_rms}). Altogether the sources exhibit a large range of peak flux density from as low as 1.25\,Jy/beam up to 600.4\,Jy/beam, with an average of 52.0\,Jy/beam and a median of 9.0\,Jy/beam at 350\,$\mu$m. In agreement with our initial selection from ATLASGAL, this shows that we have a large number of relatively faint and a few extremely bright clumps.

\subsection{Ancillary data}

To be able to determine the physical properties of the clumps, 
we rely on the far-infrared emission of the dust. For this, we use level 2.5/3 archival data from Herschel PACS and SPIRE at 70/160 $\mu$m, 250/350 $\mu$m, respectively, from the Hi-Gal survey (\citealt{Molinari10, Molinari16}). The zero point offsets of the SPIRE data have been corrected by {\it{Planck}}-HFI data via cross-calibration\footnote{The procedure is elaborated in \url{http://herschel.esac.esa.int/Docs/SPIRE/spire_handbook.pdf}}; this accounts for the thermal background produced by the instrument itself. There are 6 fields not covered in Hi-Gal or at the edges of the maps that were excluded from the following analysis. 

\begin{figure*}
\begin{tabular}{p{0.235\linewidth}p{0.235\linewidth}p{0.235\linewidth}p{0.235\linewidth}}
\includegraphics[scale=0.21]{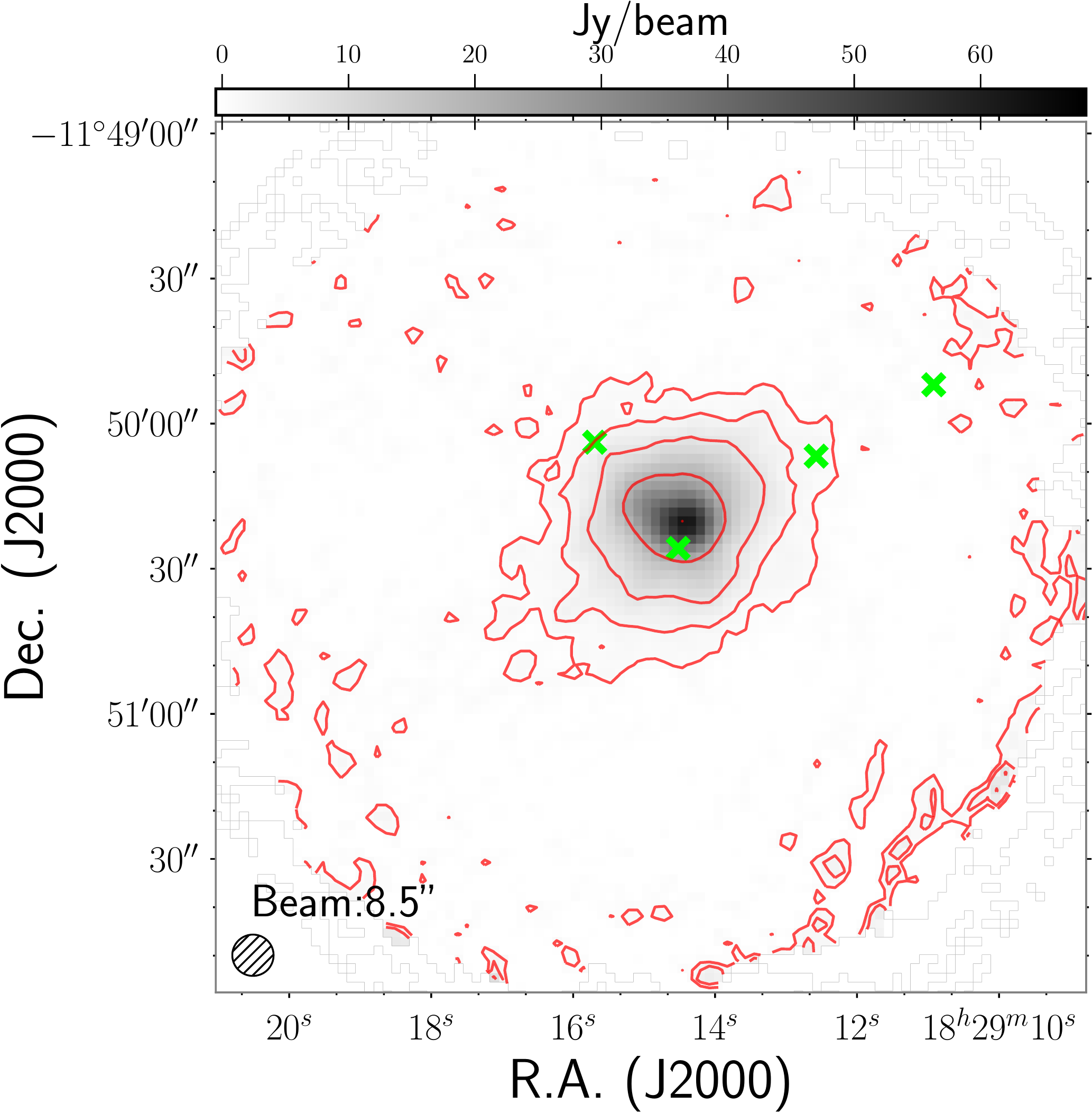}&\includegraphics[scale=0.20]{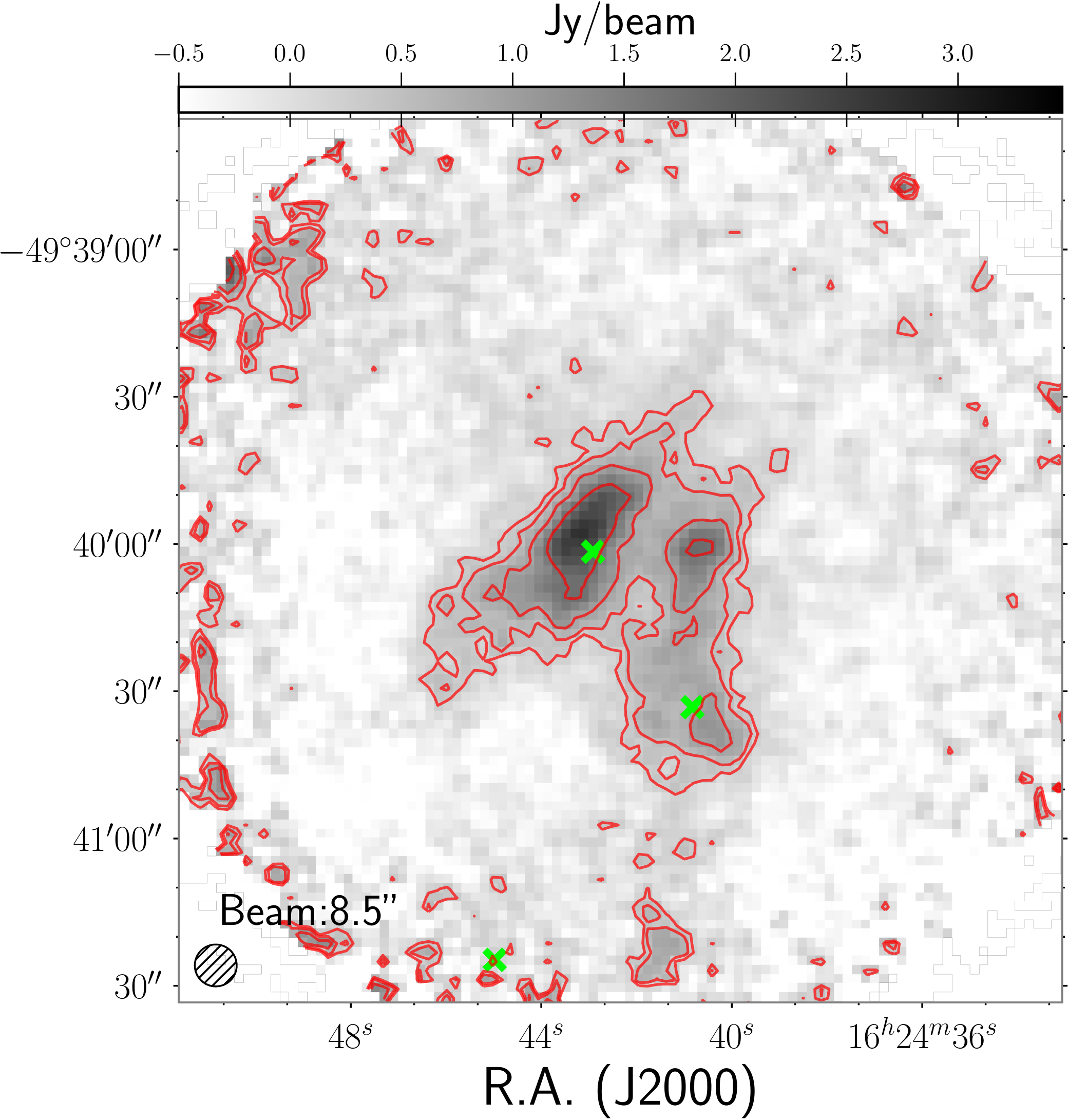}&\includegraphics[scale=0.20]{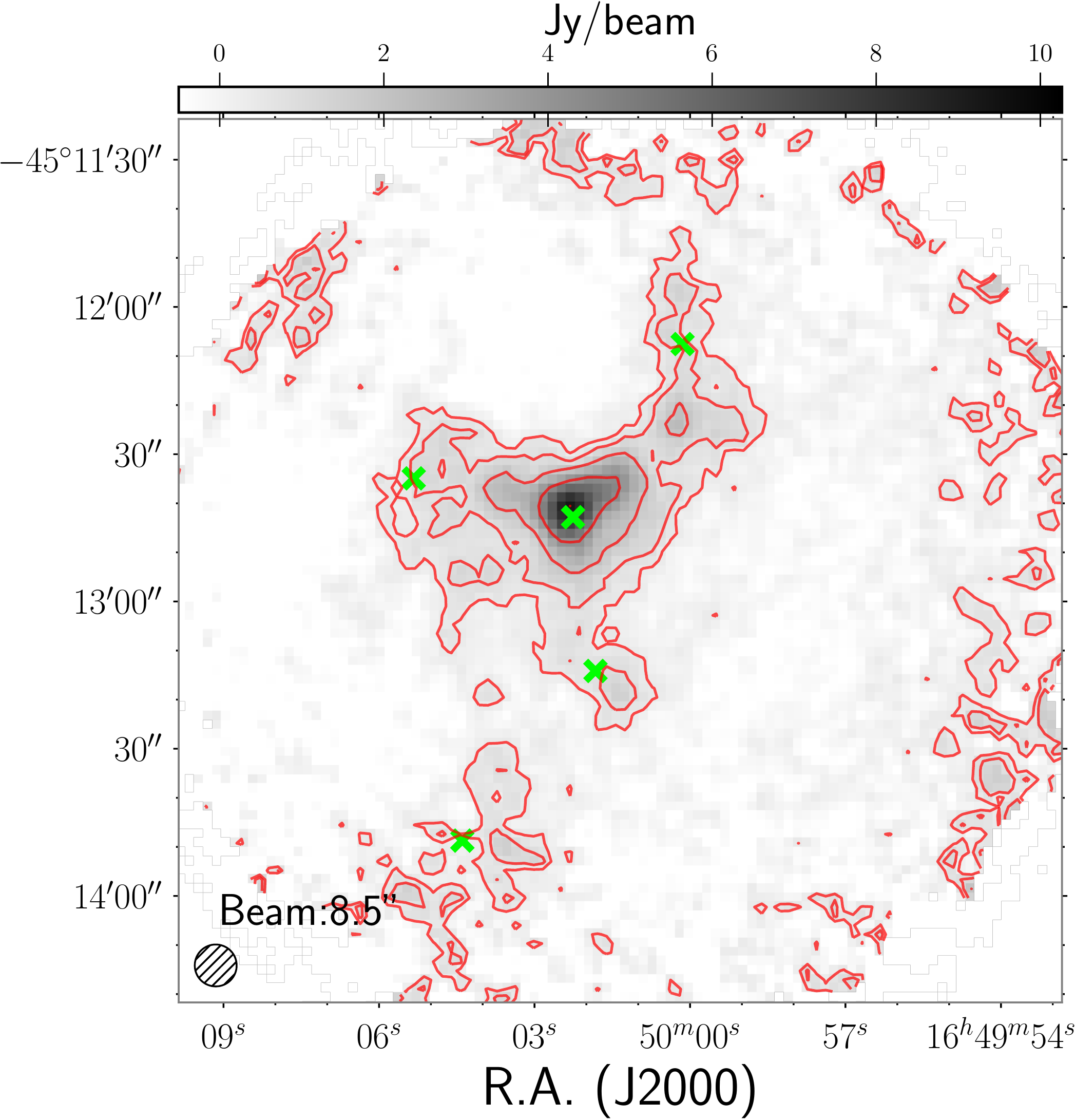}&\includegraphics[scale=0.20]{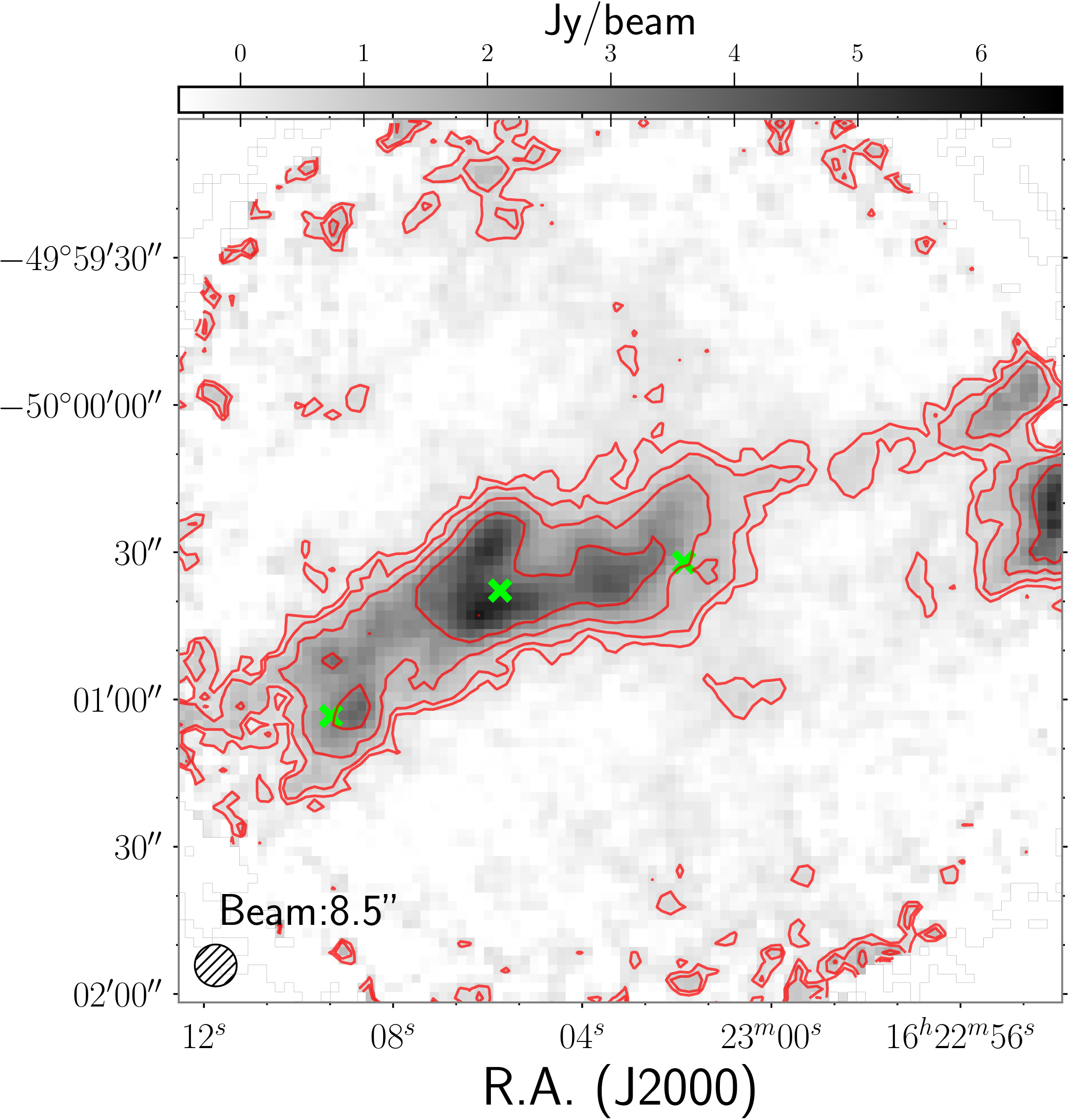}\hspace{-0.2cm}\\
\includegraphics[scale=0.245]{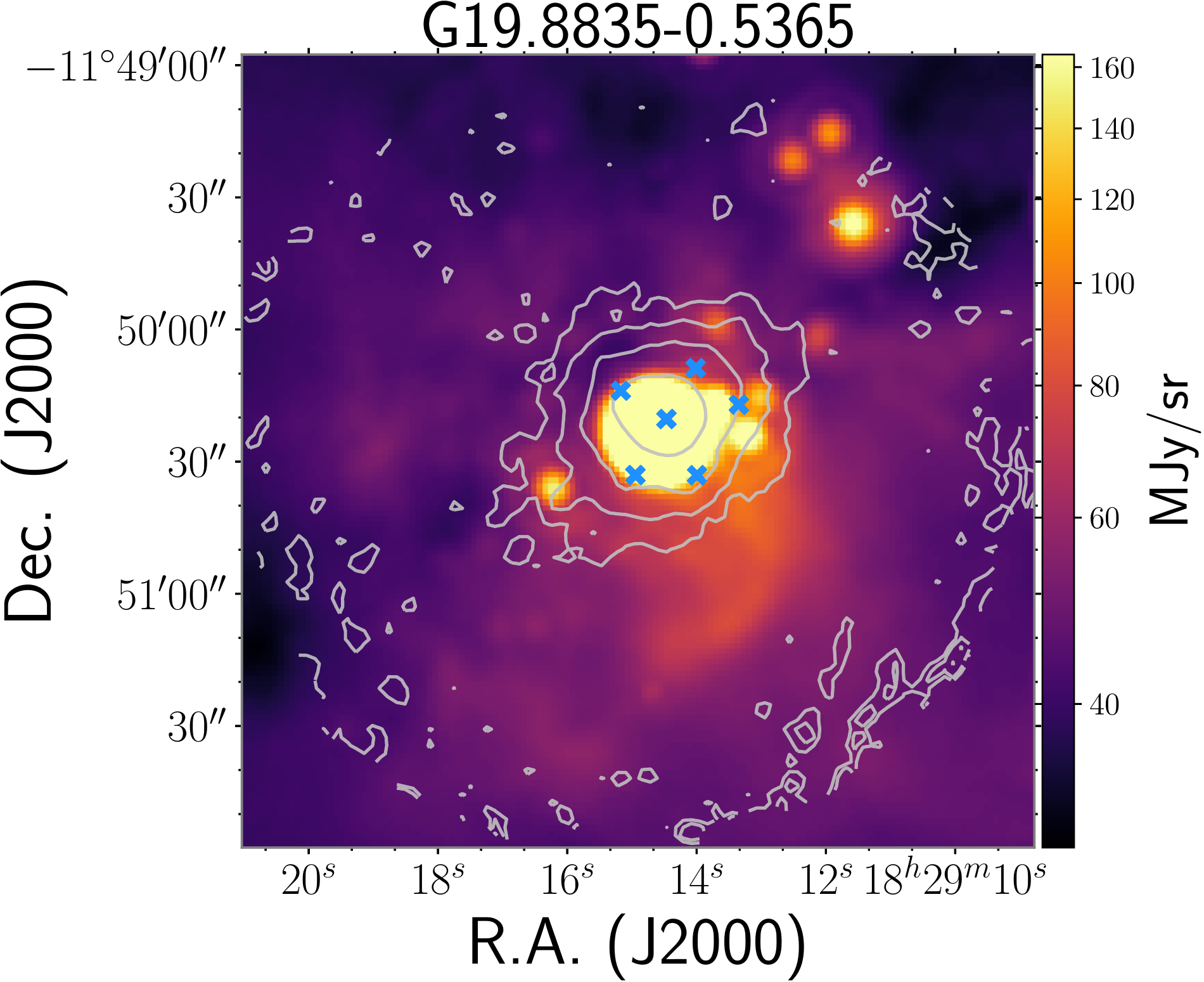}&\includegraphics[scale=0.245]{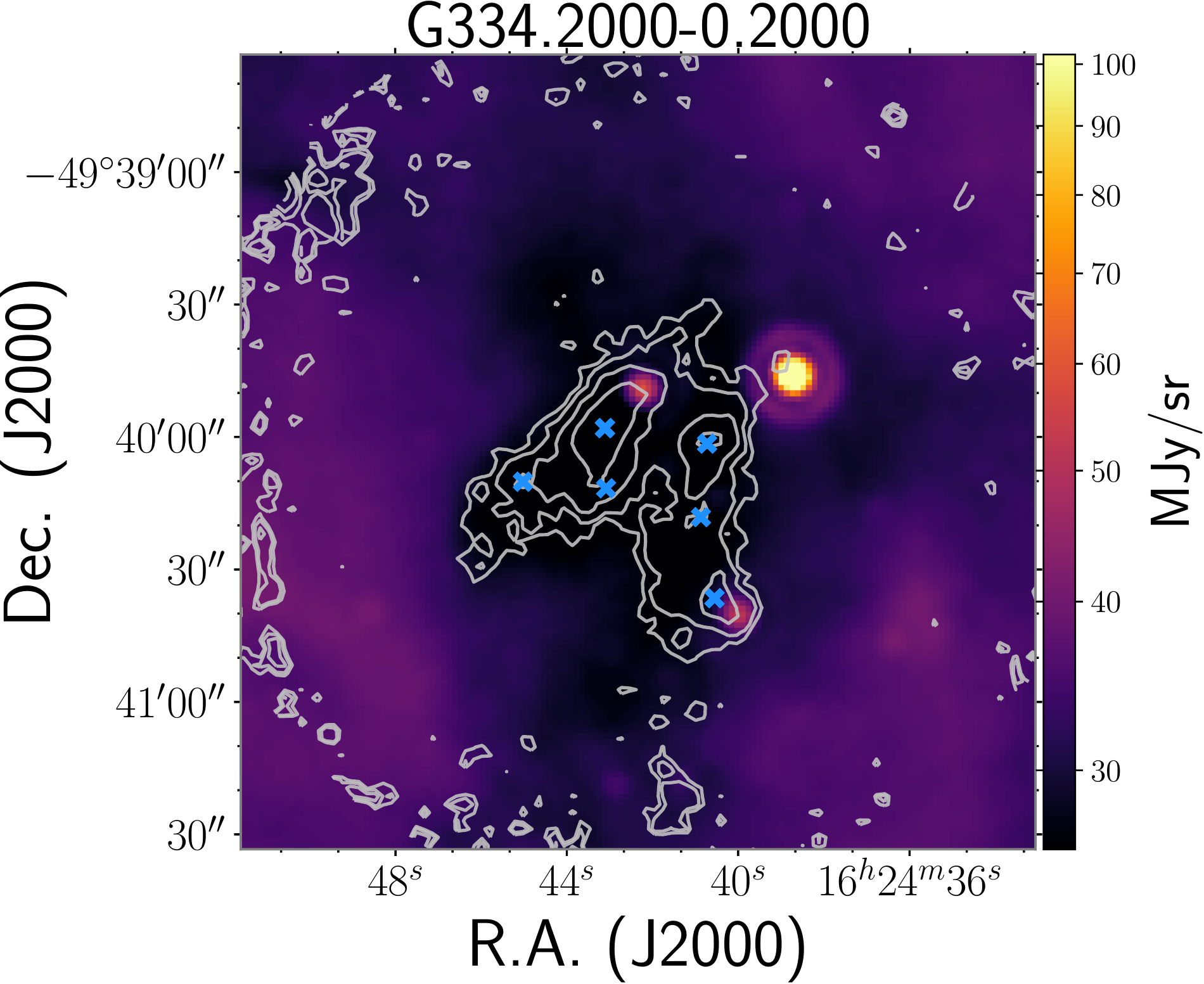}&\includegraphics[scale=0.245]{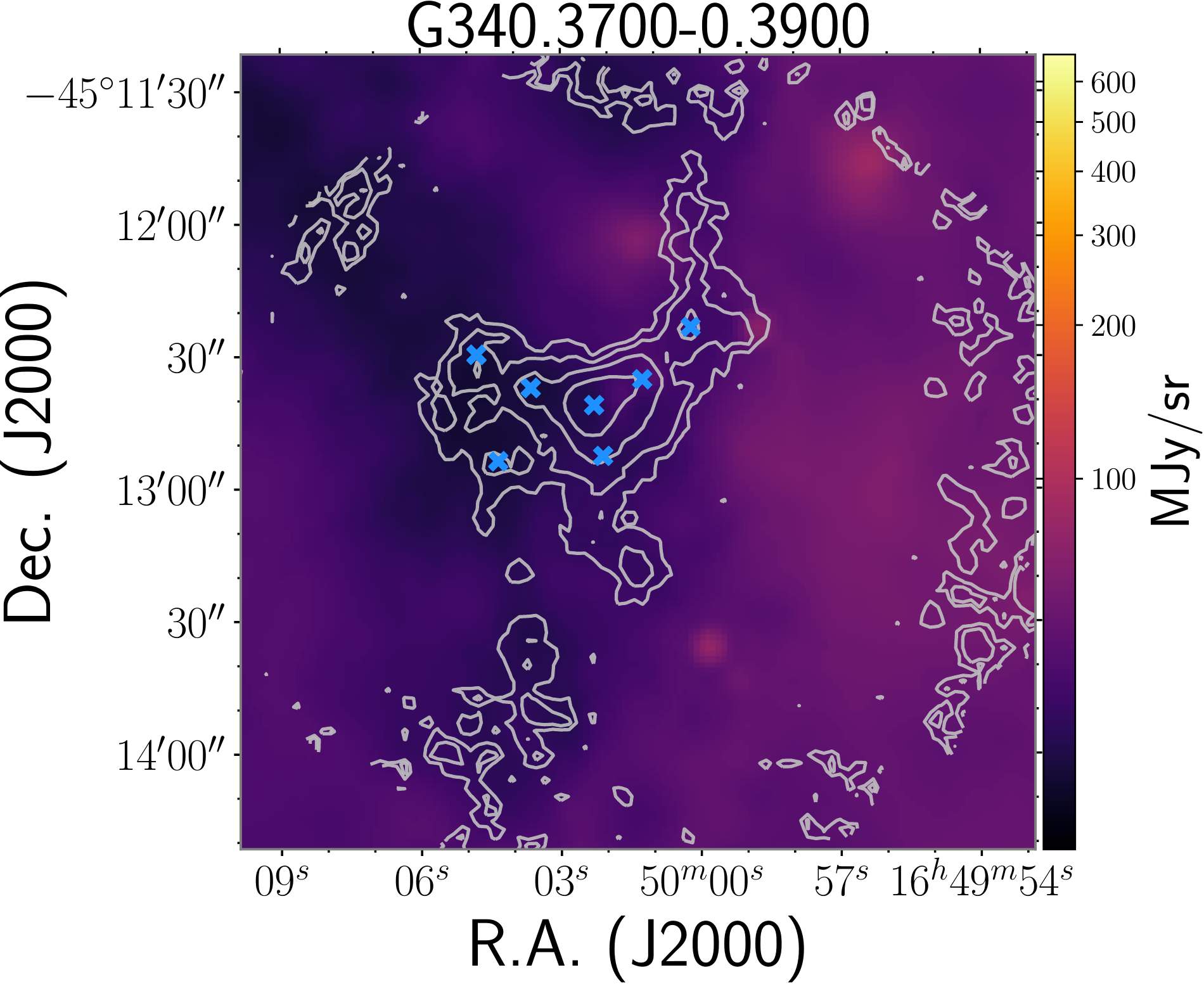}&\includegraphics[scale=0.245]{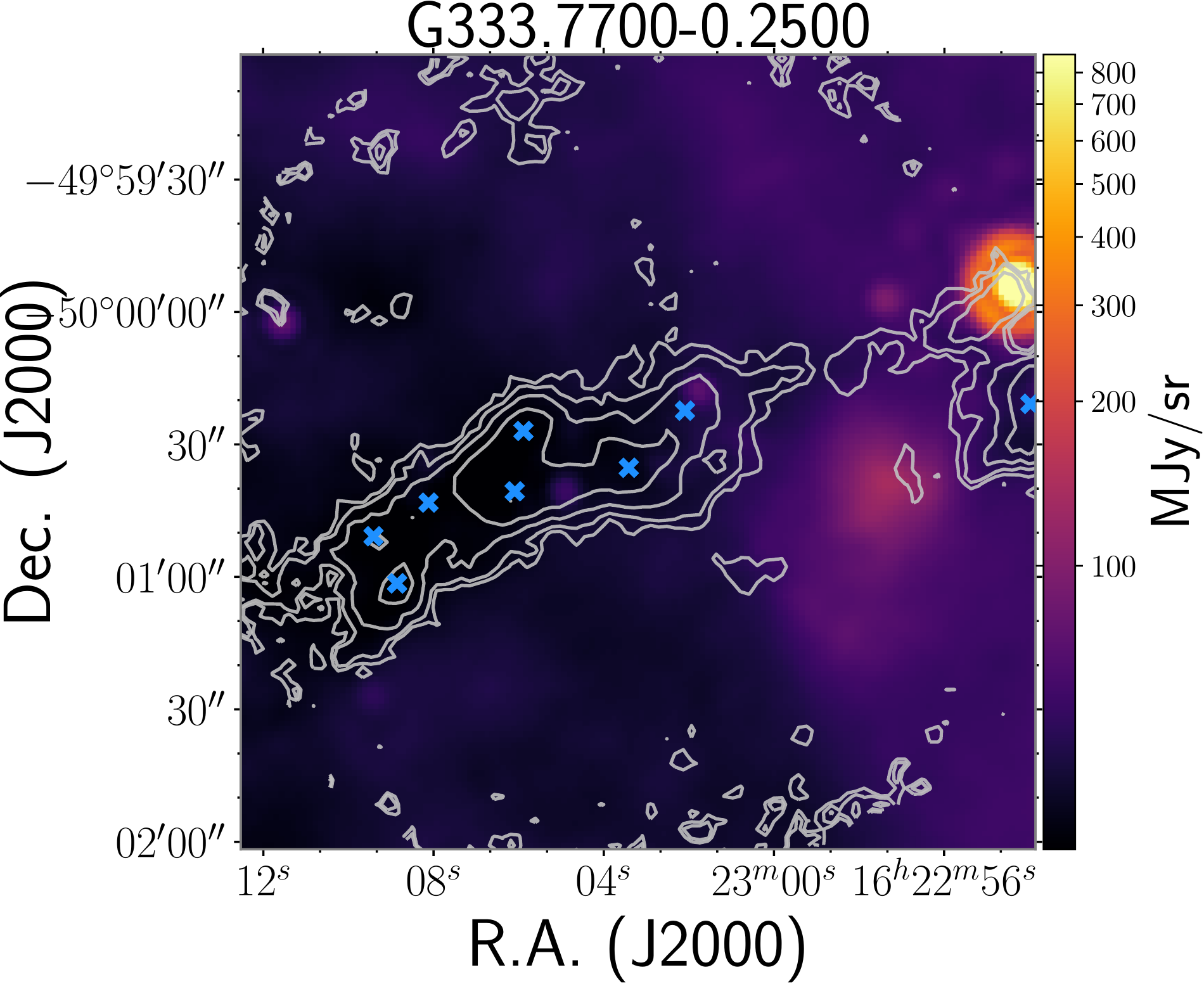}\\
\end{tabular}
\centering\caption{{\it{Top panels}}: Examples of 350\,$\mu$m emission towards ATLASGAL selected massive clumps obtained with APEX/SABOCA. Contour levels start at 5$\sigma$ and show 5 uniformly spaced intervals on a logarithmic scale up to the peak flux density in each field. The green crosses show the position of ATLASGAL sources. {\it{Bottom panels}}: Coloured images show the 24\,$\mu$m emission from MIPSGAL on a logarithmic scale, contours are the same as in the top panel of the 350\,$\mu$m emission. red crosses mark the positions of the compact sources identified in the SABOCA maps. The target name from the ATLASGAL Gaussclumps catalog is given in each plot. 
}
\label{fig:sab_cont}
\end{figure*}

We complement the HiGAL data with the combined ATLASGAL and {\it{Planck}}-HFI 870\,$\mu$m data (\citealt{Csengeri16b}). This data product has a considerably larger spatial dynamic range than the ATLASGAL data alone allowing the recovery of large-scale structures, which can be directly compared with the emission probed by Herschel. To associate the sub-millimetre emission with mid-infrared sources, we also make use of the MIPSGAL survey and its point source catalog (\citealt{Gutermuth15}).

\section{Results}\label{sec:results}

\subsection{The SABOCA 350 $\mu$m view of ATLAGSAL clumps: filaments and cores}\label{subsec:sabocamaps}

We show examples of massive clumps at 350\,$\mu$m observed by APEX/SABOCA in Fig.\,\ref{fig:sab_cont}, where we also directly compare them with the 24\,$\mu$m emission from the Spitzer/MIPSGAL survey probing their star formation activity at a comparable angular resolution of 6$\arcsec$. We detect emission at  350\,$\mu$m towards all fields, typically tracing emission from the ambient cloud material, as well as from compact sources. The images for all the targeted sources are shown in App. \ref{app:sab_maps_all}.  

The observed fields have a range of peak flux densities spanning two orders of magnitude at 350 $\mu$m with a typical dynamic range of $\sim$100, however, this is noticeably lower for fields containing brighter sources. The high dynamic range and improved resolution allows us to conduct a detailed analysis of the overall morphology of the sample in a homogeneous way for the majority of the sources; this is particularly true for the mid-infrared weak sources due to the better sensitivity and dynamic range.

As revealed in Fig.\,\ref{fig:sab_cont}, the morphology of the 350 $\mu$m emission exhibits a large variety including isolated compact spherical structures (Fig.\,\ref{fig:sab_cont}a) and prominent elongated structures, which are often referred to as filaments (Fig.\,\ref{fig:sab_cont}b, c, d). The structure of the source presented in Fig.\,\ref{fig:sab_cont}d reveals a relatively straight filamentary morphology, however,  we also find a network of connecting filamentary-like and branches, which are illustrated in Fig.\,\ref{fig:sab_cont}b, c, for example. These structures are reminiscent of the large-scale filamentary structure of the ISM (e.g. \citealt{Li16}). Visual inspection of all the maps in App. \ref{app:sab_maps_all} indicates that the majority of the fields exhibit elongated, filamentary(-like) emission (67$\%$). This either corresponds to low-intensity uniformly distributed material or consists of brighter emission with several compact sources associated with it. These sources are marked in Table\,\ref{tab:gauss} together with the properties of the identified substructures (see more detail in Sec.\,\ref{subsec:ppt}). In general, fields dominated by a single, bright source at 350\,$\mu$m represent only a minor fraction of the sample.

\subsection{Extraction of compact sources}\label{subsec:extract}

We use the Gaussian decomposition algorithm \texttt{Gaussclumps} to identify compact structures embedded within the ATLASGAL clumps.\footnote{This is implemented in the GILDAS software package that can be found at http://www.iram.fr/IRAMFR/GILDAS}. Originally developed for identifying structures in 3D, position-position-velocity data cubes, \texttt{Gaussclumps} assumes that the resolved structures are Gaussian-shaped and iteratively subtracts the fitted structures from the map (\citealt{Stutzki90}). \texttt{Gaussclumps} is the same tool used to extract the compact sources from the ATLAGSAL survey (\citealt{Csengeri14}).

We set the initial guesses of aperture cutoff and aperture \emph{$FWHM$} to $8\as5$ and 1.1 times the angular resolution respectively. The initial guess for the source \emph{$FWHM$} is also set to 1.1 times the angular resolution, and we require the identified substructure by \texttt{Gaussclumps} to have a peak flux larger than 3 times the rms. There are several other control parameters in the least-square fitting algorithm. The stiffness parameters s$_{0}$, s$_{c}$ and s$_{a}$ adjust the tolerance between the observed intensity, maximum intensity and its position, respectively, with the corresponding fitted parameters. We performed a series of tests to see how the choices for the stiffness parameters affect the identified sources. We found that using 1, 1, 10 for s$_{0}$, s$_{c}$ and s$_{a}$  results in a more stringent constraint on the fitted peak flux position, and seems to more robustly recover the prominent structures while avoiding  the detection of spurious sources at relatively low signal-to-noise levels.

In total, we identify 1120 structures with \texttt{Gaussclumps} within our SABOCA maps, yielding on average 4-5 compact sources within the 4$'$ maps.  We list the parameters of the extracted compact sources in Table\,\ref{tab:gauss}. In the following, we use the nomenclature commonly adopted in the literature (e.g., \citealt{Williams00}, \citealt{Motte07}, \citealt{Zhang09}, \citealt{Liu12b, Liu12a}), where clumps refer to structures with sizes of $\sim$0.5-1 pc, and cores refer to $\sim$0.1\,pc structures embedded within a clump. Since most of our analysis and discussions are based on the distance limited sample, in the following we use the term \emph{cores} to refer to the compact sources extracted from the SABOCA 350\,$\mu$m observations, since the majority of them have a $FWHM$ size $\lesssim$0.2 pc. When discussing the properties of SABOCA 350\,$\mu$m compact sources together with clumps, we use the term \emph{fragments} instead, to indicate their relation.

\begin{table*}
\centering
\begin{threeparttable}
\caption{Properties of SABOCA sources from {\it{Gaussclumps}} source extraction.}
\label{tab:gauss}

\begin{tabular}{llllllllll}
\toprule
        \it{SABOCA} name &RA (J2000)&Dec (J2000)&$FWHM_{maj}$ & $FWHM_{min}$ &     PA & $FWHM_{g}$ & S$_{peak}$ & S$_{int}$ &Filament \\
     &(h m s) &(d m s) & ($''$)& ($''$)&($^{\circ}$)& ($''$)&(Jy/beam)&(Jy)&flag\\
\midrule
 GS301.1338$-$0.2285 &  12:35:33.64 &  -63:02:41.2 &          25.2 &          13.7 &  147.4 &        18.6 &       28.1 &     138.8 &           $-$ \\
 GS301.1350$-$0.2220 &  12:35:34.51 &  -63:02:18.2 &          21.6 &           9.7 &  245.0 &        14.5 &       40.0 &     120.3 &           $-$ \\
 GS301.1364$-$0.2252 &  12:35:35.11 &  -63:02:29.7 &          13.0 &          11.9 &   55.4 &        12.4 &      236.5 &     521.6 &           $-$ \\
 GS301.1400$-$0.2233 &   12:35:37.00 &     -63:02:02.0 &          23.8 &          13.7 &  159.7 &        18.0 &       33.3 &     155.3 &           $-$ \\
 GS302.1493$-$0.9497 &  12:44:20.68 &  -63:48:38.5 &          12.2 &          11.2 &   65.7 &        11.7 &        5.2 &      10.1 &           $-$ \\
 GS303.1176$-$0.9717 &  12:53:07.36 &  -63:50:34.0 &          12.6 &          10.4 &  180.6 &        11.5 &        9.6 &      18.1 &           $-$ \\
 GS303.9991$+$0.2799 &  13:00:42.43 &  -62:34:21.3 &          11.5 &           9.7 &   60.6 &        10.6 &        4.5 &       7.1 &           $-$ \\
 GS304.7120$+$0.6001 &  13:06:43.36 &  -62:13:07.6 &          11.5 &           9.4 &   95.0 &        10.4 &        1.7 &       2.6 &           $-$ \\
 GS305.0947$+$0.2501 &  13:10:13.03 &  -62:32:33.7 &          20.2 &          10.4 &  101.9 &        14.5 &        3.1 &       9.3 &             Y \\
 GS305.0997$+$0.2484 &  13:10:15.67 &  -62:32:38.7 &          13.0 &          10.1 &  129.8 &        11.4 &        2.0 &       3.8 &           $-$ \\
 \bottomrule
\end{tabular}
    \begin{tablenotes}
      \small
      \item \textbf{Notes.} The columns are defined as follows: Name: the SABOCA catalog source name. RA: Right ascension of the source center. Dec: Declination of the source center. $FWHM_{maj}$ and $FWHM_{min}$: Major and minor $FWHM$ size of the source. PA: Position angle of the Gaussian structure measured from north to east. $FWHM_{g}$: Geometrical mean of the source size calculated by $\sqrt{FWHM_{maj}\times FWHM_{min}}$. S$_{peak}$: The fitted peak value of the Gaussian emission structure from SABOCA 350 $\mu$m map. S$_{int}$: The integrated source flux assuming a Gaussian distribution of flux calculated by S$_{peak}$$\times$$($FWHM$_{source}/$FWHM$_{beam})$$^{2}$. Filament flag: The SABOCA sources originate from filamentary(-like) clouds are marked with "Y".\\ 

     \item Full table will be available on CDS.

    \end{tablenotes}
  \end{threeparttable}
\end{table*}

\section{Physical properties of clumps and cores}\label{sec:properties}

\subsection{Dust temperature and H$_2$ column density maps on the clump-scale}\label{sec:clump_sed}

We performed a pixel-by-pixel fitting of the far-infrared spectral energy distribution (SED) to compute maps of the dust temperature, $T_{\rm_d}$, and the H$_2$ column density, $N(H_{\rm_2})$. For this, we used PACS 160 $\mu$m, SPIRE 250 $\mu$m, and 350 $\mu$m maps and the combined APEX/LABOCA-{\it{Planck}}-HFI 870 $\mu$m data. We first convolved the maps to a common resolution of $\sim$25\arcsec and then gridded the data on to the same pixel grid. We performed SED fit of each pixel, using a modified-blackbody model: 

\begin{equation}
I_{\nu} = B_{\nu}(T_{d})(1-e^{-\tau_{\nu}})
\end{equation}

\noindent where in $\tau_{\nu} = \mu m_{H}N(H_{2})\kappa_{\nu}$, m$_{H}$ is the mean molecular weight and we adopt a value of 2.8, $N(H_{\rm 2})$ is the gas column density. We adopted a dust opacity law of $\kappa_{\nu} = \kappa_{300\,\mu m}(\frac{\nu}{1000\,{\rm GHz}})^{\beta}$, where $\kappa_{300\,\mu m} = 0.1$\,cm$^{2}$\,g$^{-1}$ is the dust opacity per unit mass (gas and dust) at a reference wavelength 300\,$\mu$m and $\beta$ is fixed to 1.8, which is the average value found towards the Galactic plane (\citealt{Planck11}). These parameters have been used for the Herschel Gould Belt Survey (\citealt{Andre10}), the HOBYS (\citealt{Motte10}) and Hi-Gal  (e.g. \citealt{Elia17}) key programs. This dust emissivity relation yields $\kappa_{870\,\mu m} = 0.0147$\,cm$^{2}$\,g$^{-1}$, while the previous ATLASGAL studies, e.g. \citet{Csengeri14}, \citet{Konig17} use a  $\kappa_{870\,\mu m}$ of 0.0185\,cm$^{2}$\,g$^{-1}$, which corresponds to a factor of 1.25 difference in the mass estimates. In addition, for the clumps which have a fitted temperature larger than 45\,K we additionally include the PACS\,70\,$\mu$m measurements in the final fits since in this temperature regime inclusion of the shorter wavelength better constrains the fit. We note that there is emerging evidence that $\beta$ and $T_{\rm_d}$ are anti-correlated (e.g. \citealt{Juvela13}), which means fixing $\beta$ to a certain value will cause an uncertainty of $T_{\rm_d}$ which propagates into the mass estimate. More sophisticated modelling methods are required to accurately recover this anti-correlation (\citealt{Juvela13}; \citealt{Galliano18}) by properly taking into account observational uncertainties, line-of-sight mixing of different components, etc. In addition, the dust opacity (reference value) can also increase in dense and cold cloud, due to grain growth (dust coagulation, e.g. \citealt{OH94}, \citealt{Planck11},\citealt{Ysard13}). In general, the mass estimates may have an uncertainty up to a factor of 2-3 due to possible variations of the dust opacity and emissivity index $\beta$.

\begin{figure*}[htb!]
 \begin{tabular}{p{0.49\linewidth}p{0.49\linewidth}}
\hspace{-1.2cm} \includegraphics[scale=0.36]{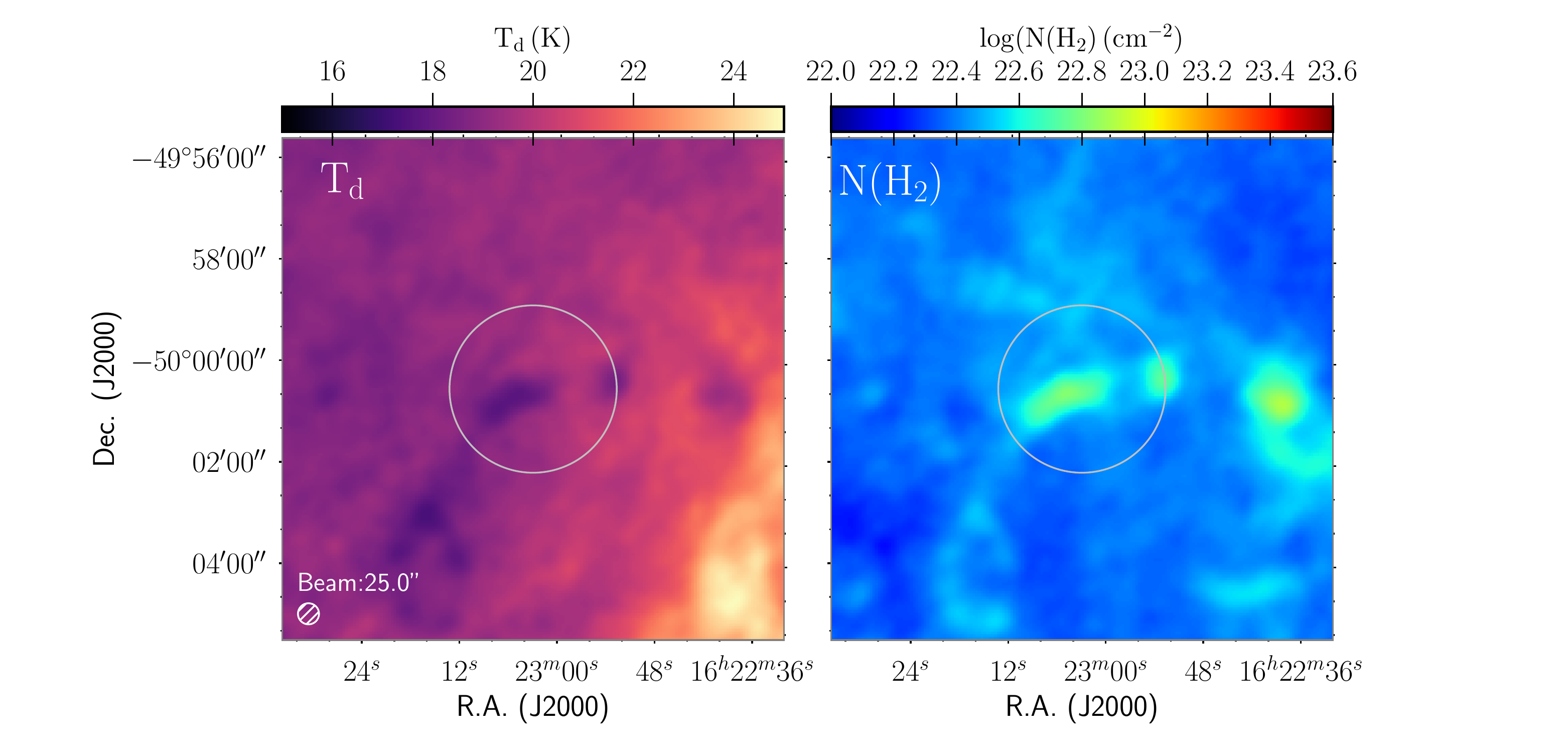}&\hspace{-0.7cm}  \includegraphics[scale=0.36]{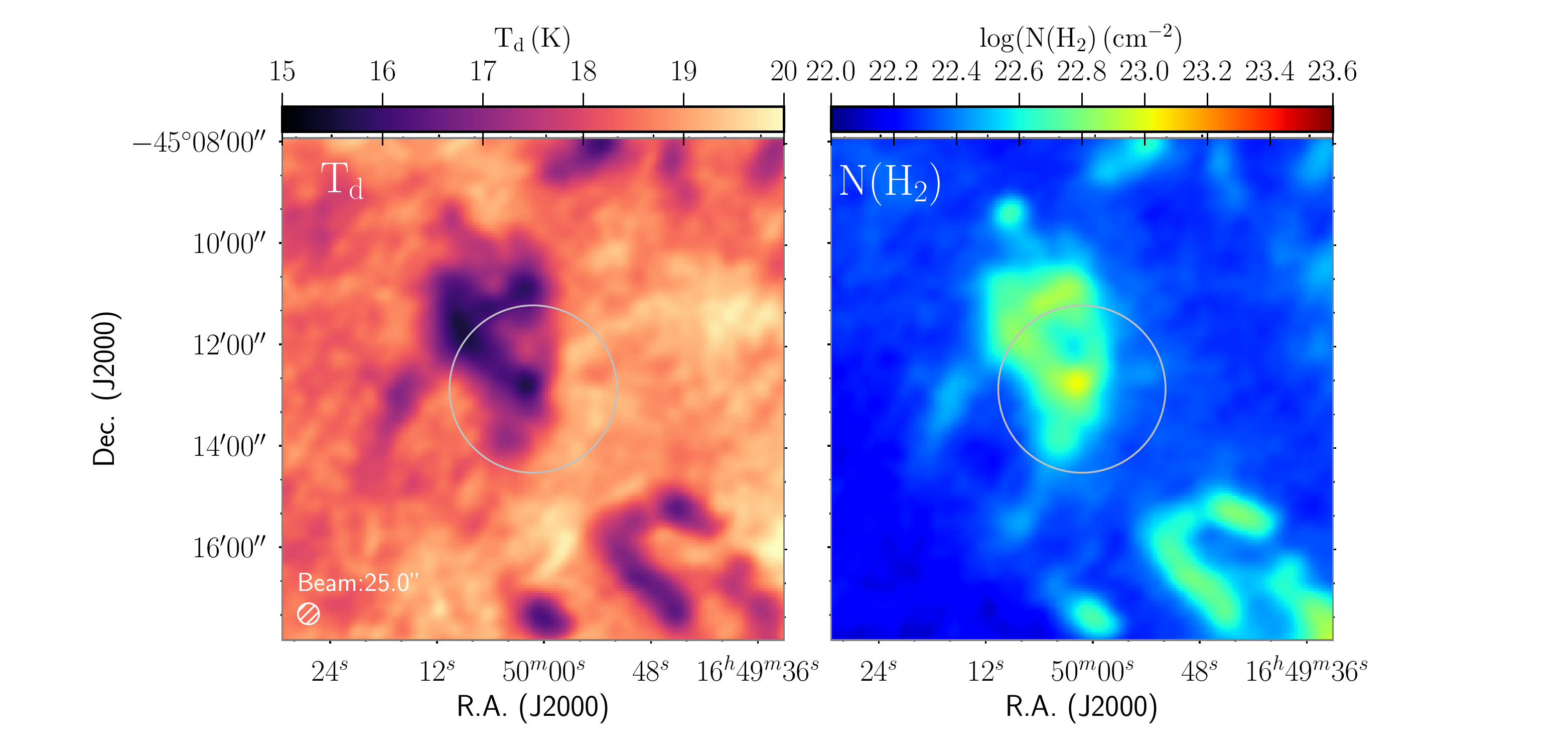}\\
\hspace{-0.8cm}  \includegraphics[scale=0.32]{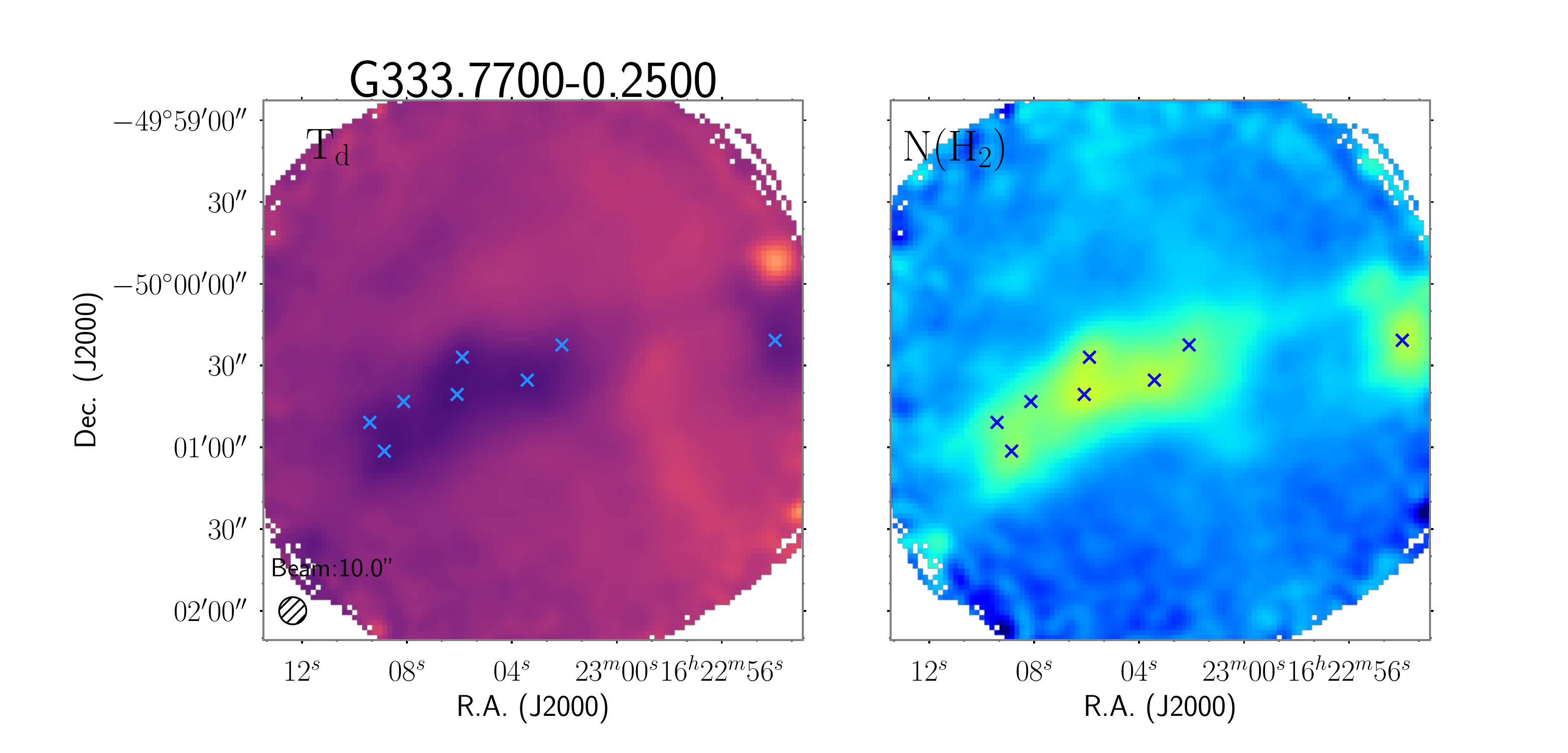}&\hspace{-0.2cm}  \includegraphics[scale=0.32]{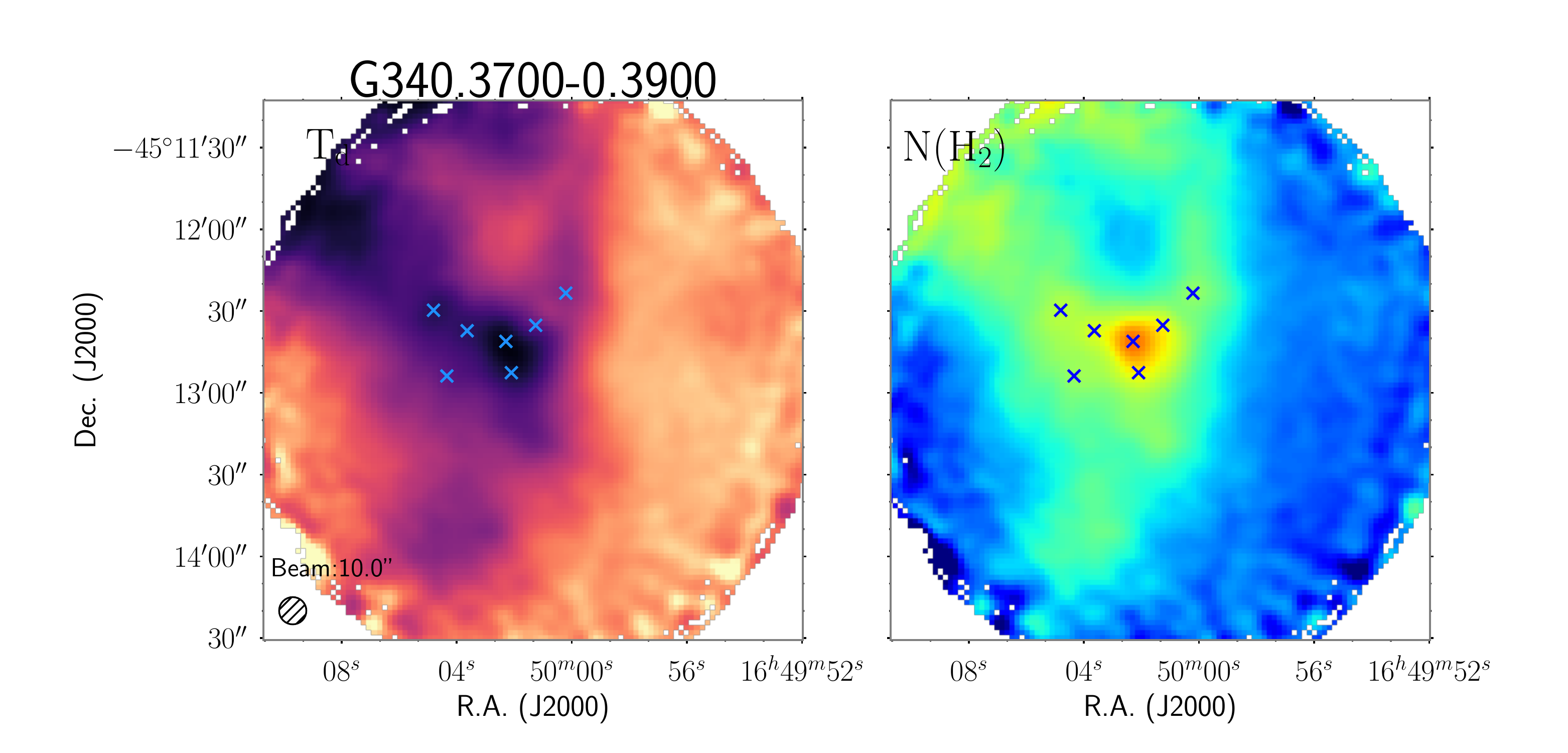}\\

 \end{tabular}
 \caption{{\it{Top panels}}: Dust temperature ($T_{\rm d}$) and H$_2$ column density ($N_{\rm H_2}$) maps obtained with a pixel-by-pixel SED fitting of PACS\,160\,$\mu$m, SPIRE\,250, 350\,$\mu$m and combined LABOCA and \emph{Planck}\,870\,$\mu$m data at 25$\arcsec$ resolution. {\it{Bottom panels}}: As top panels but using the PACS\,70\,$\mu$m and the SABOCA and SPIRE\,350\,$\mu$m combined maps at 10\arcsec\ resolution. Top panel and lower panel are in same color-scale for dust temperature and column density maps, respectively. The circles in the top panel indicate the region mapped by SABOCA.}
 \label{fig:eg_T_N}
\end{figure*}

Using ATLASGAL and Herschel data, \citet{Konig17} determined the properties of 109 ATLASGAL selected massive clumps from SEDs of mid- to far-infrared data. That study is based on the clump averaged integrated properties, while here we map the spatial distribution of the physical properties (e.g. dust temperature and gas column density). Examples of the resulting dust temperature and column density maps on the 25$''$ grid are shown in the top panels of Fig.\,\ref{fig:eg_T_N}.

\subsection{Dust temperature and H$_2$ column density maps on the core-scale}\label{sec:core_sed}

\begin{figure}
\includegraphics[scale=0.395]{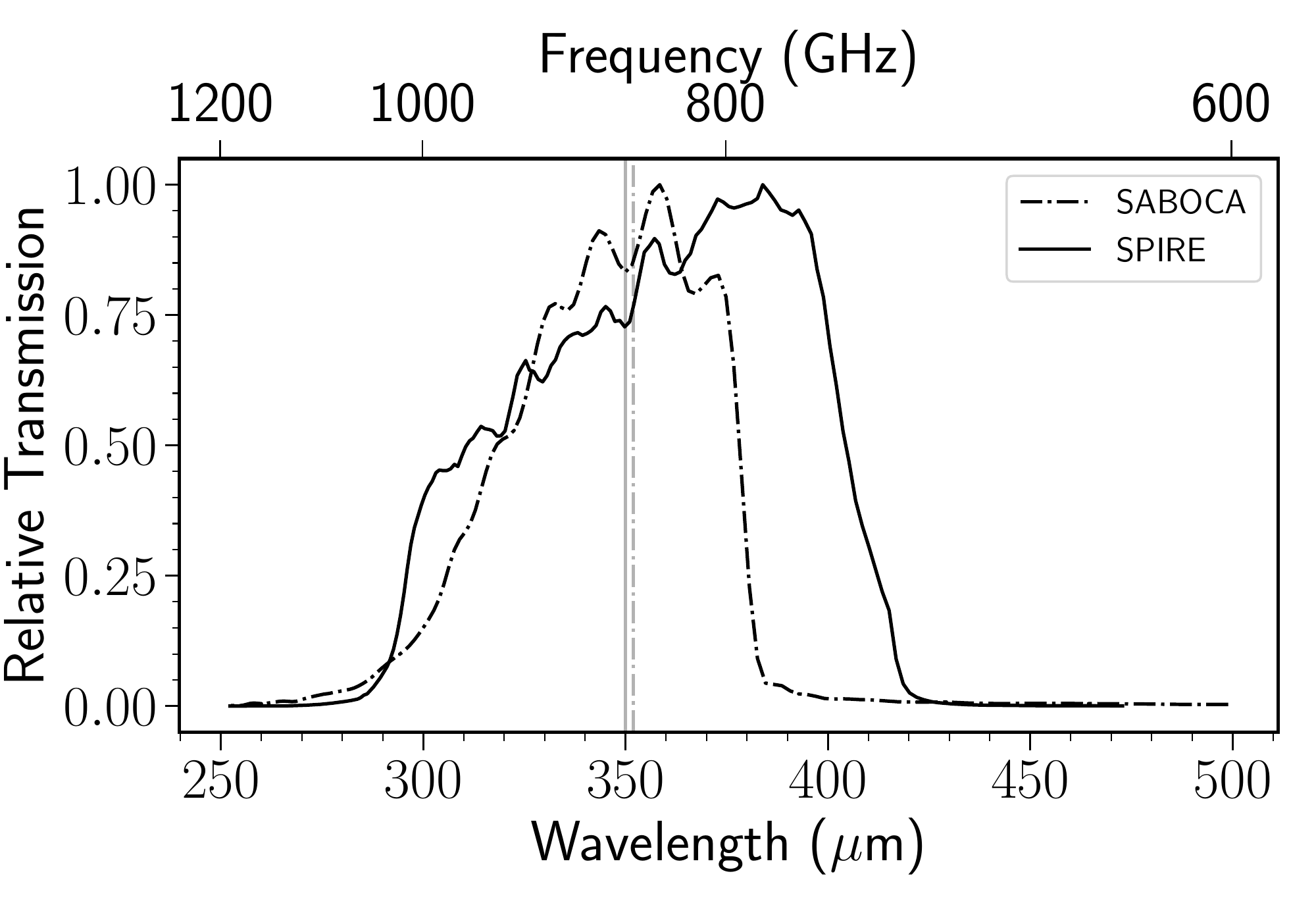}
\includegraphics[scale=0.395]{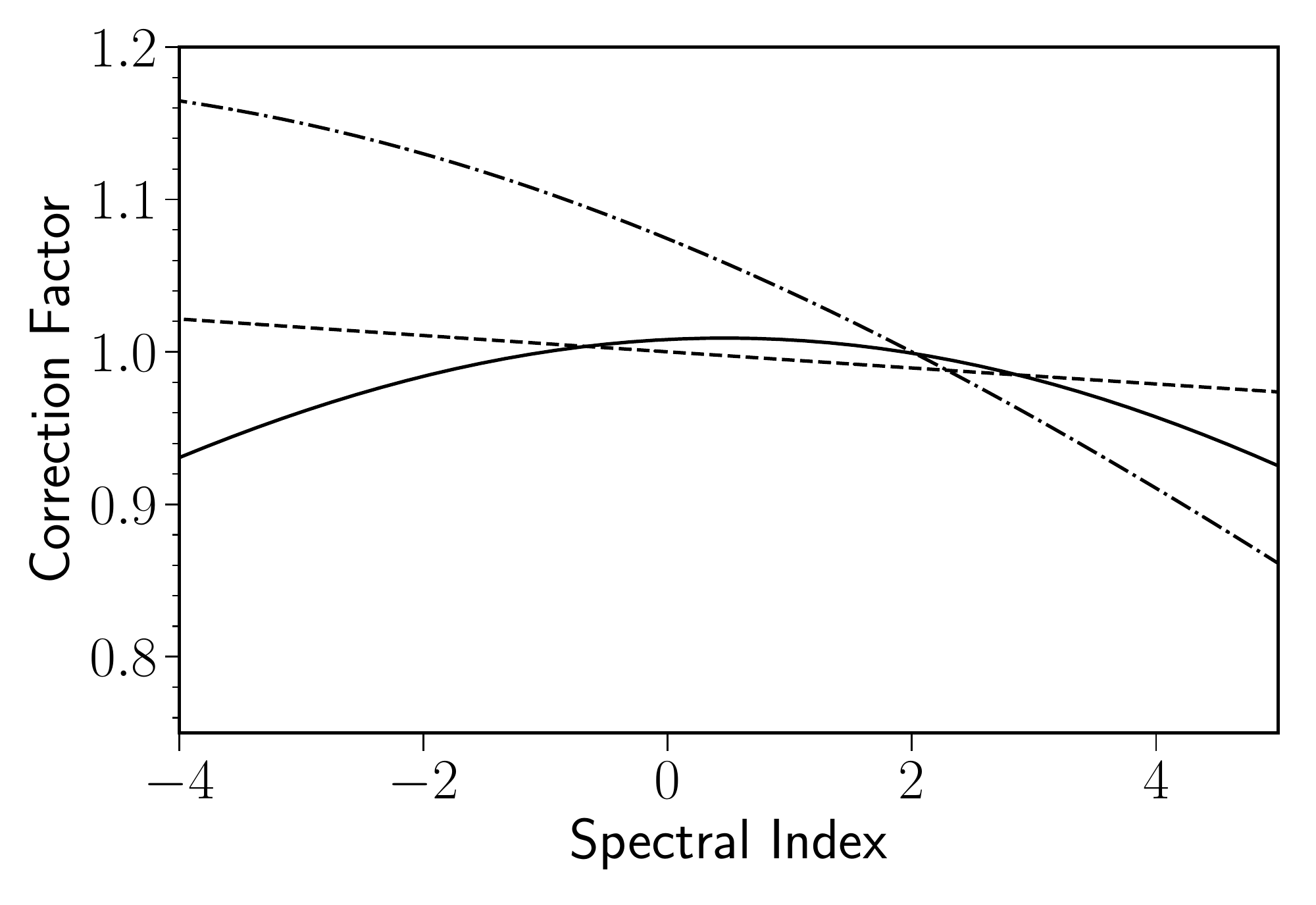}
\caption{{\it{Top panel}}: Transmission of the {\it{Herschel}}/SPIRE  and the APEX/SABOCA instruments. {\it{Bottom panel}}: Color correction factors for SABOCA (dashed line) and SPIRE (solid line) as a function of the spectral index of the source. The dash-dotted line shows the conversion factor to scale the SPIRE intensities to the nominal center wavelength of SABOCA.}
\label{fig:transmission}
\end{figure}

To obtain the same parameters at the scale of cores, we make use of the high-resolution SABOCA observations for the SED fitting. Since ground-based submillimeter observations are not sensitive to extended emission due to sky noise subtraction, the SABOCA maps do not recover emission from scales larger than $\sim$0.9-1.0$'$ ($\sim$ 60-70$\%$ instrument's field-of-view of SABOCA) (\citealt{Siringo09}). Since the Herschel SPIRE\,350\,$\mu$m covers a similar frequency range (Fig.\,\ref{fig:transmission}) we can use these data to recover the extended emission. The sampled angular scales of the two data sets have an overlap between the $\sim$25$''$ resolution Herschel 350\,$\mu$m and the largest angular scales recovered by SABOCA.

For the combination of the SABOCA and the SPIRE 350\,$\mu$m data, we first calculate the color corrections due to the different shape of the transmission curves and nominal frequencies of SABOCA and SPIRE instruments. Following a similar procedure described in \citet{Csengeri16a}, we determine the central wavelength correction factor from SPIRE to SABOCA (F = 0.97) and color correction factors for both instruments (C$\rm_{SPIRE}=0.93$, C$\rm_{SABOCA}=0.86$). These factors were applied to maps made with each instrument before the combination procedure. The transmission curves and the relation between the assumed spectral index and the correction factors are shown in Fig.\,\ref{fig:transmission}. 

An additional complication is that 10 of the brightest sources are saturated in the SPIRE\,350\,$\mu$m images. We correct for this by interpolating between the saturated pixels assuming a 2-dimensional Gaussian flux distribution around the center. We find that the interpolated images are, in general, consistent with the SABOCA images after they are smoothed to the SPIRE 350\,$\mu$m resolution of 25$''$. 
Finally, for all sources, we linearly combine the SABOCA and SPIRE\,350\,$\mu$m images in the Fourier domain using the {\tt{immerge}} task of the Miriad software package (\citealt{Sault95}). Before the combination, we calculate a weighting factor from the overlapping spatial scales between the two dataset, which is then applied to scale the SABOCA flux to match that of the SPIRE\,350\,$\mu$m measurements. For more details on the procedure of the combination, we refer to \citet{Lin16}.

We use these combined maps together with the PACS\,70\,$\mu$m band, at a similarly high resolution of $\sim$10$''$ after smoothing, to estimate the physical properties of the SABOCA sources, such as H$_2$ column density $N_{\rm H_2}$, mass, and average dust temperatures $T_{\rm d}$. We do this by fitting SEDs to each pixel in the maps to trace the local heating sources and estimate the core-scale temperature and column density. The formulations used are the same as when fitting the 4 bands at a coarser resolution (Sec.\,\ref{sec:clump_sed}). Assuming that the 4-band SED fits, in general, describe the bulk of the cold component in the coarser resolution maps, we scaled the 70\,$\mu$m flux densities in order to remove the contribution of a second, hotter gas component often seen towards deeply embedded protostars and massive young stellar objects (MYSO). The scaling factor is defined as the flux ratio of the observed 70\,$\mu$m emission map and the extrapolated 70\,$\mu$m flux based on the SED fit with the 4 bands. This step is necessary
because the majority of our sources are mid-infrared weak, thus dominated by cold dust which mainly emits at wavelengths longer than $\sim$100\,$\mu$m. Hence our method of using only the 70\,$\mu$m and the 350\,$\mu$m data would lead to an underestimate of core masses, with flux at 70\,$\mu$m predominantly originating from the deeply embedded heating sources.

We find that the scaling factors are generally less than 1.0 for most of our mid-infrared dark and relatively bright evolved sources. However, for the extremely bright OB cluster forming regions such as G10.6-0.4, G12.8-0.2 (ATLASGAL names: G010.6237-0.3833, G012.7914-0.1958) we find a ratio larger than one for the predicted 70\,$\mu$m flux compared to the observed flux density. These sources are at the edge of prominent compact\,70 $\mu$m emission features, and due to their more evolved nature, including the 70\,$\mu$m flux in the SED fits without any scaling provides a better constraint on their temperature fit (e.g. \citealt{Lin16}).  Therefore, for those cases where the scaling factor exceeds 1, we use the original 70\,$\mu$m fluxes. In the App.\,B we present two examples including one of these extremely bright OB clusters and the statistics of the scaling factors with the derived properties. The difference between the derived values using the original and scaled 70\,$\mu$m flux density in the SED fits are generally within 20$\%$ for temperature and 50$\%$ for column density depending on the flux scaling factor, which are discussed in detail in Appendix B.  The uncertainties of the fitted column density and dust temperature are within 20$\%$. Examples of the 10$\arcsec$ temperature and column density maps are shown in bottom panels of Fig.\,\ref{fig:eg_T_N}. Altogether we could perform a pixel-by-pixel SED fits for 198 of our maps, which includes 971 cores identified in the SABOCA maps.

\begin{figure*}[th!]
\begin{tabular}{p{0.45\linewidth}p{0.45\linewidth}}
\includegraphics[scale=0.44]{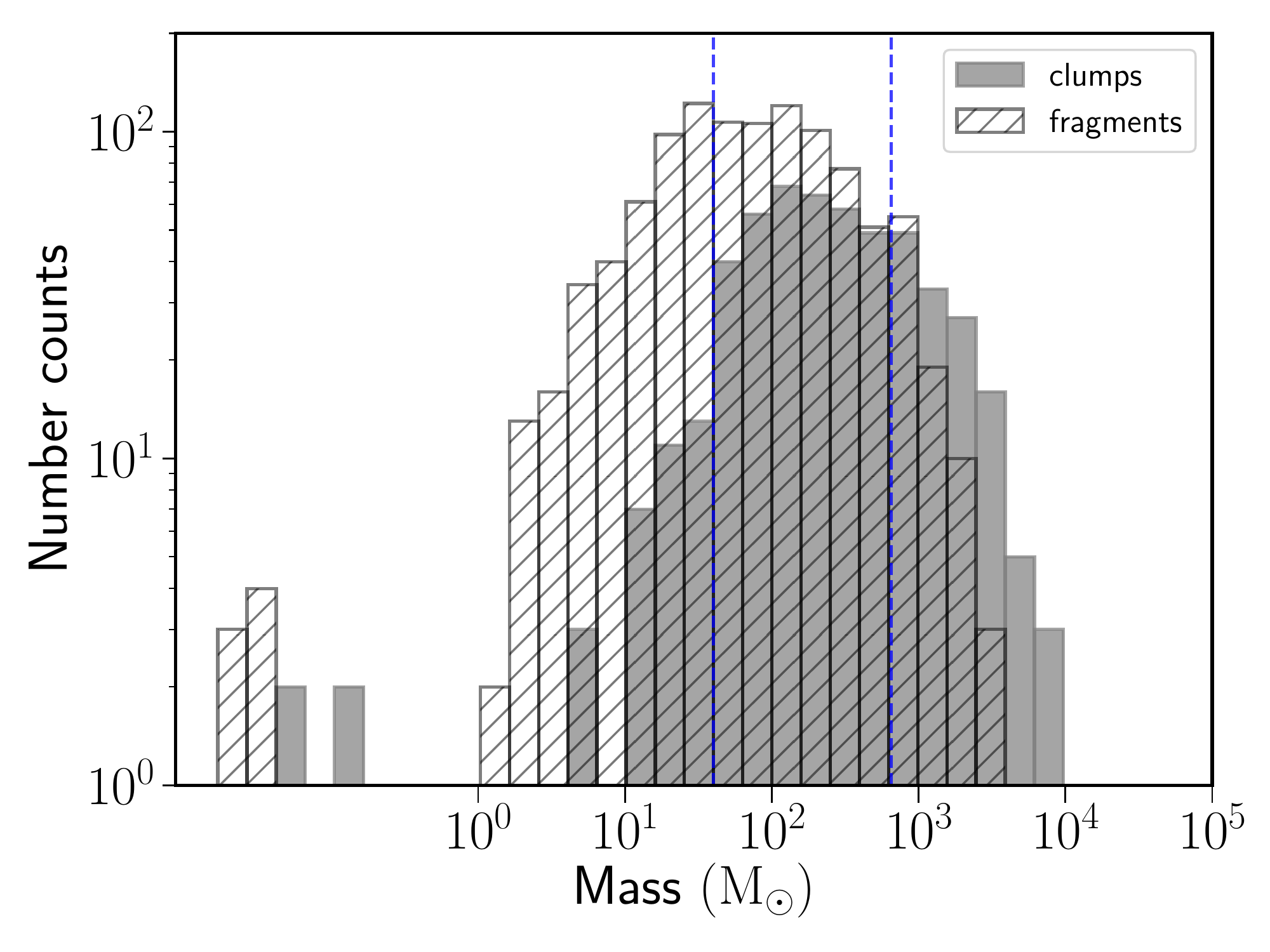}&\includegraphics[scale=0.44]{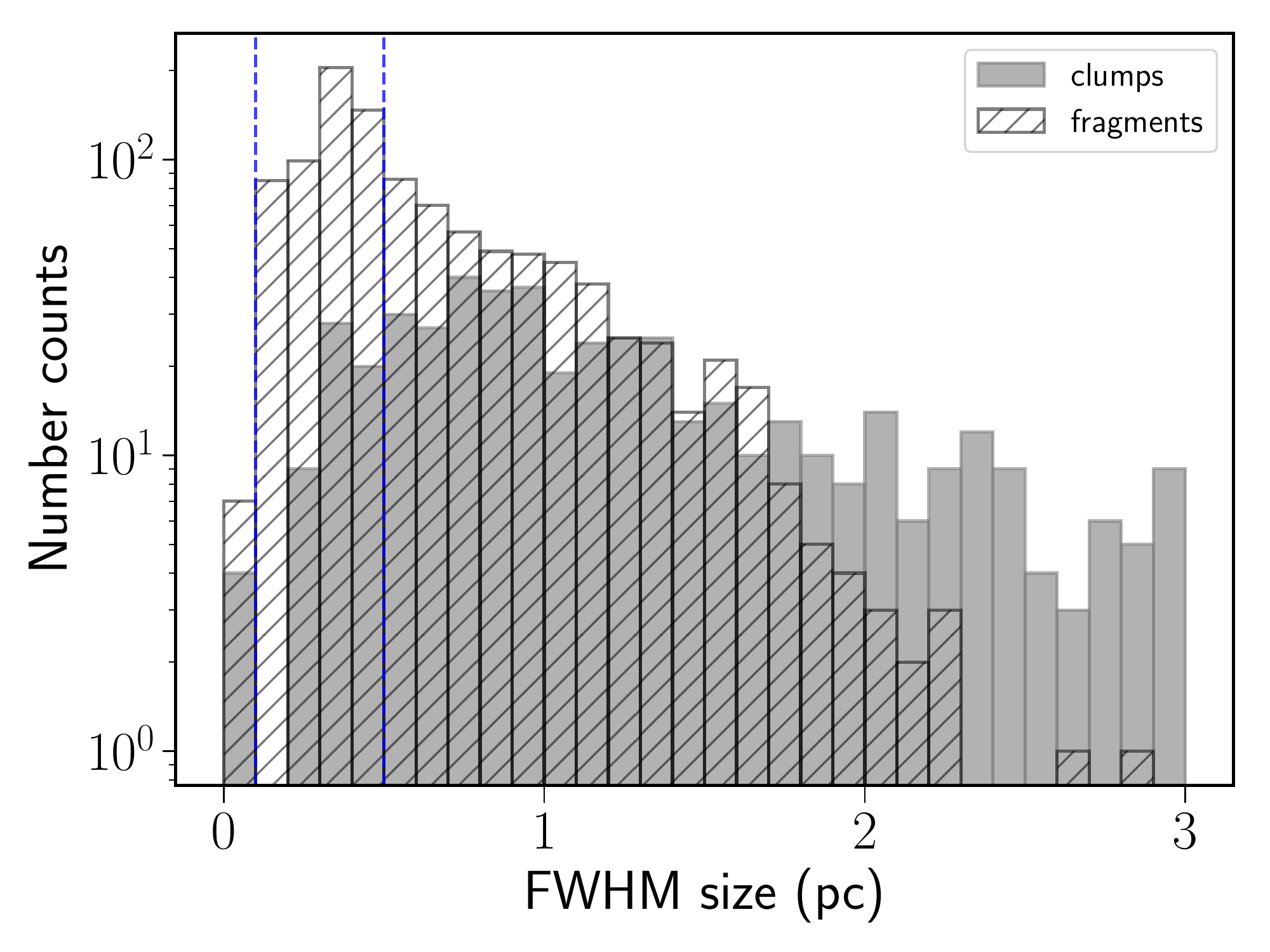}\\
\includegraphics[scale=0.44]{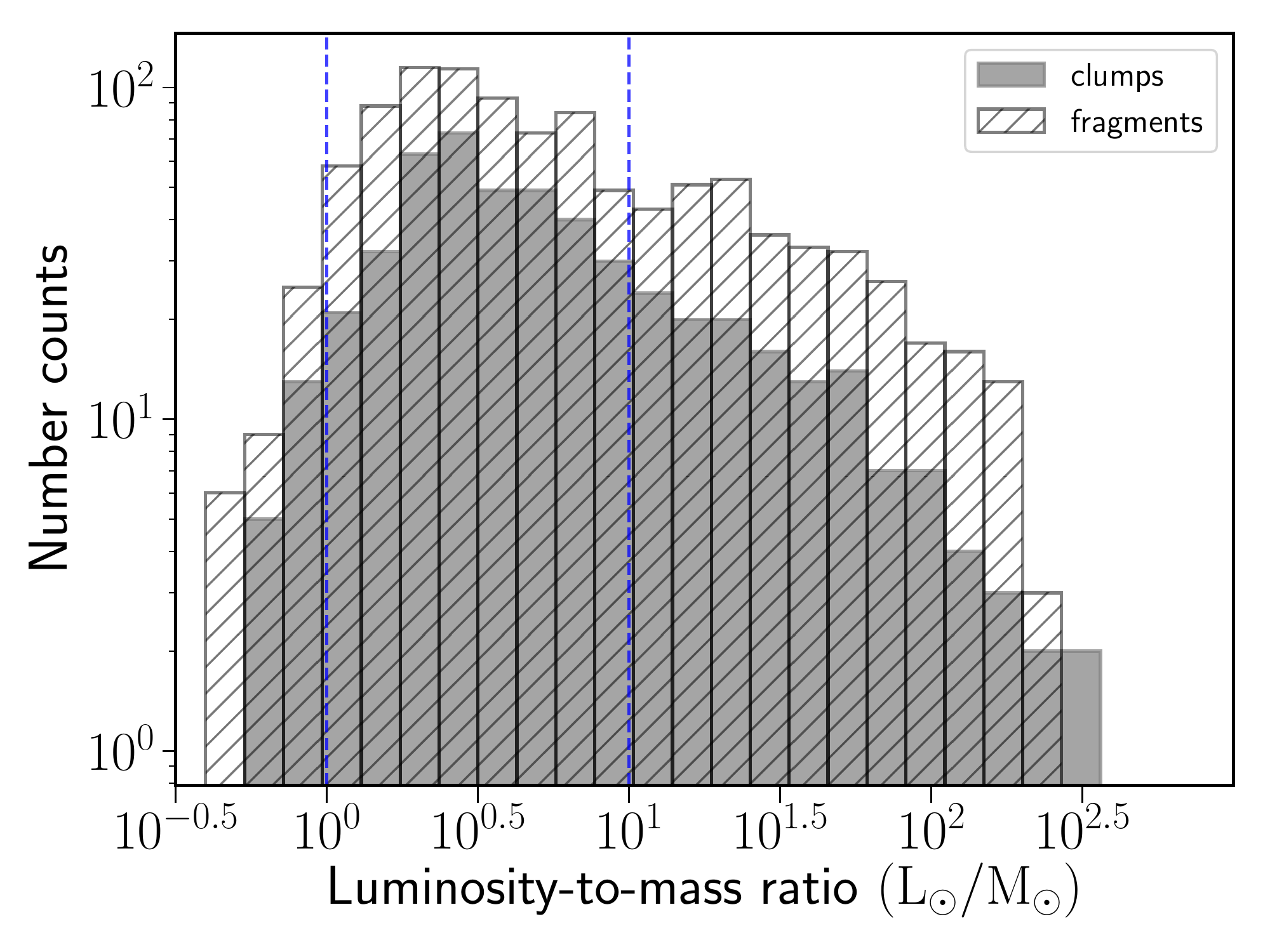}&\includegraphics[scale=0.44]{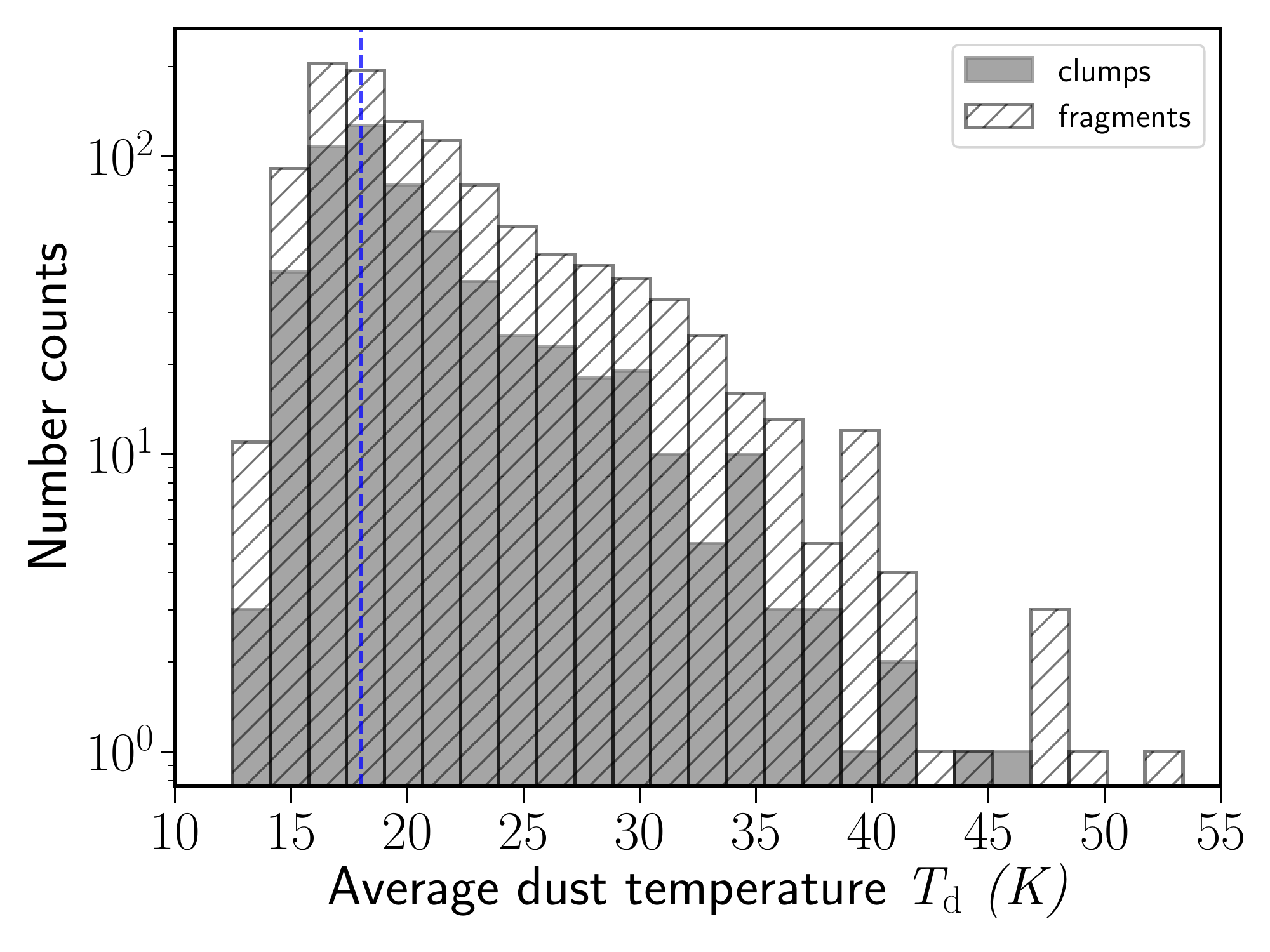}\\
\end{tabular}
\caption{Distribution of the physical properties of all the observed ATLASGAL clumps and compact sources identified at 350\,$\mu$m with SABOCA. {\it{Top left}}: The mass distribution of clumps and SABOCA compact sources. The vertical lines show mass limits of 40\,$M_{\odot}$ and 650\,$M_{\odot}$. {\it{Top right}}: Distribution of size which are the measured FWHM size from the {\it{Gaussclumps}} source extraction. Vertical lines show the 0.1 and 0.5\,pc range for definition of cores and clumps. {\it{Bottom left}}: The luminosity-to-mass ratio distribution of the clumps and SABOCA compact sources. The vertical dotted lines indicate luminosity-to-mass ratios of 1 and 10, which are taken as empirical evolutionary boundaries in \citet{Molinari16} for massive clumps. {\it{Bottom right}}: The temperature distribution with a vertical line showing 18 K, which is the average temperature for galactic background.}
\label{fig:clump_frag_hist}
\end{figure*}

\subsection{Statistics of the physical properties of ATLASGAL clumps and SABOCA cores}\label{subsec:ppt}

We derive the physical properties, such as average dust temperature, mass, bolometric luminosity of the identified SABOCA compact sources, and their ATLAGSAL host clumps, from the pixel-by-pixel SED fitting described in Sec.\,\ref{sec:core_sed}.

The average dust temperature is calculated from the mean pixel values within the identified Gaussian structures based on the $FWHM$s reported in \citet{Csengeri14}. Similarly,  we calculate the clump masses by summing up the column densities ($N_{\rm{H_2}}$) of each pixel over the source area at distance, $d$ (see Sec.\,2.1), 

\begin{equation}\label{eq:eq_M}
M_{clump} = \mu m_{H}\overline{N_{\rm{H_2}}}Ad^{2}
\end{equation}

\noindent where A stands for the clump area defined by 
$\pi R^{2}$. The radius is defined as $R=\sigma$, where $\sigma = \frac{FWHM}{2\sqrt{2ln2}}$. We estimate the total bolometric luminosity by integrating the fitted SEDs, 

\begin{equation}\label{eq:eq_lbol}
 L_{bol} = 4\pi d^2 \int_0^{\infty} S_{\nu}d\nu \qquad
\end{equation}

\noindent where $S_{\nu}$ is derived by the fitted SED curve described in Sec.\,3.3.1, and is summed up within $R = \sigma$, as well. 

\begin{table*}
\centering

\begin{threeparttable}
\caption{ATLASGAL clump properties.}
\label{tab:clumps}
\begin{tabular}{rlrrrrllll}
\toprule
  ATLASGAL name & $FWHM_{maj}$ &  $FWHM_{min}$ &  S$_{int}$ (870\,$\mu$m) &   Distance&                 Mass &             Luminosity & Average $N(H_{\rm 2}$) & Average $T_{\rm d}$ \\
 &($''$)&($''$)&(Jy)& (kpc)&(M$_{\odot}$)&(L$_{\odot}$)&(cm$^{-2}$)&(K)\\
\midrule
 G300.9506$+$0.8944 &          70.0 &          22.0 &       4.2 &       3.3 &   308 &                    297 &  4.5 $\times$ 10$^{22}$ &  15.1 \\
 G301.1292$-$0.2278 &          50.0 &          19.0 &       2.1 &       4.6 &   301 &  1.7 $\times$ 10$^{4}$ &  4.8 $\times$ 10$^{22}$ &  30.7 \\
 G301.1365$-$0.2256 &          26.0 &          23.0 &      34.5 &       4.6 &   494 &  3.9 $\times$ 10$^{4}$ &  1.0 $\times$ 10$^{23}$ &  33.9 \\
 G301.1438$-$0.2227 &          42.0 &          19.0 &       2.1 &       4.6 &   260 &  7.1 $\times$ 10$^{3}$ &  6.1 $\times$ 10$^{22}$ &  27.1 \\
 G302.1487$-$0.9494 &          29.0 &          26.0 &       1.8 &      11.1 &   474 &  5.3 $\times$ 10$^{3}$ &  1.5 $\times$ 10$^{22}$ &  22.9 \\
 G303.1169$-$0.9724 &          24.0 &          20.0 &       1.8 &       1.6 &     6 &                    133 &  1.4 $\times$ 10$^{22}$ &  25.7 \\
 G303.9978$+$0.2804 &          28.0 &          19.0 &       0.8 &      11.6 &   417 &  3.2 $\times$ 10$^{3}$ &  1.9 $\times$ 10$^{22}$ &  21.5 \\
 G304.0209$+$0.2919 &          27.0 &          22.0 &       2.3 &      11.6 &   961 &  4.6 $\times$ 10$^{3}$ &  3.5 $\times$ 10$^{22}$ &  19.8 \\
 G304.7127$+$0.6005 &          29.0 &          22.0 &       0.8 &       2.3 &    24 &                     30 &  2.1 $\times$ 10$^{22}$ &  15.7 \\
 G305.0943$+$0.2510 &          60.0 &          20.0 &       2.1 &       6.0 &   378 &  1.4 $\times$ 10$^{3}$ &  2.3 $\times$ 10$^{22}$ &  18.9 \\
 \bottomrule
\end{tabular}
    \begin{tablenotes}
      \small
      \item \textbf{Notes.} The columns are defined as follows: ATLASGAL name: clump name from the ATLASGAL catalog. $FWHM_{maj}$ and $FWHM_{min}$: Major and minor $FWHM$ size of the source from \citet{Csengeri14}. Distance: distance of the source from Urquhart et al (2018). S$_{\rm int}$: the integrated flux density at 870\,$\mu$m. Mass, luminosity, average $N(H_2)$ and $T_{\rm d}$ are based on the SED fitting in Sec.\,\ref{sec:clump_sed}, calculated with 1$\sigma$ source size.\\

     \item Full table will be available on CDS.

    \end{tablenotes}
  \end{threeparttable}
\end{table*}

\begin{table*}

\begin{threeparttable}
\caption{Properties of SABOCA sources, their 24\,$\mu$m and 70\,$\mu$m associations and parental clumps.}
 \label{tab:cores}

\begin{tabular}{llllllllll}

\toprule
              Name & Distance & $M_{\rm 1\sigma}$  &    $T_{\rm d}$ & $N(H_{\rm2}$) &              $L_{\rm1\sigma}$ &  MIR type &Parental clump\\
                        & (kpc)&(M$_{\odot}$)&(K)&(cm$^{-2}$)&(L$_{\odot}$)&&\\
\midrule
\midrule
 GS301.1350$-$0.2220 &      4.6 &  208.0 &    31.1 &  1.56 $\times$ 10$^{23}$ &  9.87 $\times$ 10$^{3}$ &            24sat &  G301.1365$-$0.2256 \\
 GS301.1364$-$0.2252 &      4.6 &  162.3 &    38.9 &  1.66 $\times$ 10$^{23}$ &  2.40 $\times$ 10$^{4}$ &  22ct,70ct,24sat &  G301.1365$-$0.2256 \\
 GS301.1400$-$0.2233 &      4.6 &  217.2 &    29.1 &  1.05 $\times$ 10$^{23}$ &  7.94 $\times$ 10$^{3}$ &            24sat &  G301.1438$-$0.2227  \\
 GS302.1493$-$0.9497 &     11.1 &   96.0 &    26.0 &  1.90 $\times$ 10$^{22}$ &  2.23 $\times$ 10$^{3}$ &  22ct,70ct,24sat &  G302.1487$-$0.9494  \\
 GS303.1176$-$0.9717 &      1.6 &    1.9 &    30.1 &  1.92 $\times$ 10$^{22}$ &  1.04 $\times$ 10$^{2}$ &  22ct,70ct,24sat &  G303.1169$-$0.9724  \\
 GS303.9991$+$0.2799 &     11.6 &  182.2 &    21.1 &  4.04 $\times$ 10$^{22}$ &  1.23 $\times$ 10$^{3}$ &        22ct,70ct &  G303.9978$+$0.2804 \\
 GS304.7120$+$0.6001 &      2.3 &    5.2 &    15.5 &  3.07 $\times$ 10$^{22}$ &  6.19 $\times$ 10$^{0}$ &             70ct &  G304.7127$+$0.6005\\
 GS305.0947$+$0.2501 &      6.0 &  115.7 &    18.1 &  5.10 $\times$ 10$^{22}$ &  3.17 $\times$ 10$^{2}$ &              70q &  G305.0943$+$0.2510 \\
 GS305.0997$+$0.2484 &      6.0 &   54.4 &    18.9 &  3.86 $\times$ 10$^{22}$ &  1.95 $\times$ 10$^{2}$ &              70q &                 $-$ \\
 GS305.2566$+$0.3162 &      6.1 &  145.4 &    18.7 &  8.50 $\times$ 10$^{22}$ &  4.70 $\times$ 10$^{2}$ &              70q &                 $-$ \\
\bottomrule
\end{tabular}
    \begin{tablenotes}
      \small
      \item \textbf{Notes.} The columns are defined as follows: Name: the SABOCA catalog source name. Distance: distance of the source in kpc. M$_{1\sigma}$: Mass of the source by summing up all the pixel values in the column density map according to Eq.\,\ref{eq:eq_M}, using a radius of 1$\sigma$.  T$_{d}$: Average dust temperature of the source. N(H$_{2}$): Average column density of the source. L$_{sab}$: Luminosity of the source by integrating the SED from 1$\mu$m to 1\,mm according to Eq. \ref{eq:eq_lbol}. MIR type: Source mid-infrared association as described in Sec. 3.5. (24sat: associated with saturated 24\,$\mu$m emission. 70ct: associated with PACS\,70\,$\mu$m compact source. 24ct:associated with MIPSGAL\,24\,$\mu$m compact source. 22ct: associated with WISE 22\,$\mu$m compact source. HII: associated with (UC)HII regions. 70q: associated with extended 70$\mu$m emission or 70\,$\mu$m quiet). Parental clump: The name of the ATLASGAL clump that the SABOCA core is associated with (Sec.\,\ref{sec:fmtt}). \\

     \item Full table will be available on CDS.

    \end{tablenotes}
  \end{threeparttable}
\end{table*}

To estimate the physical properties of the 350\,$\mu$m SABOCA compact sources we follow the same procedure as described above using the corresponding  parameters fitted by {\it{Gaussclumps}} and using the high-resolution column density and dust temperature maps  (10\arcsec). The physical parameters of the individual ATLASGAL clumps and the SABOCA sources are listed in Table\,\ref{tab:clumps} and Table\,\ref{tab:cores}, respectively. We list the general statistics of the clump and core properties in Table\,\ref{tab:general} for all the sources and the distance limited sample.

\begin{table*}
\scriptsize
\caption{Physical properties of ATLASGAL clumps and SABOCA compact sources.}
\begin{tabular}{ccccc|ccccclccccl}
\toprule 
&&All clumps (507)&&&&Clumps at 2-4 kpc (205, 41$\%$)&&&&\\
\midrule
&     Max &    Min&  Mean &    std  &     Max & Min &  Mean &    std &&&& \\
\midrule
$FWHM$ size (pc) &     2.9 &   0.13 &  0.68 &    0.45 &    0.87 &   0.19 & 0.42 & 0.11  \\
Mass (M$_{\odot}$)      & 9.71$\times10^{3}$ &    0.05 & 598 & 1.03$\times10^{3}$ & 1.42$\times10^{3}$  & 3.0 &172 & 156 \\
$T_{\rm d}$ (K)       &      45.6 &    14.1 &    20.8 &     5.0   &    45.5 &    14.1 &  21.3 &  6.3 \\
$\Sigma$ (gcm$^{-2}$)     &     3.6 & 0.004 &  0.33 &    0.32 &       1.8 & 0.005 & 0.4 & 0.4  \\
\midrule
&&All SABOCA sources (971) &&&&SABOCA sources at 2-4 kpc (405, 42$\%$)\\
\midrule
&     Max &    Min &  Mean &    std &   Max & Min &  Mean &    std  \\
\midrule
$FWHM$ size (pc) &     1.17 & 0.052 & 0.32 &    0.21  &    0.40 & 0.08 & 0.19 & 0.05  \\

Mass (M$_{\odot}$)       & 5.43$\times10^{3}$ &   0.017 &   198 & 374 &   435 &   5 &   52  &    50 \\
$T_{\rm d}$ (K)           &     53.3 &   13.3 &  21.6 &     6.0&   53.3   &   13.7 &  21.8 &   7.0  \\

$\Sigma$ (gcm$^{-2}$)      &     4.3 &  0.05 & 0.4 &    0.3  &     2.0 &  0.06 & 0.4 &  0.3 \\
 \bottomrule
\end{tabular}
  \label{tab:general}
\end{table*}

The distribution of the physical properties of the clumps are shown in Fig.\,\ref{fig:clump_frag_hist}.  The average mass of all of  the clumps is 600\,$\rm M_{\odot}$. In total, 65$\%$ of the sources have $>100\,{\rm M_{\odot}}$ and the most massive clump has a mass of $1 \times 10^{4}$\,M$\rm_{\odot}$, which is one of the more distant sources in our sample. In addition, there are 130 clumps above 650\,M$_{\odot}$, which is a mass limit used by \citet{Csengeri14} as a threshold for a massive clump to potentially form high-mass stars. The average mass of all the SABOCA sources is 198\,$M\rm_{\odot}$, and for the distance limited sample it is 50\,$\rm M_{\odot}$, which is $\sim$3 times lower than the values of the clumps (Table\,\ref{tab:clumps}).

 The mean clump $FWHM$ size of all the clumps and those in 2-4\,kpc is $0.68\pm0.45$\,pc and $0.42\pm0.11$\,pc, corresponding to the typically observed scales for Galactic clumps. The mean 
 $FWHM$ size is $0.31\pm0.21$\,pc for all the SABOCA sources, and for the sources in 2-4\,kpc the value is $0.19\pm0.05$\,pc, corresponding to a typical core-scale.

We also calculate the surface density based on the average column density derived from SED fits, where $\Sigma = \mu m_{H}N(H_{\rm2})$\,(g\,cm$^{-2}$). In Table\,\ref{tab:general}, we can see that SABOCA sources have typically larger surface densities on average than the clumps, indicating a concentration of dense gas at smaller scales. The majority of the clumps (65$\%$) and cores (78$\%$) have a mass surface density larger than 0.2\,g\,cm$^{-2}$ which is used to identify promising massive star progenitors by \citet{ButlerTan12}.

We find dust temperatures between $\sim$14\,K and $\sim$50\,K for the clumps, and on average we find similar values for clumps and cores. This is expected from our SED fitting method which is mainly sensitive to the cold component. The most active star-forming clumps have an average dust temperature of $>40$\,K and correspond to the most massive OB cluster-forming regions in the Galaxy.

Due to our source selection strategy, we have a large fraction of clumps with low bolometric luminosity and a luminosity-to-mass ratio of less than 10 (74$\%$). Since a low bolometric luminosity together with a large mass is indicative of an early evolutionary stage (e.g. \citealt{Motte07}, \citealt{Molinari16}), a large number of our targeted clumps are likely to be at early evolutionary stages. We indicate the $L_{\rm bol}/M$ = 1 and $L_{\rm bol}/M$ = 10 vertical lines in the luminosity-to-mass ratio distribution plot (Fig.\,\ref{fig:clump_frag_hist});  which show empirical limits from \citet{Molinari16} for massive clumps in a starless or pre-stellar stage, and clumps hosting high-mass ZAMS stars, respectively. The $L_{\rm bol}/M$ $>$ 10 ratio can also correspond to a qualitatively different heating source reflecting the presence of deeply embedded UC-H\textsc{ii} regions (see Sec.\,\ref{sec:cat}; see also \citealt{Kim18}). The most evolved clumps, with luminosity-to-mass ratios above 100, e.g. G34.3+0.2, G10.6$-$0.4, G333.6+0.2 (\citealt{Campbell04}; \citealt{Liu10}; \citealt{Lo15}, respectively), are clumps residing in well-known extreme Galactic massive star-forming regions.

\subsection{Signposts of star-forming activity of the 350 $\mu$m SABOCA sources}\label{sec:cat}

We use  the MIPSGAL point source catalog at 24\,$\mu$m ($\sim$6$''$; \citealt{Gutermuth15}) and the Herschel/Hi-GAL point source catalog at 70\,$\mu$m  ($\sim$8\rlap{.}{\arcsec}4; \citet{Molinari16}) to assess the star formation activity of the SABOCA 350\,$\mu$m compact sources. These catalogs have a comparable angular resolution to the SABOCA observations. We also make use of the WISE point source catalog (11$\arcsec$ at 22\,$\mu$m; \citealt{Cutri12}) to complement the mid-infrared surveys as MIPSGAL is saturated at $>$2\,Jy, while the WISE point source photometry is reliable to $\sim$330.0\,Jy flux density at 22\,$\mu$m. We also used the existing RMS (\citealt{Lumsden13}; \citealt{Urquhart13}) and the CORNISH (UC-)H\textsc{ii} region catalogs (\citealt{Kalcheva18}) to identify the sources associated with H\textsc{ii} regions that are likely to be saturated in mid-infrared images. These catalogs have a coordinate range that covers in total $\sim$30$\%$ of our SABOCA (295 out of 971) cores. The higher angular resolution of the SABOCA data allows a better matching with these higher angular resolution ancillary catalogs compared to ATLASGAL, which is necessary to assess the star-forming activities in the cores.

\begin{figure*}[t]
\centering
\begin{tabular} {p{0.45\linewidth}p{0.45\linewidth}}
\includegraphics[scale=0.4]{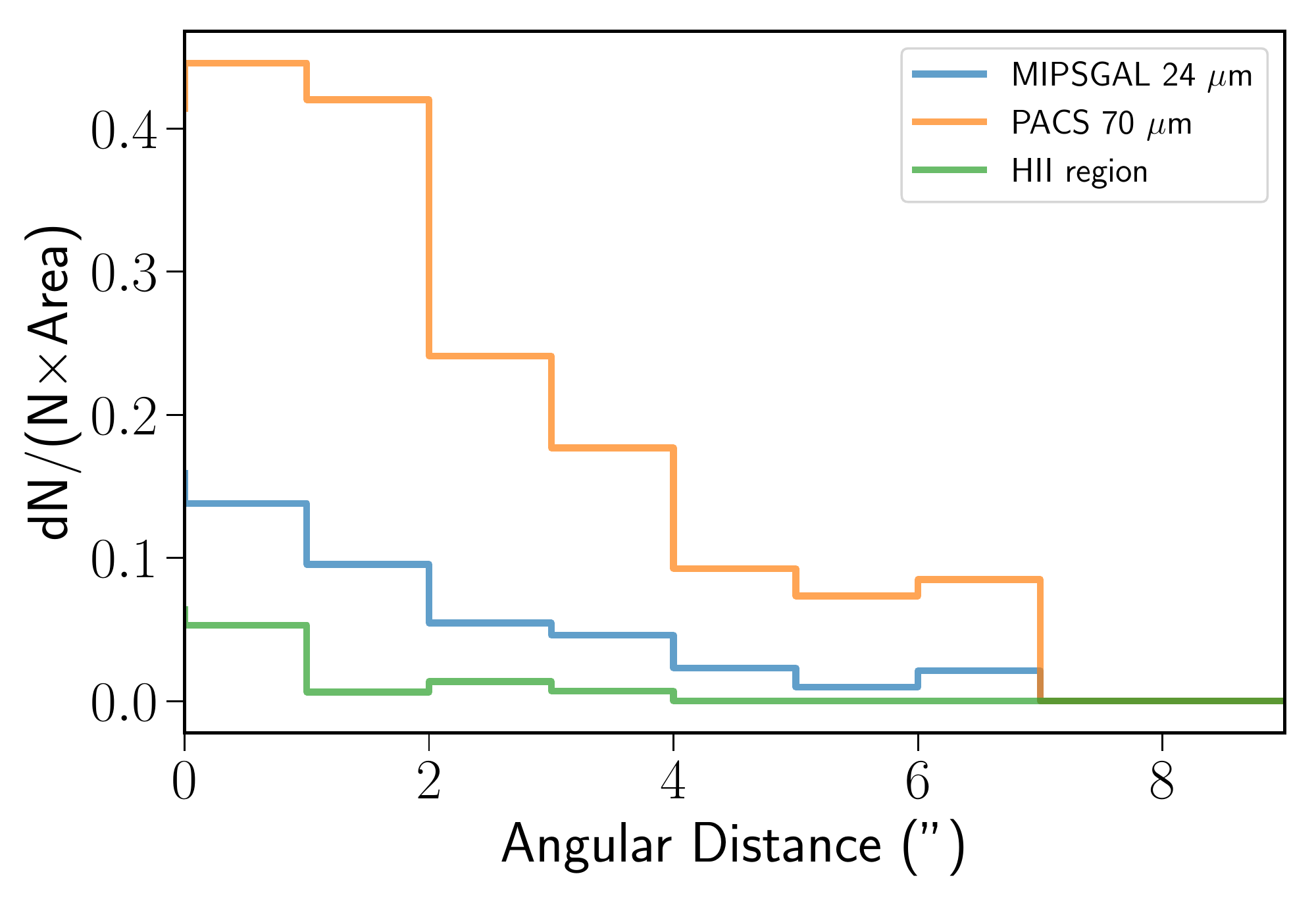}&\includegraphics[scale=0.4]{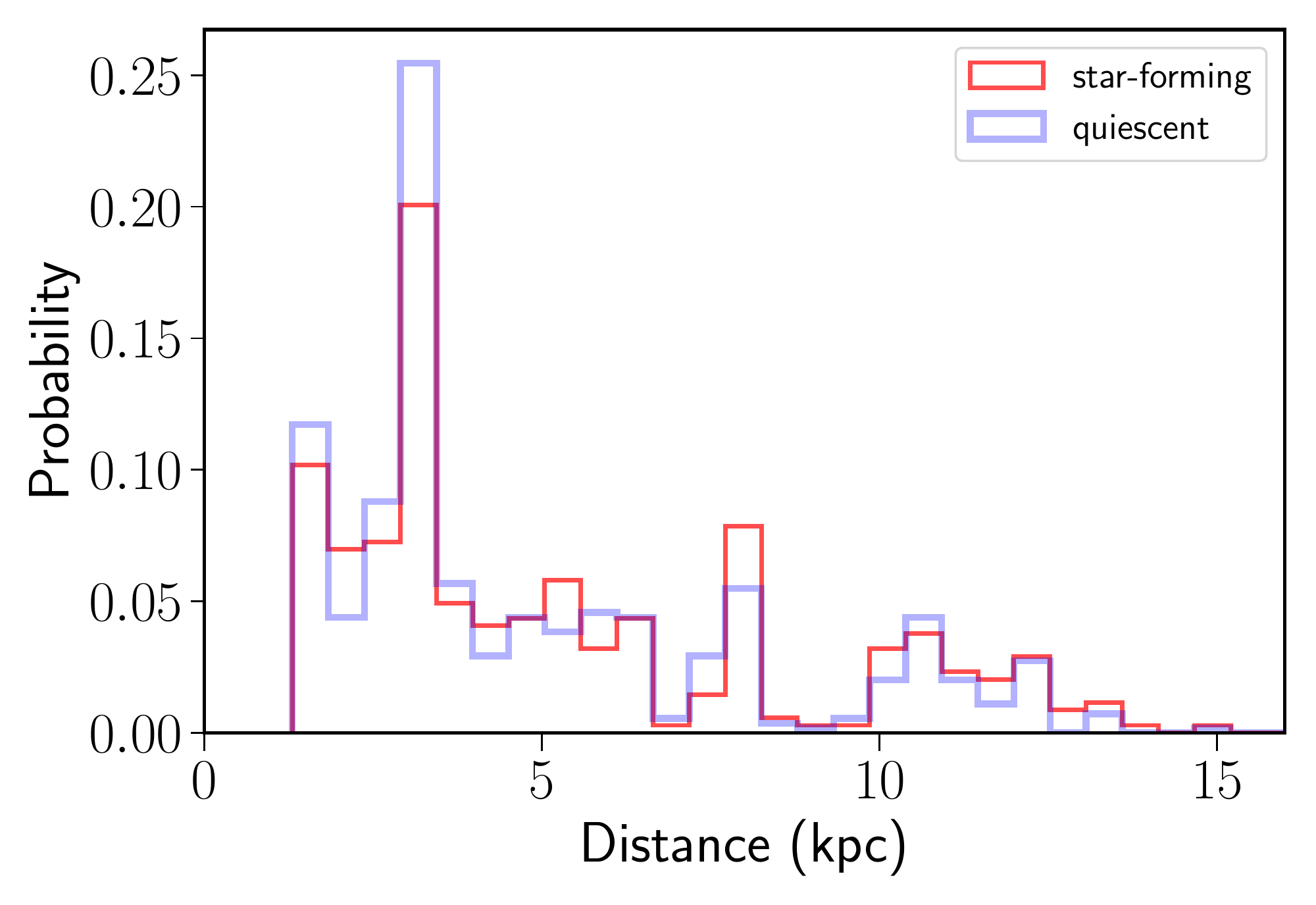}\\
\end{tabular}
\caption{{\it{Left}}: Distribution of angular distances between SABOCA sources and their MIR compact source or/and H\textsc{ii} region associations. In each bin, the probability of dN/N is normalised by the annular area defined by adjacent angular bins as dN/(N$\times$Area); {\it{Right}}:  Distribution of distances of star-forming and quiescent SABOCA compact sources.}
\label{fig:cat_dist}
\end{figure*}

We performed a cross-match between the position of the SABOCA sources and these catalogs with an angular separation limit of $8\as5$ corresponding to the angular resolution of SABOCA maps. The distribution of the angular offsets between the SABOCA compact sources and their mid-infrared emission (Fig.\,\ref{fig:cat_dist}, left panel) suggests a good positional correlation between the mid-infrared sources and the 350\,$\mu$m SABOCA peak positions. The dispersion of offsets in the range of 2-4\arcsec, which is consistent with the pointing accuracy of the telescope.


We use the association with UC-H\textsc{ii} regions, 24/22\,$\mu$m or 70\,$\mu$m point sources to distinguish between star-forming and quiescent cores and assess the statistics of their physical properties. Due to the varying diffuse mid-infrared background emission, the completeness of these catalogs is, however, not homogeneous. The sensitivity limit of the MIPSGAL\,24\,$\mu$m survey (5$\sigma$, $\sim$10\,mJy) can probe down to 18\,L$_{\odot}$ protostars within 15\,kpc following the calculation of \citet{Motte07} based on IRAS colors from \citet{Wood89}. This suggests that even intermediate-mass protostars could be revealed throughout our sample. However, as discussed in \citet{Gutermuth15}, the MIPSGAL catalog completeness is strongly dependent on the spatial variation of the local emission, therefore the sensitivity limit does not reflect the level of the completeness towards all sources. The sensitivity limit of the PACS\,70\,$\mu$m band of 0.2\,Jy (\citealt{Molinari16}) corresponds to a luminosity of $\sim$25$\,\rm L_{\odot}$ at 5\,kpc (\citealt{Dunham08}). This similarly suggests that at least intermediate-mass protostars should be detected at 70\,$\mu$m, if neglecting the spatial variation of the completeness level and source confusion.

In all the 1120 compact sources identified from the SABOCA maps, of which 971 have physical parameters based on their SED fitting, 13 (1.3$\%$) have UCH\textsc{ii} region counterparts. In the remaining sources, 184 (19.0$\%$) are directly associated with 24/22\,$\mu$m compact sources. In addition, we also have 87 (8.9$\%$) sources that spatially coincide with regions of saturated 24\,$\mu$m emission, but without 22\,$\mu$m or 70\,$\mu$m counterparts which are ignored in the following analysis. Of the 184 sources that have 24/22\,$\mu$m compact sources, 18 do not have a 70\,$\mu$m compact source counterpart. Visual inspection of these 18 sources shows that they are all associated with faint or extended 70\,$\mu$m emission, and hence they are likely to be below the completeness limit of the PACS\,70\,$\mu$m catalog. Among the remaining sources, 142 (14.6$\%$) have a compact source counterpart only at 70\,$\mu$m.  The remaining 545 (56.1$\%$) sources are either dark at 24/22\,$\mu$m or are associated with extended or diffuse 70\,$\mu$m emission which is not included in the compact source catalog. To summarise, we have used the various surveys described above to classify 884 cores;  545 cores are found to be quiescent (64\%) and 339 cores to be as star-forming (36\%). 

Dust extinction may prohibit detection of deeply embedded 24/70\,$\mu$m compact sources. We calculate the dust opacities at 70\,$\mu$m based on the derived column densities for quiescent cores with our assumed opacity law (see in Sec. \ref{sec:clump_sed}). We find that 90$\%$ of the calculated opacities are lower than 1 with a 95$\%$ quartile level of opacity $\sim$2.4, indicating that dust extinction does not significantly affect our classification. However, the effect of source geometry and the uncertainty of dust emissivity may also influence the actual extinction.

To check whether the non-uniform distances of our sources affect the classification, we show the distance distribution of the star-forming and quiescent cores in Fig.\,\ref{fig:cat_dist} (right panel). The two distributions are similar with a KS-test yielding p-value $=$ 0.31 and D $=$ 0.08 (p-value $>$ 0.05 and D $<$ D$_{\alpha}$ at a significance level of $\alpha$ = 0.05), which means that there are not more quiescent cores appearing at nearer distances than star-forming cores, suggesting that the different distances do not bias our classification. However, we caution that this similarities in the distributions could be due to the presence of a mutual cancellation of two opposite biases: at larger distances, the objects
identified will tend to be more massive and hence more likely to be star-forming already. This effect will be counteracting the decreasing likelihood of detecting mid-infrared compact sources. In the following, we restrict our analysis to distance-limited sample, or samples in certain distance bins, so that the results are less affected from the classification bias due to distance non-uniformity.

\section{Discussion}

In this section we compare the physical properties of the star-forming and quiescent SABOCA cores (Section\,\ref{sec:comparepp}), and investigate their capability of forming high-mass stars (Section\,\ref{sec:hm}). We study the fragmentation properties and the clump structure in Section\,\ref{sec:fragmentation} based on a distance limited sample of 2-4 kpc. In Section \ref{sec:irquietmassive} we identify a sample of massive quiescent SABOCA cores at $<5$\,kpc that are potential high-mass pre-stellar core candidates.

\subsection{Physical properties of star-forming and quiescent cores}\label{sec:comparepp}

In Sec.\,\ref{sec:cat} we classify the SABOCA cores into two categories according to their radio, mid- and far-infrared compact source associations: star-forming and quiescent. As discussed in the following paragraphs, our approach is more sensitive to the deeply embedded low- to intermediate mass protostars than using a single wavelength.

A 24\,$\mu$m point source associated with a compact 350 $\mu$m source suggests the presence of a local heating source, corresponding to one or more protostars. However, the absence of a 24\,$\mu$m point-source does not necessarily mean that there is no star-forming activity, a high optical depth or 
mid-infrared confusion may inhibit the detection of compact 24\,$\mu$m sources.  In that case, being more sensitive to the outer regions of internally heated cores, the PACS\,70\,$\mu$m compact source association (e.g. \citealt{Dunham08}; \citealt{Ragan13}) is used to further pinpoint the star-formation activity. While it is clear that the absence of a 24-70\,$\mu$m point source appears to exclude the presence of a massive YSO, early stage (even up to high-mass) protostars have been detected towards infrared quiet massive clumps (\citealt{Bontemps10}; \citealt{Feng16}; \citealt{Csengeri18}), and dense cores with luminosities as low as 400\,$L_{\odot}$ (\citealt{Duarte-Cabral13}).

\begin{figure*}[t]
\centering
\begin{tabular} {p{0.45\linewidth}p{0.5\linewidth}}

\includegraphics[scale=0.4]{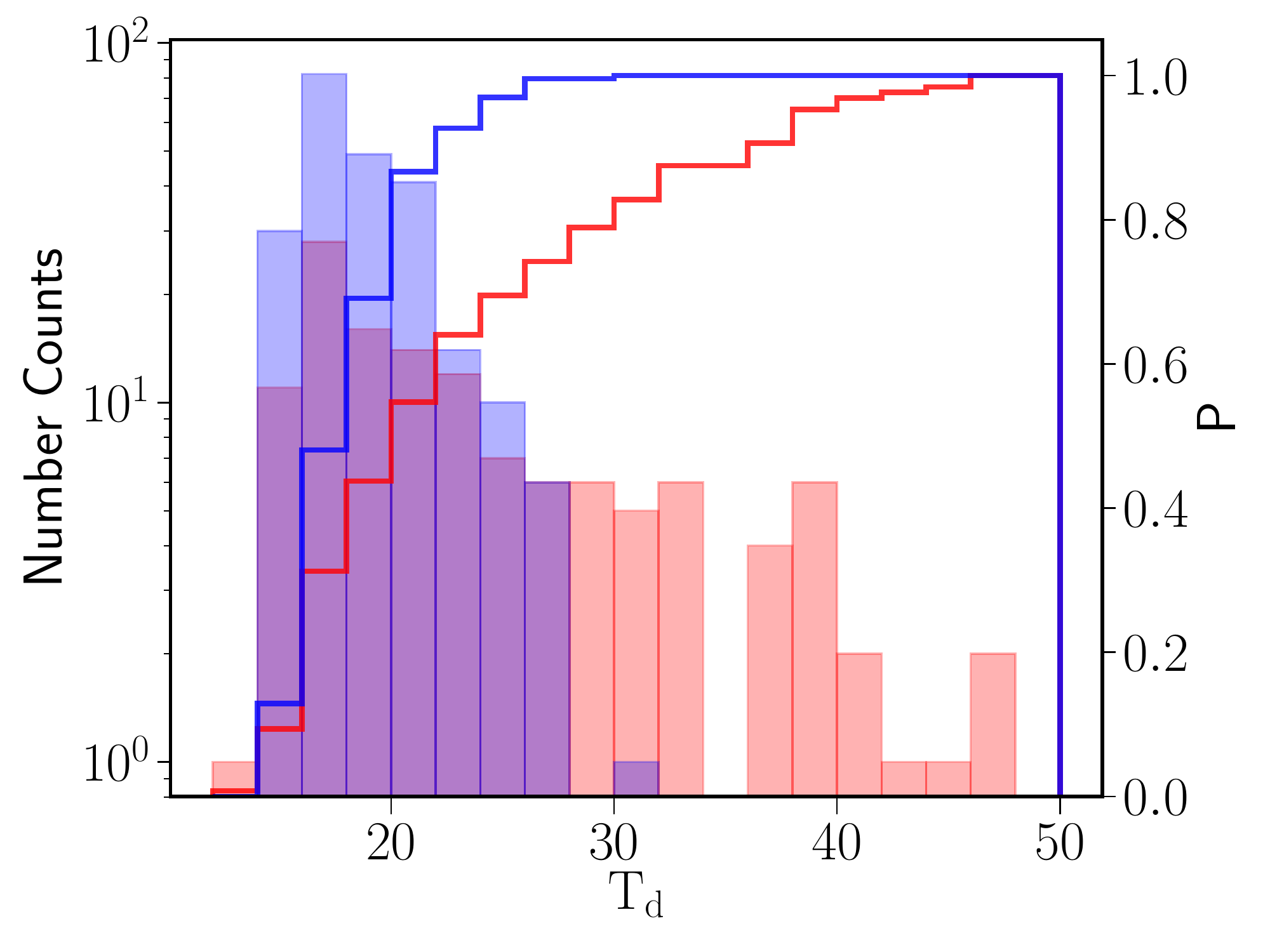}&
\includegraphics[scale=0.4]{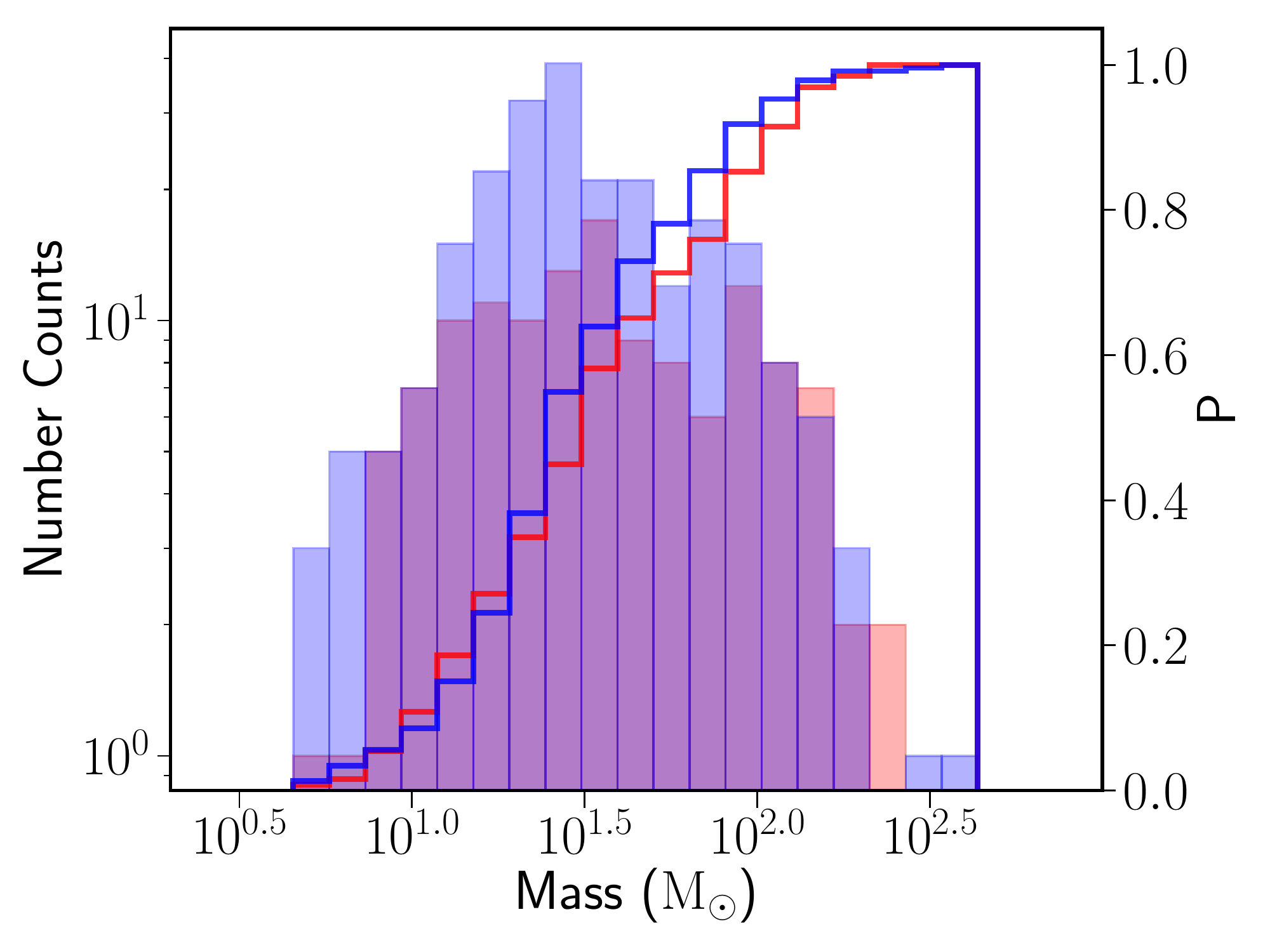}\\

\includegraphics[scale=0.4]{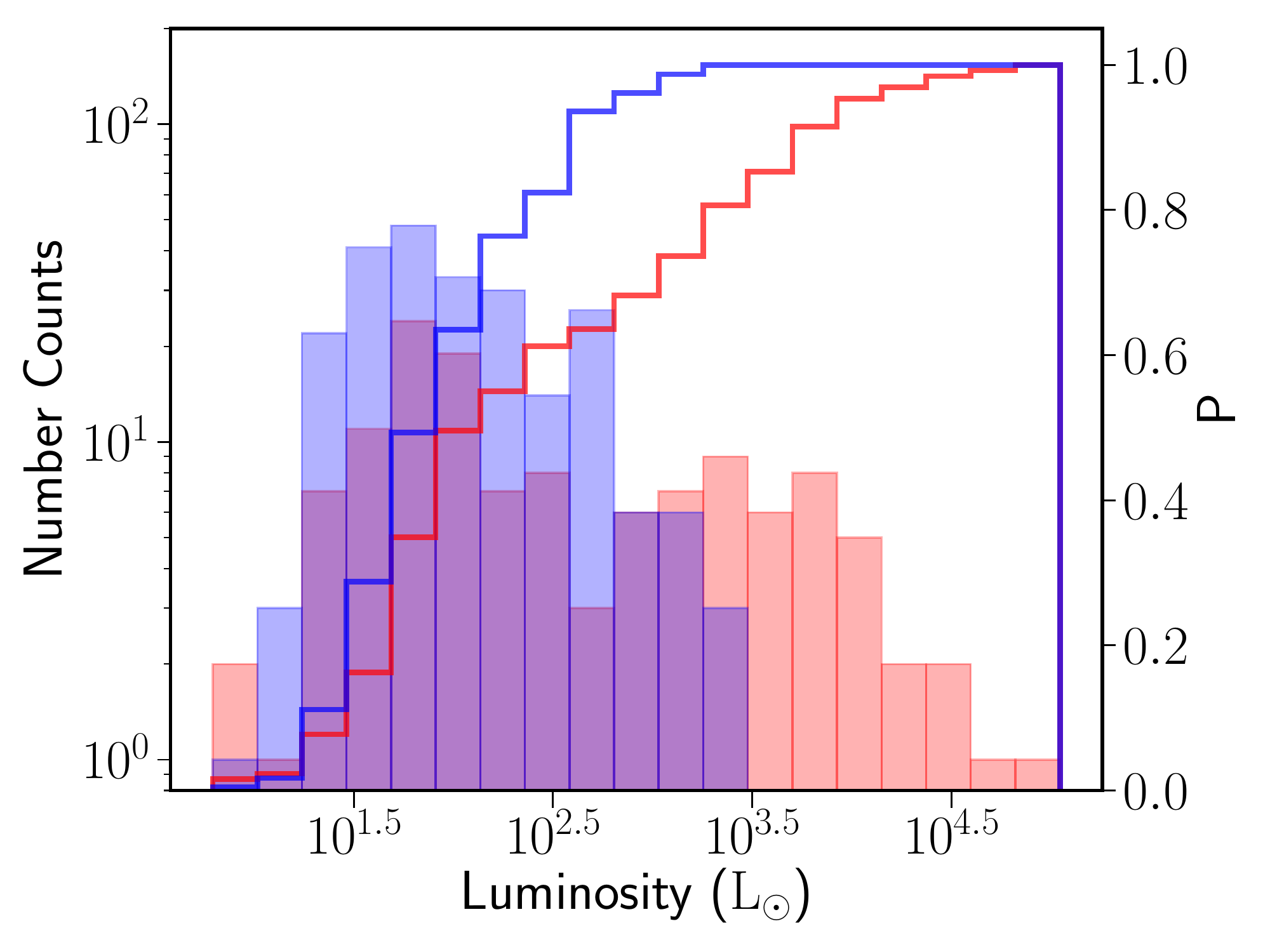}&
\includegraphics[scale=0.4]{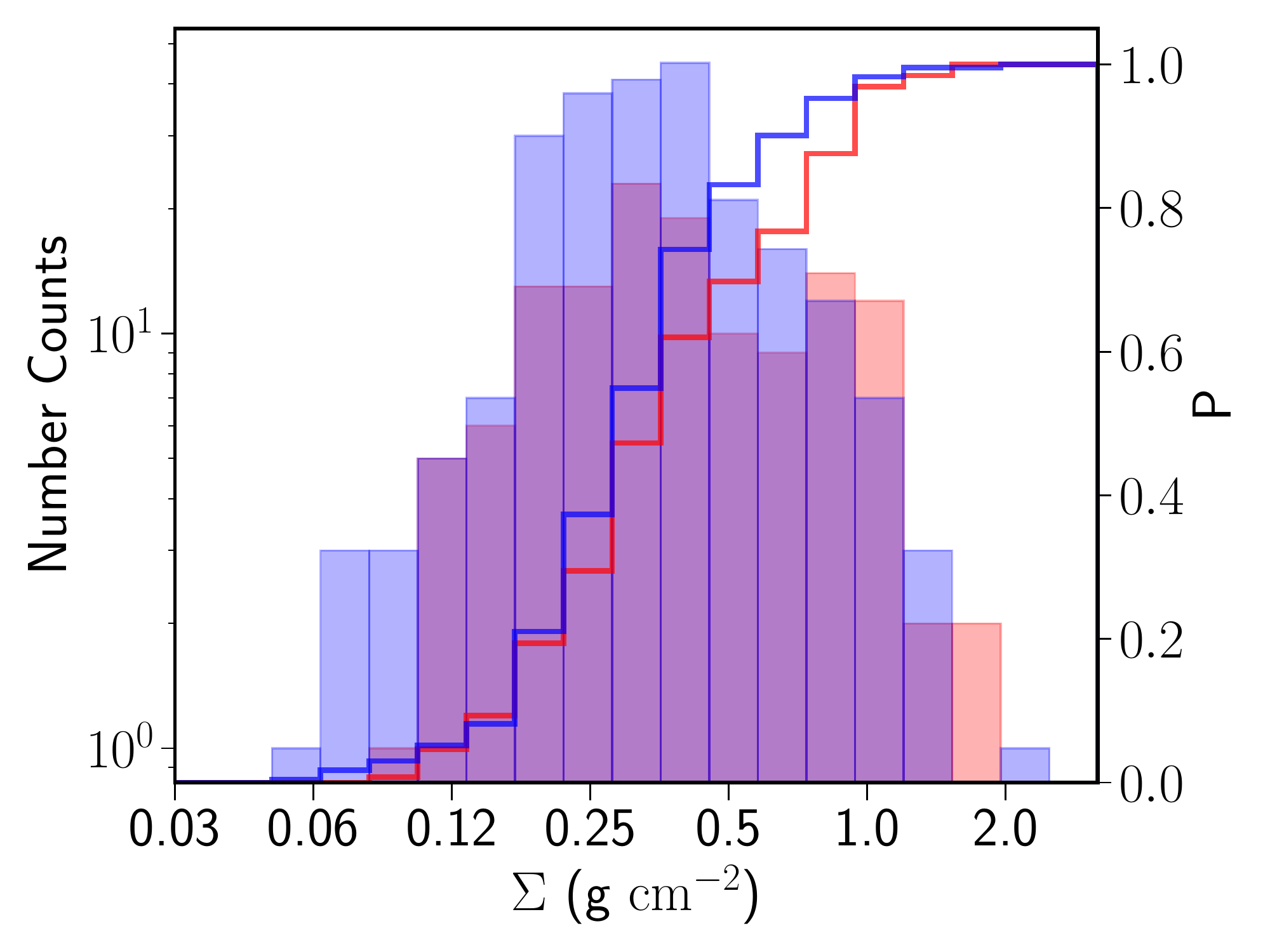}\\

\end{tabular}
\caption{Distribution of the physical properties of star-forming and quiescent SABOCA compact sources in the 2-4 kpc distance bin, in red and blue color, respectively. {\it{Top left}}: The temperature distribution of two categories.  {\it{Top right}}: The mass distribution. {\it{Bottom left}}: The luminosity distribution. {\it{Bottom right}}: The surface density distribution. In each plot, the cumulative probability distribution is drawn with respective colors according to the right y-axis.}
\label{fig:cat_hist}
\end{figure*}

In Fig.\,\ref{fig:cat_hist} we show the distribution of the physical parameters of  cores located at a distance of 2-4 kpc. This sample consists of 233 quiescent and 129 star-forming cores. We find that the star-forming cores have, in general, a modestly higher temperature (24.0\,K$\pm$0.75\,K) than the quiescent cores (18.9\,K$\pm$0.19\,K). We obtain a $p$-value $<$ 0.001 for a KS-test confirming that the temperatures of the two populations are statistically different. However, we also note that there is a large fraction (43$\%$) of star-forming cores have a relatively low dust temperature ($<20$\,K). There is also a small number of quiescent cores with temperatures larger than 25\,K (5$\%$). Visual inspection of these cores reveals that they are located close to extended nebulosities at 70\,$\mu$m; these cores may, therefore, be star-forming cores that remain undetected due to confusion from the nearby bright source or genuinely quiescent cores with the warmer temperature caused by external heating.

The luminosity distribution of the SABOCA cores is presented in the bottom left panel of  Fig.\,\ref{fig:cat_hist}. This plot shows a similar picture as the dust temperature with the star-forming cores having higher luminosities. The KS test results in $p$-value $<$ 0.001, again indicating a distinct distribution. However, we find a large number of quiescent sources with high luminosities of $>$10$^3$\,L$_{\odot}$. The number of quiescent sources decreases to zero above $L_{\rm bol}$$\sim$$10^{3.5}$\,L$_{\odot}$, which is equivalent to the ZAMS luminosity of a B2-B4 star. As discussed in the previous paragraph, these large values of the bolometric luminosities towards quiescent cores are either due to external heating or to a higher level of confusion and blending with nearby sources. So it is possible that some of the quiescent cores might be protostellar, however, the lack of a corresponding bright 24\,$\mu$m source suggests that the star formation is still at a very early stage. 

The mass distribution of the quiescent and star-forming sources is remarkably similar. Together with their difference in bolometric luminosities, this may suggest that the cores classified as quiescent are in an earlier stage of star formation, and would follow the same evolutionary path as the star-forming cores (\citealt{Csengeri17a}, see also discussion in Sec. \ref{sec:hm}). The range of typical core masses is between 10 and 320\,M$_{\odot}$, with an average mass of star-forming and quiescent cores of $53.1\pm$4.2\,M$_{\odot}$ and $45.5\pm$3.1\,M$_{\odot}$, respectively. The KS test yields $p$-value $=$ 0.26, indicating a similar mass distribution. The typical core masses for star-forming and quiescent cores are not significantly different. The star-forming cores have on average smaller size (0.15$\pm$0.002 pc in FWHM) compared to quiescent cores (0.16$\pm$0.003 pc in FWHM). The mean mass surface density of star-forming cores (0.50$\pm$0.03\,g\,cm$^{-2}$) is higher than that of the quiescent cores (0.40$\pm$0.02\,g\,cm$^{-2}$). The KS-test yields $p$-value $\sim$ 0.012 ($\sim$1$\%$), indicating that the two distributions are different, with star-forming cores having higher surface density. 

The similar mass distribution and relatively different surface density distribution of quiescent and star-forming cores indicate that, at the typical scale of $\sim$0.2\,pc the quiescent cores have enough mass assembled to evolve into protostellar cores when contracting. 
Cores can be gravitationally contracting ever since they first appear, allowed by the presence of an unstable background (\citealt{Naranjo-Romero15}). In the framework of dynamical evolution cores are clump-fed (see also \citealt{Wang10}), which is supported by growing observational evidence for  gas replenishment beyond the core-scale (e.g. \citealt{Galvan09}, \citealt{Csengeri11}, \citealt{Peretto13}, \citealt{Liu15}, \citealt{Chen19}). In this case cores prior to star formation may be expected to have a lower mass compared to star-forming cores, however, the fact that quiescent cores are relatively more extended, and their aspect ratio (from {\it{Gaussclump}}) showing them to be less roundish compared to star-forming cores, may indicate the similar mass distributions do not contradict with such dynamical evolution scenario. 
In addition, some of the star-forming cores may be in a state of gas dispersal (Fig. \ref{fig:cat_ml}). 
Among the quiescent cores, the most luminous ones ($L>10^{2.5}$\,L$_{\odot}$) have larger surface densities of 0.69$\pm$0.13\,g\,cm$^{-2}$ compared to their less luminous counterparts with 0.33$\pm$0.01\,g\,cm$^{-2}$. The number of quiescent cores decreases significantly above $\sim$0.5\,g\,cm$^{2}$. This may suggest that when reaching a higher mass surface density regime the cores necessarily form high-mass stars, for which the formation time-scales are shorter than for low- and intermediate mass star-formation (\citealt{Csengeri14}, \citealt{Urquhart18}).

\subsection{Fraction of massive star-forming cores}\label{sec:hm}

\begin{figure}[h]
\centering
\includegraphics[scale=0.42]{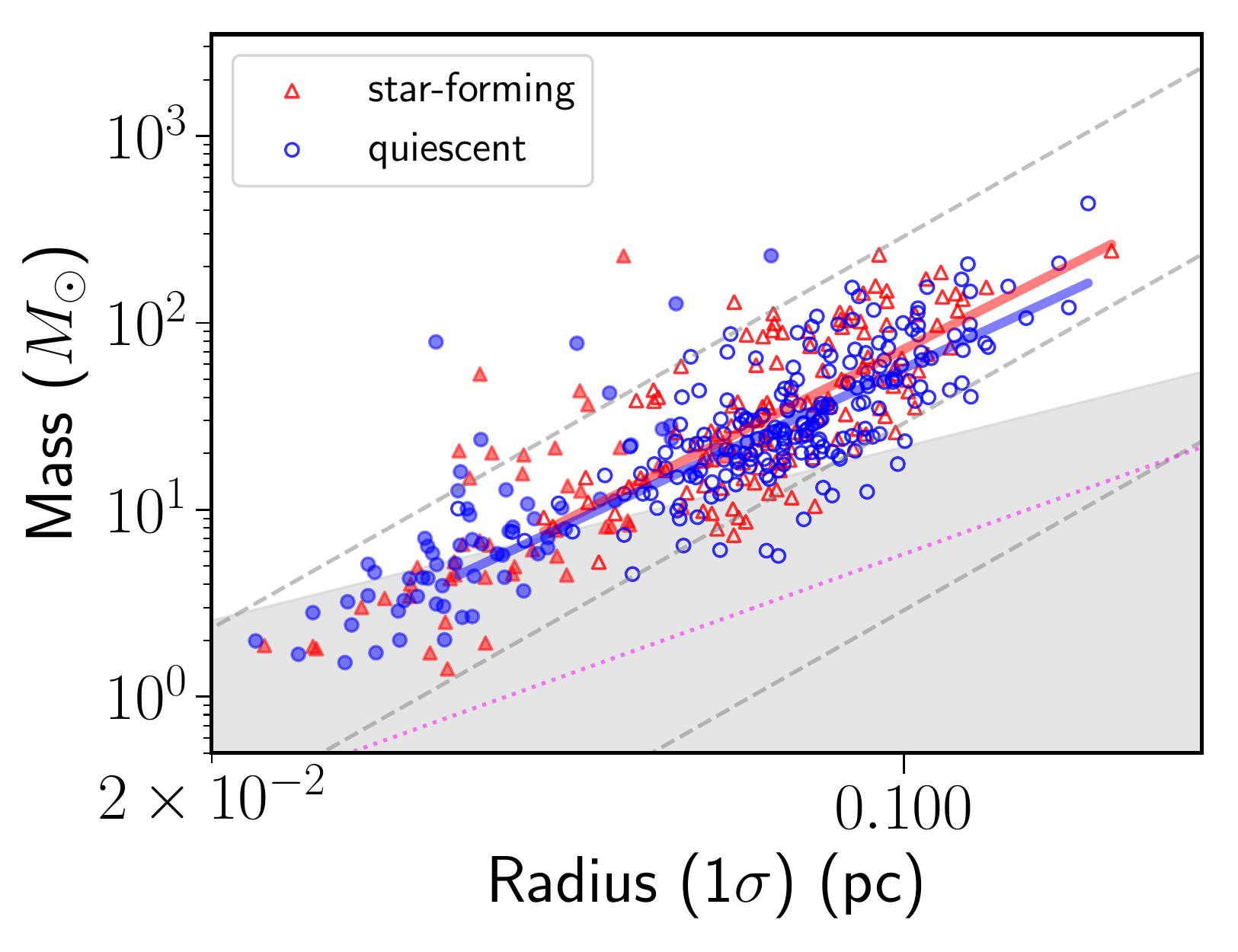}\\
\caption{Mass-radius diagram of SABOCA compact sources. The quiescent cores and star-forming cores are indicated with blue and red colors, in 2-4 kpc (hollow markers) and 1-2 kpc (filled markers) distance bins. The dashed lines show the constant volume density of 10$^{4}$, 10$^{5}$ and 10$^{6}$\, cm$^{-3}$. Magenta dotted line indicate Larson's third relation of approximately 10$^{21}$\,cm$^{-2}$ column density. Gray filled region is below the empirical threshold of massive star-formation from Kauffmann \& Pillai (2010) scaled to match the dust opacity values we used in this work. Red and blue lines indicate the fitted linear regression for the quiescent cores and star-forming cores in 2-4 kpc.}
\label{fig:cat_mr}
\end{figure}

\begin{figure}[h]
\centering
\includegraphics[scale=0.42]{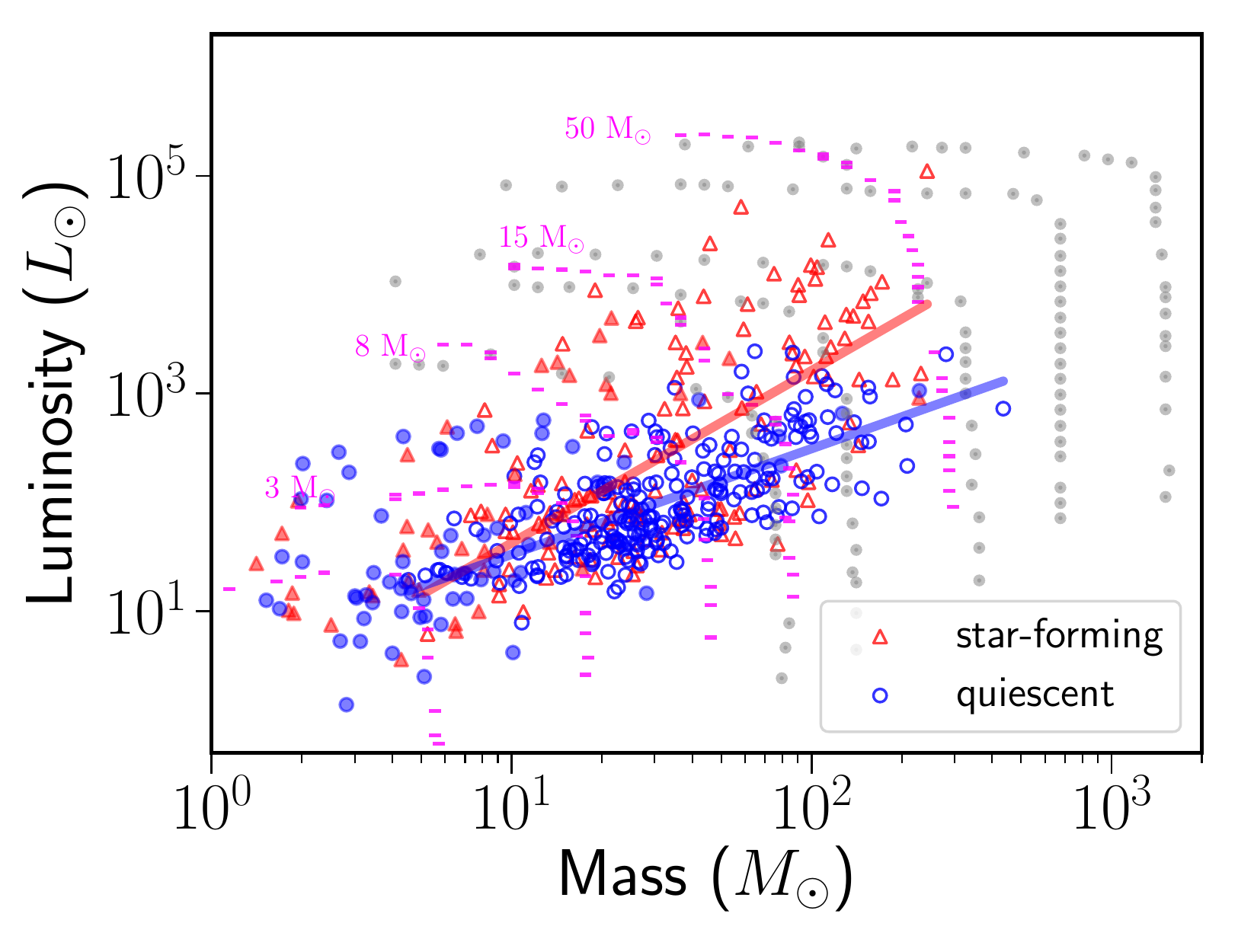}\\
\caption{Luminosity-mass diagram of SABOCA compact sources.  The quiescent cores and star-forming cores are indicated with blue and red colors, in 2-4 kpc (hollow markers) and 1-2 kpc (filled markers) distance bins. The empirical evolutionary tracks for massive clumps of envelope masses 80, 140, 350, 700, and 2000\,M$_{\odot}$ in \citet{Molinari08} are shown in grey dots. \citet{Andre08} tracks for final stellar masses of 1, 3, 8, 15 and 50\,M$_{\odot}$ are shown in magenta dashed lines. Red and blue lines indicate the fitted linear regression for the quiescent cores and star-forming cores in 2-4 kpc.}
\label{fig:cat_ml}
\end{figure}

To estimate what fraction of the SABOCA cores are potential precursors or in the early stage of forming massive stars, we show the mass-radius plot and mass-luminosity diagram of all the samples in Fig.\,\ref{fig:cat_mr} and Fig.\,\ref{fig:cat_ml}. Both of these plots show the distribution for the star-forming and quiescent cores in red and blue, respectively. For this analysis we only use sources located at 1-2 kpc, and 2-4 kpc distance ranges, which allows us to compare sources with similar size-scales. We briefly discuss sources more distant for comparison purposes.

The mass-radius plot (Fig.\,\ref{fig:cat_mr}) shows no clear difference between the star-forming and the quiescent cores in mass scale. A substantial number of the cores are found above the empirical threshold for massive star-formation introduced by \citet{KauffmannPillai10} suggesting that a large fraction of them are capable of, or are already in the process of forming high-mass stars. In particular, we cover a large fraction of quiescent cores that have a radius of $\sim$0.2\,pc with masses of $\sim$100\,M$_{\odot}$ corresponding to a mass surface density of $\sim$0.2\,g\,$\rm\,cm^{-2}$, thus representing potentially young regions where massive star formation could occur (e.g. \citealt{ButlerTan12}). 
 
The derived power-law relation of $M\propto r^{\alpha}$ gives $\alpha$=2.67$\pm$0.25 and $\alpha$=2.60$\pm$0.16 for star-forming and quiescent cores in a distance range of 2-4 kpc,  respectively. For sources in 1-2 kpc, the indices are somewhat larger, with $\alpha$=3.00$\pm$0.62 and $\alpha$=3.07$\pm$0.32 for the star-forming and quiescent sample, respectively.  Moving to sources at 4-8, 8-12 kpc, more shallower slopes of $\sim$2.0 are found. These slopes are generally steeper than observed on the even larger clump-scale where a relation of M$\propto$R$^{1.4-1.7}$ is derived (e.g. \citealt{Urquhart18}). The trend of a steepening mass-radius relation at small scale (or smaller distance range) is also seen in, e.g., \citet{Ragan13} towards cores in infrared dark clouds, and in \citet{Beuther18} towards young massive star-forming regions at thousands of au scale (as reflected in their Fig. 11).

{Observationally, the Larson's density-size relation (M$\propto$R$^{2}$) could be due to a selection effect of a certain column density threshold, as frequently pointed out by previous works (e.g. \citealt{Larson1981}, \citealt{Ballesteros-Paredes02}, \citealt{Camacho16}).  In our case, the more distant objects having smaller slope of power-law dependence on radius could be simply due to the fact that they correspond to larger physical sizes, 
and thus a larger contribution from their more diffuse envelope. 
The steepening of the mass-radius relation from slope$\sim$2 to $\sim$3 correspond to a change from constant column density to constant volume density. Considering that cores are usually embedded in a gaseous environment, the slopes of $\alpha$=3, 2, 1 can result from different environments such as (spherically) uniform medium, a two-dimensional layer or a centrally concentrated filament, respectively (\citealt{Myers09}). This again indicates that at larger distances the amount of envelope gas might be influencing the slope, i.e.  cores appear more "concentrated" if a diffuse layer is involved.} On the other hand, the different power-law forms of M$\propto$R$^{2}$ or M$\propto$R$^{3}$ might indicate different modes of accretion (\citealt{Hennebelle12}), either dominated by turbulence from larger scale or the proto-cluster's self-gravity. Hence the steepening of the mass-radius relation may be reflective of the successively prevailing role of self-gravity at smaller scale (see also \citealt{Camacho16}).

The luminosity and mass distribution for the star-forming versus quiescent cores is shown in Fig.\,\ref{fig:cat_ml}. 
The evolutionary tracks for different final stellar masses and initial envelope masses (\citealt{Andre08}; \citealt{Molinari08}) are also included in this plot. 
These tracks have two main components, a vertical track that represents the main accretion phase and a horizontal track that indicates the clump gas dispersal phase. 
Where these lines join forms an apex that effectively defines the start of a star's zero-age main sequence (ZAMS) lifetime. 
The majority of the quiescent cores are below this apex and are, therefore, in the early stage of the evolution. Some of the star-forming cores, on the other hand, may be in a phase of gas dispersal.
The brightest star-forming cores are clusters around or above the apex, indicating that many have reached the ZAMS stage and the envelope clean-up phase has probably begun (\citealt{Molinari08}). 
Although scattered, the two distributions for quiescent and star-forming cores are clearly distinct for core masses larger than a few tens of M$_{\odot}$, with star-forming cores having larger luminosities. 
A linear regression fit between the log-log mass and luminosity give L$\propto$M$^{0.97\pm0.06}$ and L$\propto$M$^{1.58\pm0.16}$ for quiescent and star-forming cores in 2-4 kpc, respectively. Similarly drastic differences in power-law slopes for the sources in 1-2 kpc distance are found, with L$\propto$M$^{0.76\pm0.16}$ and  L$\propto$M$^{1.37\pm0.23}$ for quiescent and star-forming cores.  Moving further to sources at 4-8 kpc, the difference in slopes gets smaller,  with quiescent cores having L$\propto$M$^{1.20\pm0.08}$ and star-forming cores L$\propto$M$^{1.40\pm0.13}$. 
Within the uncertainties, for clumps a consistent relationship is found between luminosity and mass with L$\propto$M$^{1.3}$ (e.g. \citealt{Molinari08}; \citealt{Urquhart18}). The less pronounced difference in the mass-luminosity scaling relation of these two populations with larger distances can be explained if the mid-infrared source association with the submillimeter compact source becomes less robust with increasing distance to distinguish genuine quiescent cores, but also because more distant sources have more ambient gas, as mentioned before, which contributes to the mass but not much to the luminosity. The latter case is consistent with the fact that for sources more distant than 8 kpc, shallower slopes are found for quiescent and star-forming cores as L$\propto$M$^{0.81\pm0.13}$ and L$\propto$M$^{0.91\pm0.20}$.
In line with our previous results, i.e. the mass distribution and the mass-radius relation of the quiescent cores and star-forming cores, the luminosity and mass diagram with evolutionary tracks also suggests that a significant fraction of these cores can form a high-mass star, should they eventually collapse on a single object.

\begin{figure}[h]
\includegraphics[scale=0.42]{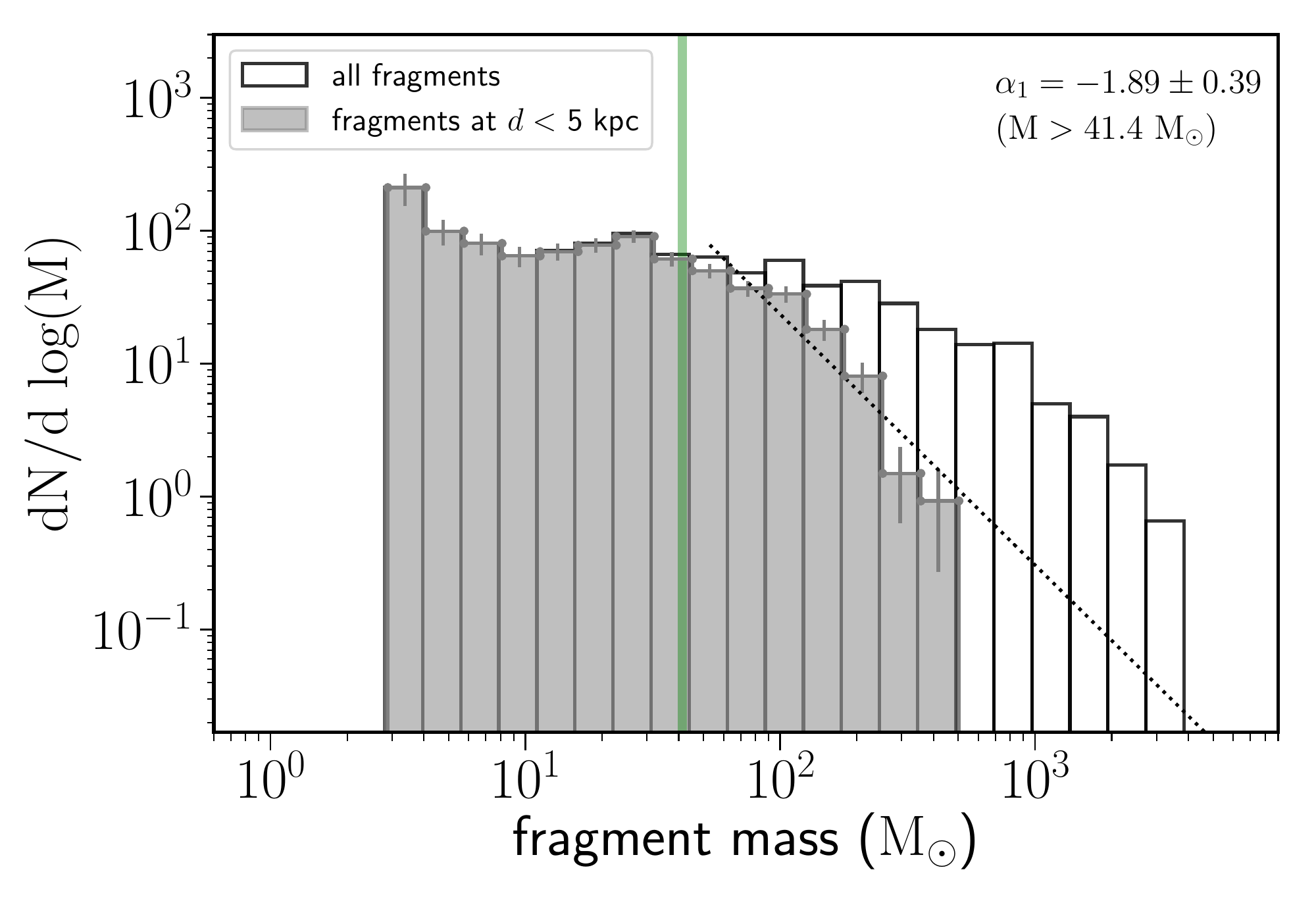}
\caption{Fragment mass distribution of all SABOCA sources. The fitted power-law index of the all the sources at $<5$\,kpc above our mass completeness ($> 41.4$\,M$_{\odot}$) is indicated by the green vertical line.}
\label{fig:frag_cmf}
\end{figure}

In Fig.\,\ref{fig:frag_cmf} we show the mass distribution of the cores ($\Delta$N/$\Delta log(M)$, \citealt{Andre14}), of the SABOCA cores at $<5$\,kpc. The completeness level of our sample is not straightforward to assess due to the varying sensitivity, and the different dust temperatures in each field. We estimate the mass completeness use the largest noise level (0.18\,Jy/beam) for sources at $d<5$\,kpc, and assume a dust temperature of 15\,K, which is the lowest value we have for the core population. This allows us to obtain a conservative estimate of the mass completeness limit of cores in the sample. With a flux density of 3$\sigma$ and a source size close to the beam size, the mass completeness limit is approximately 41.4 M$_{\odot}$. We fit a power-law to the cores at $d<5$\,kpc above 41.4 M$_{\odot}$, and find a power-law index of -1.89$\pm$0.39. 
Although the fit has large uncertainty, the slope is larger than the Salpeter IMF ($-$1.35), and is also steeper than recent findings of mass distribution of cores in higher angular-resolution studies of the early-stages of high-mass proto-clusters (e.g. \citealt{Csengeri17b}; \citealt{Motte18}).  This could be explained by the fact that our study is more limited by confusion due to the poorer angular resolution; the slope is fitted in a higher mass range than these works suggesting that there could be intrinsically less massive cores at this larger scale (0.1-0.2 pc). 
On the other hand, the derived core mass distribution may be described better by a log-normal than a power-law distribution, as shown by previous studies (e.g. \citealt{PerettoFuller10}). Indeed we also find that moving the cutoff mass to a higher value would result in a steeper fitted slope. We also note that different source extraction methods may result in different core mass distributions (e.g. \citealt{Cheng18}) and such results should be interpreted with caution.

\subsection{Clump structure and fragmentation}\label{sec:fragmentation}

\subsubsection{Association of cores to clumps: fragmentation level}\label{sec:fmtt}

Here, we investigate the hierarchical structure of the clumps by assigning the SABOCA cores (child fragments) to their parent clumps in which they are embedded.
We focus on the 971 SABOCA sources that have physical properties derived in the analysis here.
Because the Jeans length is typically resolved up to 4\,kpc, we again consider sample in 2-4 kpc distance range, while the sample located in 1-2 kpc are added in some plots for comparison purpose.

\begin{figure*}
\hspace{0.5cm}
\includegraphics[scale=0.475]{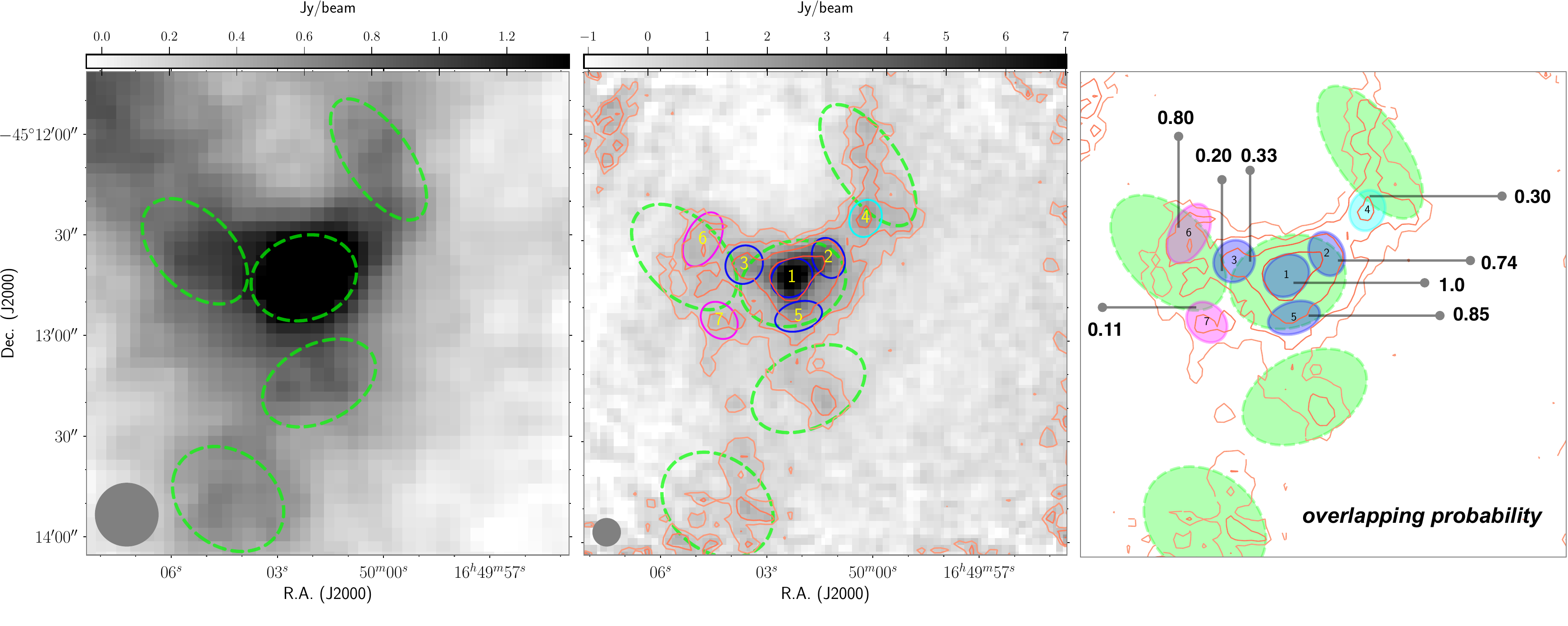}

\caption{Example of fragment allocation criterion towards an example SABOCA mapping field. The left panel shows the 870 $\mu$m LABOCA emission in grayscale, overlaid with ATLASGAL clumps in green dashed ellipses. The middle panel shows the 350\,$\mu$m SABOCA emission in grayscale with contour levels from 3$\sigma$ to the peak flux with 5 linear logarithmic spacing. Each SABOCA core is labeled with a number and will be refered to as Nr.\,x. Green dashed ellipses mark the ATLASGAL clumps as identified in Csengeri et al. (2014). Blue, magenta and cyan ellipses mark the position of three sets of fragments belonging to several ATLASGAL clumps. In the right panel (slightly zoomed in compared to the middle panel), the overlapping possibilities of each fragment with the ATLASGAL clumps are indicated with arrows. }
\label{fig:alloc_eg}
\end{figure*}

For each parent and child structure, we generate a mask based on their $FWHM$ size from \texttt{Gaussclumps} and assign the child fragments to their parental clumps based on their overlapping area as illustrated in Fig.\,\ref{fig:alloc_eg}. Clumps located at the edge of the SABOCA maps are omitted from this analysis since we may not cover the full extent of the clump. We calculate the fragmentation level in two ways: $\it{exact}$ N$_{mm}$ corresponds to the exact sum of the probability of overlapping areas of child fragments within the extent of the parent clump; the other is $\it{rounding}$ N$_{mm}$, which assigns each fragment to the clump with which it has the largest overlapping area. In short, the $\it{exact}$ N$_{mm}$ considers fragments overlapping with multiple clumps.
For example, fragment 3 in Fig.\,\ref{fig:alloc_eg}, which overlaps with two clumps giving 0.20 and 0.33 to the two clumps' $\it{exact}$ N$_{mm}$ while it is counted to the central clump as 1 in $\it{rounding}$ N$_{mm}$. In this approach, the sum of $\it{exact}$ N$_{mm}$ equals that of $\it{rounding}$ N$_{mm}$ while for some clumps, $\it{rounding}$ N$_{mm}$ may be zero if a fragment overlapping with it has a larger proportion of area inside another clump.

\begin{figure*}[!ht]
\centering
\begin{tabular} {p{0.25\linewidth}p{0.25\linewidth}p{0.25\linewidth}p{0.25\linewidth}}



\hspace{-0.93cm}\includegraphics[scale=0.31]{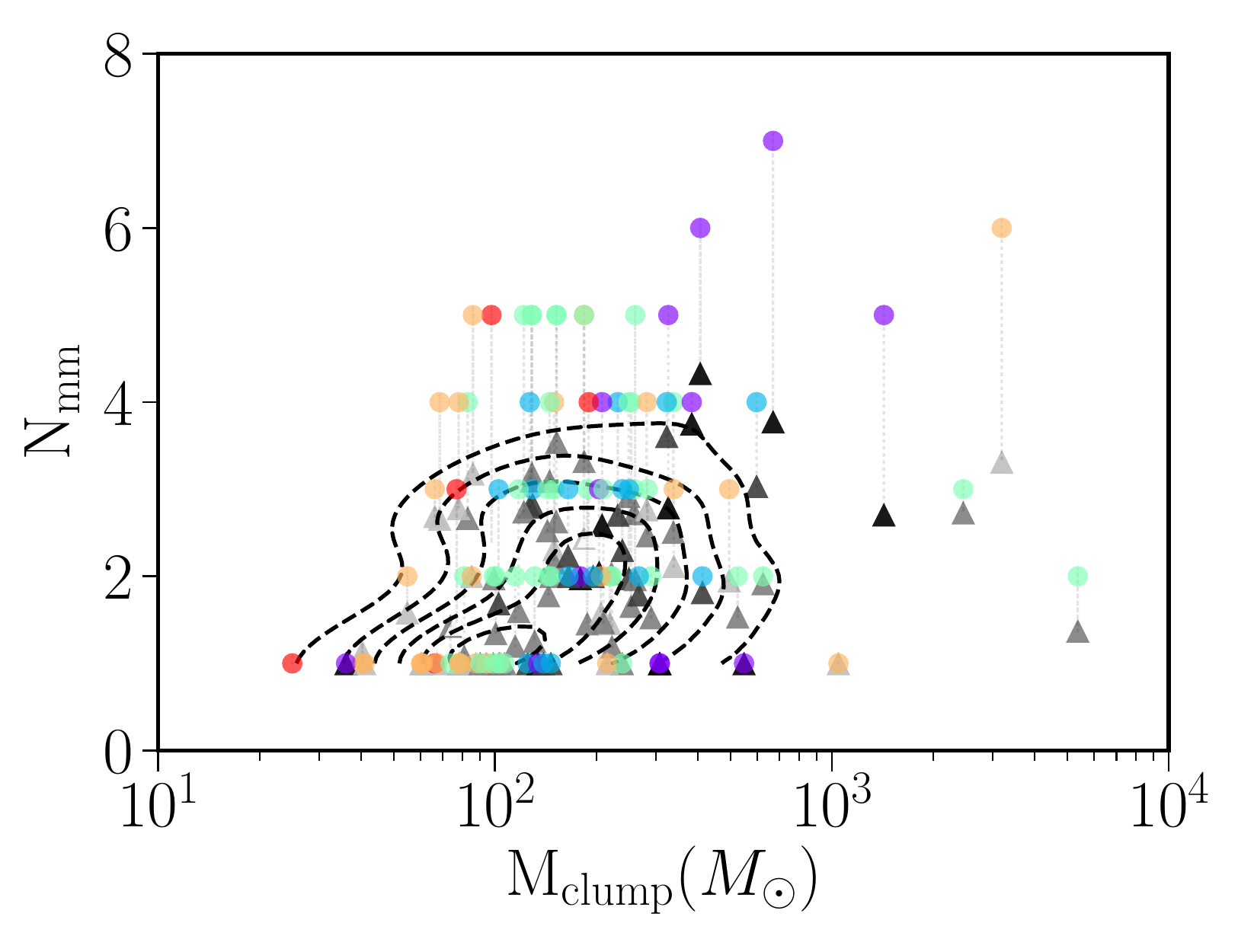}&\hspace{-0.99cm}
\includegraphics[scale=0.31]{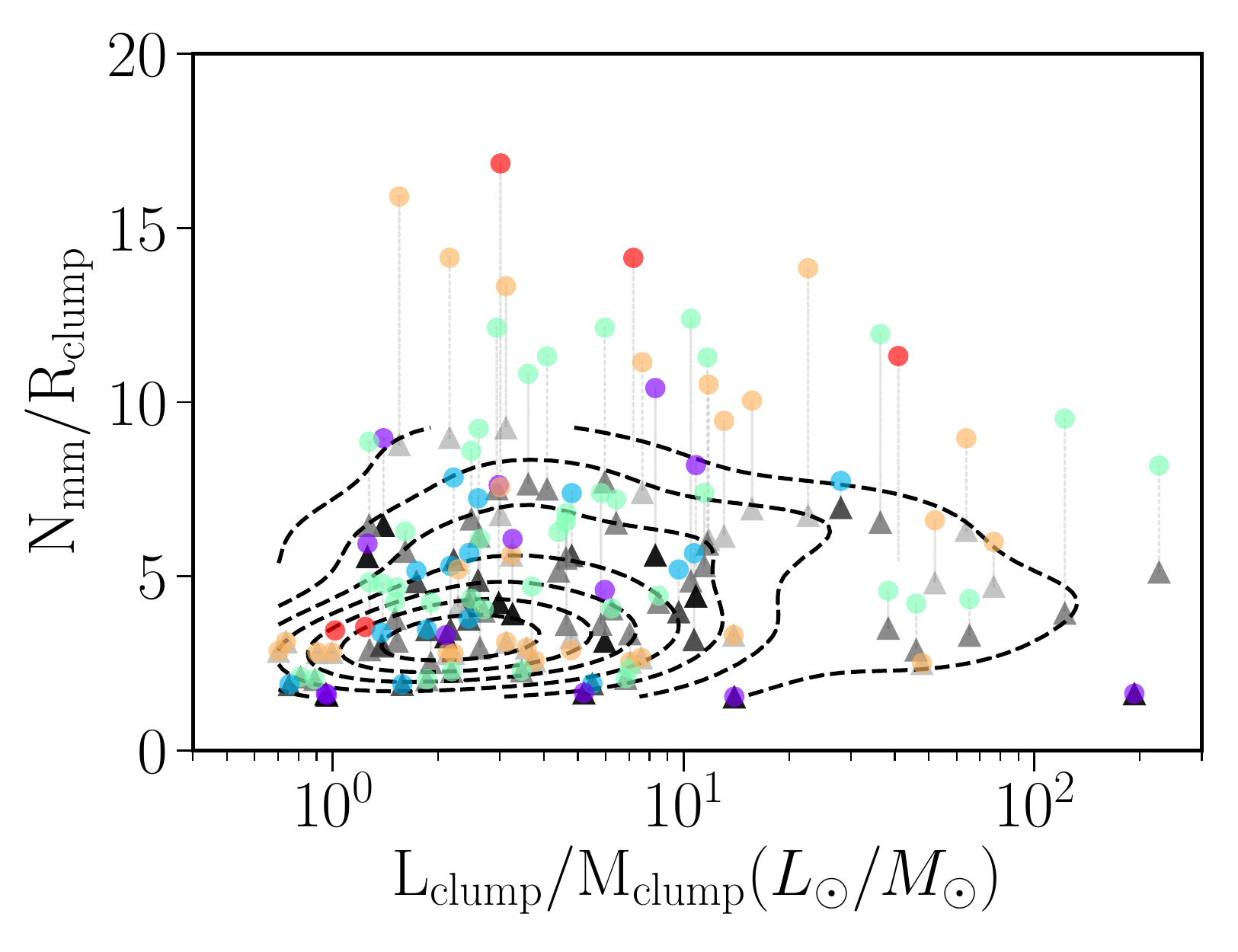}
&\hspace{-0.99cm}\includegraphics[scale=0.31]{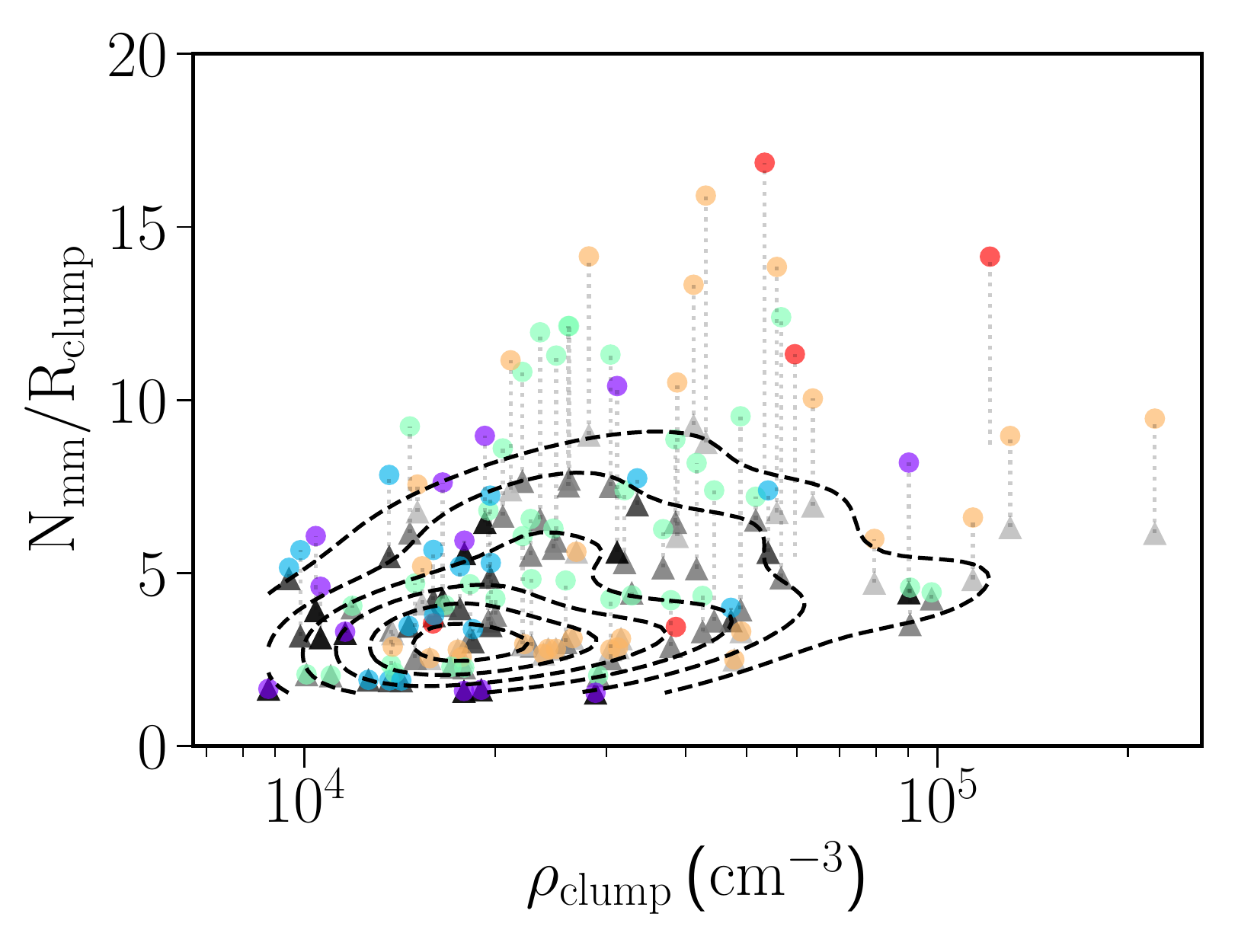}&\hspace{-0.99cm}\includegraphics[scale=0.31]{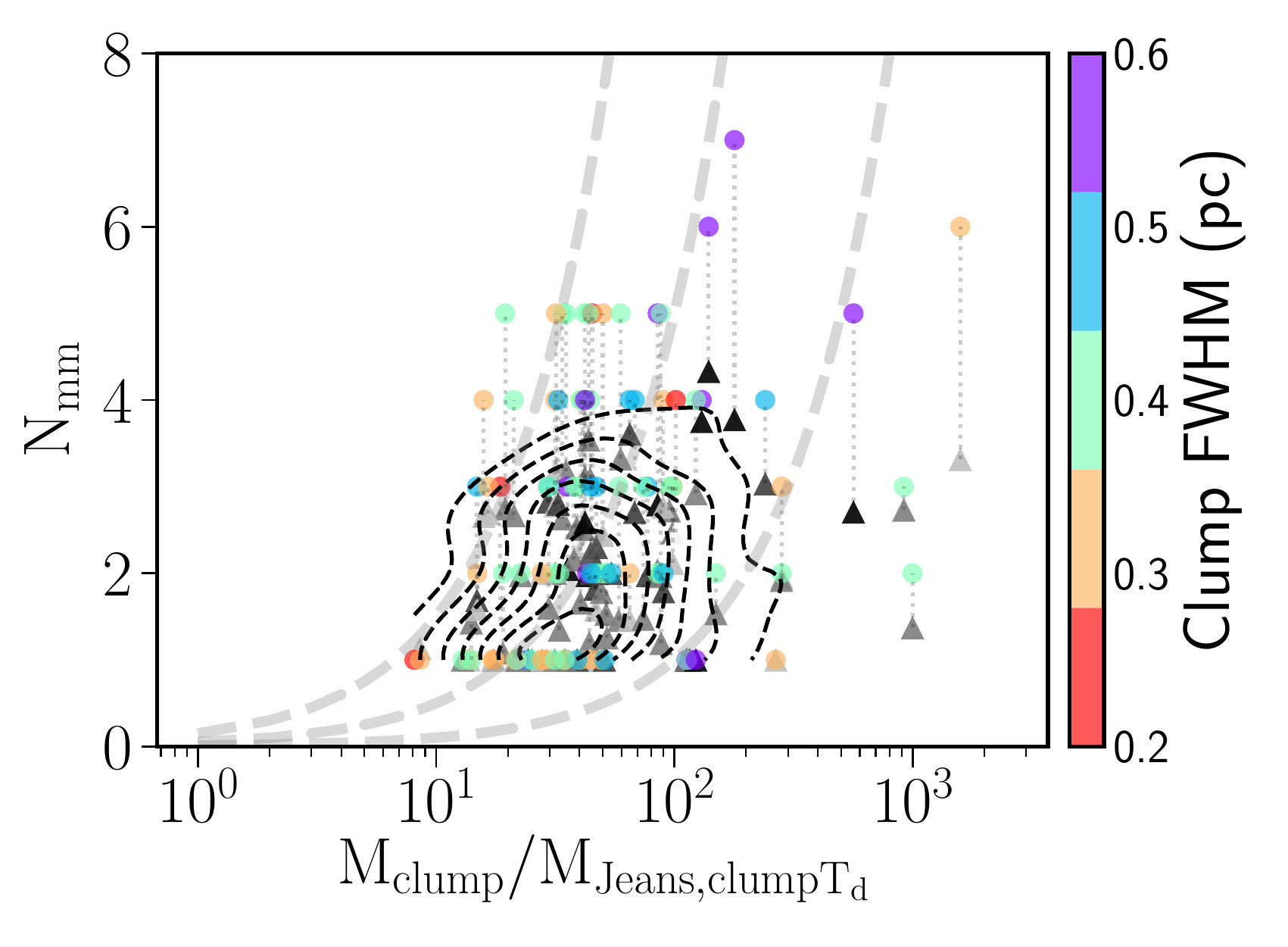}\\
\end{tabular}
\caption{Fragmentation level as a function of clump properties of sources located at a distance in 2-4\,kpc. In each plot, data points are color-coded according to the clump size.  Gray triangles show the exact fragment number {\it{exact}} N$_{mm}$; colored dots show the rounding fragment number {\it{rounding}} N$_{mm}$ (see definitions in Sec. \ref{sec:fmtt}).  Dashed contours show the distribution of {\it{exact}} N$_{mm}$ from Gaussian kernel density estimation. \emph{Left}: Number of fragments as a function of clump mass. \emph{Middle left}:  Specific fragment level ($N_{mm}$ normalised by clump radius) as a function of the clump's luminosity-to-mass ratio.  \emph{Middle right:} Specific fragment level as a function of the clump's density.  \emph{Right}:  Number of fragments as a function of predicted number of fragments based on Jeans fragmentation scenario. Gray dashed lines mark the lines of $N_{mm}$ = 0.15/0.05/0.01$\times$$M_{clump}/M_{Jeans,clumpT_{d}}$.
}
\label{fig:fmtt_level}
\end{figure*}

\begin{figure*}[htb]
\begin{tabular} {p{0.33\linewidth}p{0.33\linewidth}p{0.33\linewidth}}
\hspace{-0.5cm}\includegraphics[scale=0.40]{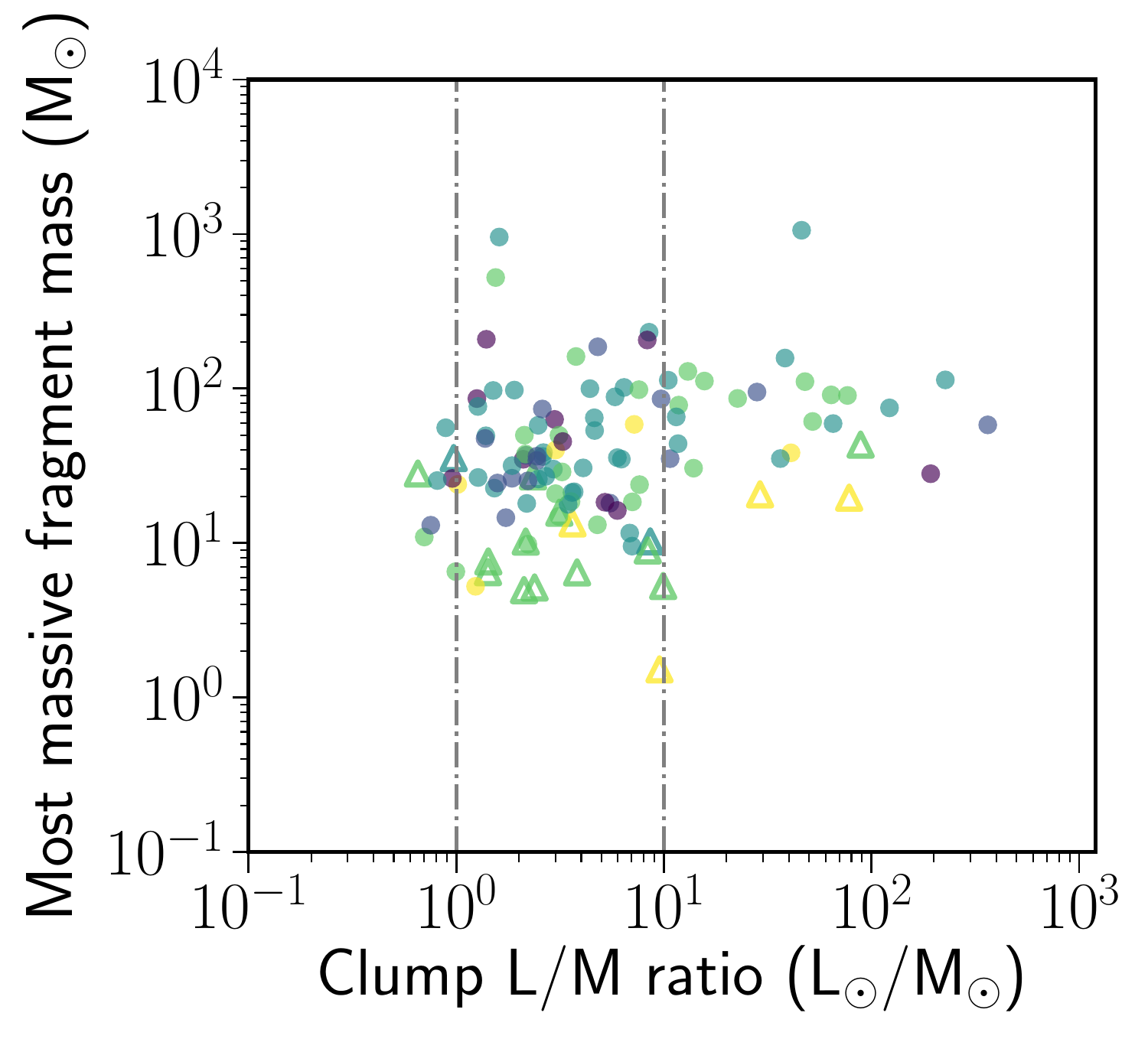}&
\hspace{-0.65cm}\includegraphics[scale=0.40]{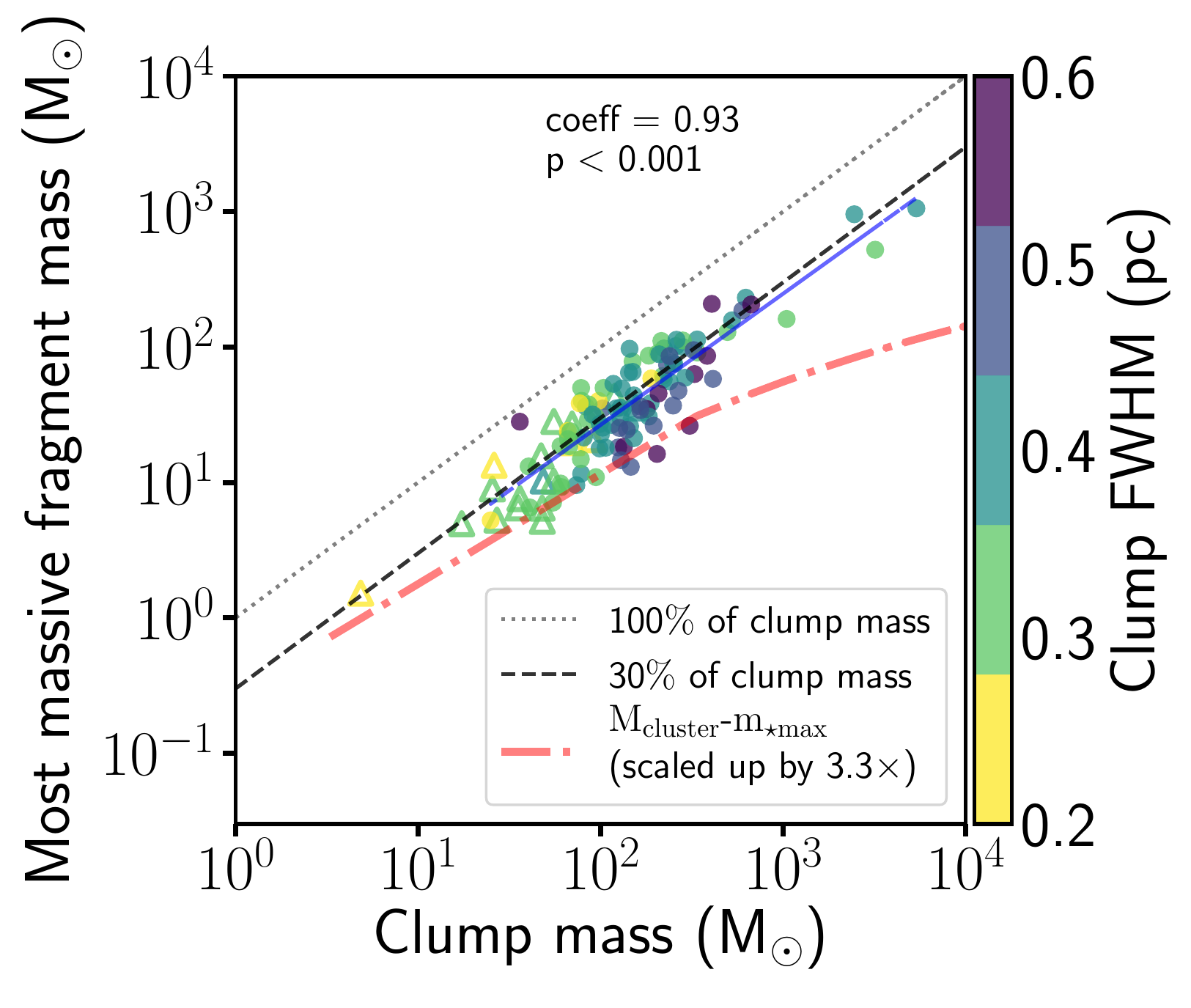}&
\hspace{-0.65cm}\includegraphics[scale=0.40]{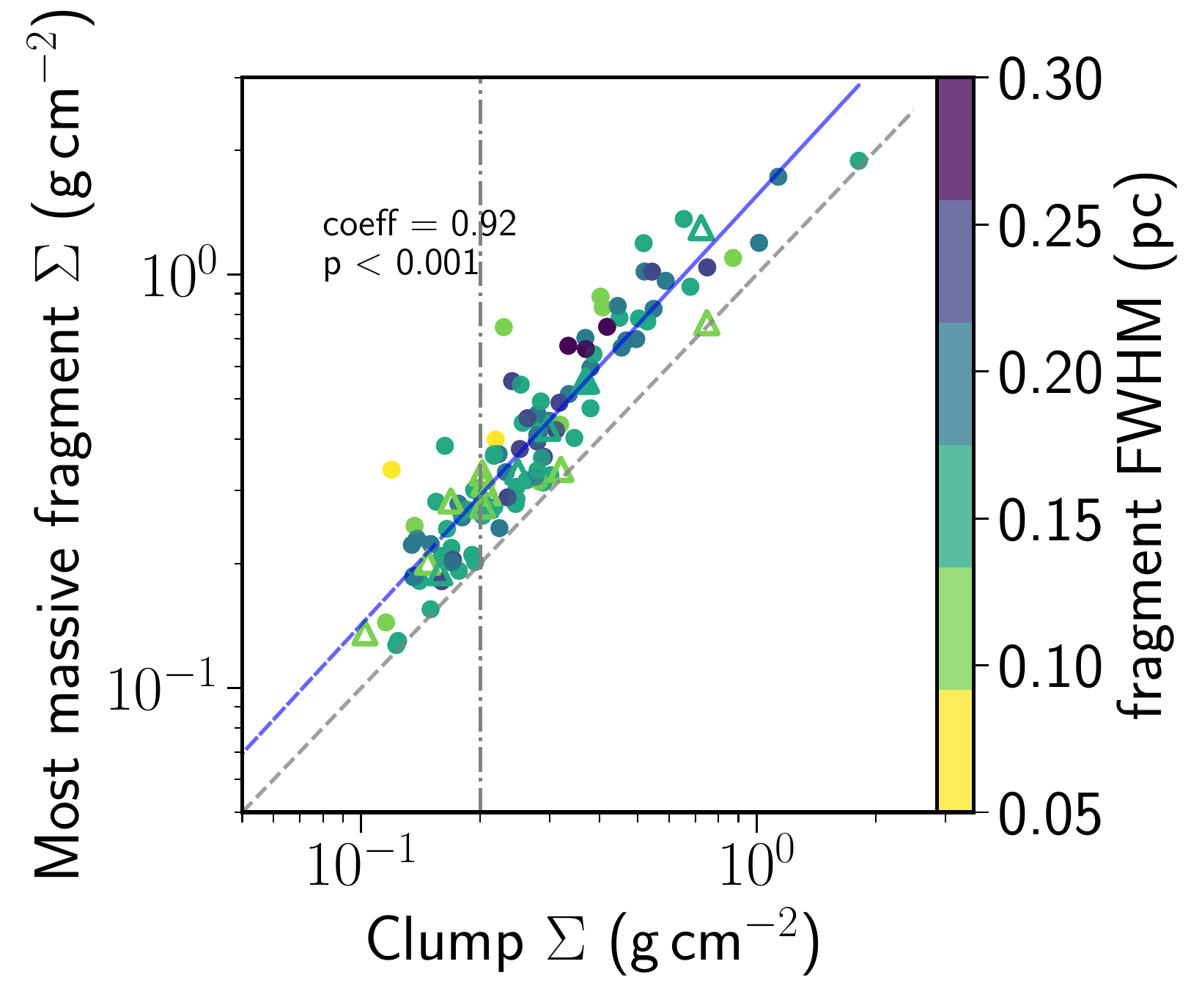}\\

\end{tabular}
\caption{Most massive fragment mass as a function of clump properties for sources at a distance 2-4\,kpc (dots) and 1-2\,kpc (triangles). \emph{Left}: Mass of the most massive fragment as a function of clump luminosity-to-mass ratio. Vertical lines mark the luminosity-to-mass ratio of 1 and 10. \emph{Middle}: Mass of the most massive fragment mass as a function of clump mass. Gray lines show the 30$\%$ and 100$\%$ proportion of clump mass; blue dashed line shows the result of a linear fit on logarithmic scale to sources in 2-4\, kpc distance range of $M_{\mathrm{fragments}}$ = 0.96log$M_{\mathrm{clumps}}-1.18$. The relation between cluster mass and maximum stellar mass in indicated in red dashed-dotted line, which is an analytical relation from \citet{Weidner13}, and is scaled up by 3.3 assuming a 30$\%$ star-formation efficiency. Left and middle plots share the same color bar, with clumps color coded according to their sizes. \emph{Right}: Surface density of the most massive fragment as a function of clump surface density. Vertical line marks the 0.2\,g\,cm$^{-2}$ empirical limit from Butler \& Tan (2012). The blue dashed line shows the result of a linear fit to sources in 2-4\, kpc distance range of log$\Sigma_{\mathrm{fragments}}$ = 1.03log$\Sigma_{\mathrm{clumps}}$+0.43.}
\label{fig:mmf}
\end{figure*}

We allocate 874 fragments to 444 clumps for the whole sample, with the remaining $\sim$100 SABOCA sources lack of parental clump structure, and 353 fragments to 161 clumps in the 2-4 kpc distance bin, 77 fragments to 37 clumps in the 1-2 kpc distance bin; consequently, each clump average $\sim$2 fragments. This is similar to what \citet{Merello15} find going from 33\arcsec\ to 8.5\arcsec\ resolution. The thermal Jeans mass and Jeans length for a clump with a temperature of 20\,K and a volume density of 10$^{4}$-10$^{5}$\,cm$^{-3}$, is 2-6\,M$_{\odot}$ and 0.1-0.3\,pc, respectively (\citealt{Bontemps10}; \citealt{Palau15}). Even at a later stage with a typical temperature of 30\,K, the Jeans mass is only a factor of two higher, and the Jeans length is only 20\% larger. Therefore, we resolve the hierarchical fragmentation of clumps to cores, on the scales of 0.1-1.0\,pc. 
Considering the typical clump properties of our sample, our results suggest that the number of fragments is broadly consistent with the Jeans length that we resolve here: the number of fragments is consistent with the number of predicted Jeans lengths within the clump's size. 
But the SABOCA cores have considerably larger masses compared to that of the thermal Jeans mass by a factor of 10-100. This has been found in similar studies of $\lesssim$0.1 pc cores or condensations (e.g. \citealt{Motte07}, \citealt{Wang11}, \citealt{Csengeri17b}).  

 
In Fig. \ref{fig:fmtt_level} we investigate in detail the fragmentation level of the clumps as a function of their physical properties for sources in 2-4 kpc distance. For the analysis, we only take into consideration parental clumps that have at least one fragment lying entirely within the clump. The coloured points mark the fragment numbers according to the {\it{rounding}} N$_{mm}$ criterion and gray points the {\it{exact}} N$_{mm}$ with each pair linked with dotted lines. We find that although the different fragment allocation methods may result in quantitatively different fragmentation levels, the general trend of fragmentation is similar. Therefore, both methods give a qualitatively similar picture.

Comparing the level of fragmentation, i.e. the number of fragments with clump mass, $L/M$ and the predicted number of fragments with Jeans scenario, we find a generally significant scatter in the fragmentation levels, while the left panel of Fig.\,\ref{fig:fmtt_level} suggests that the number of fragments increases slightly with clump mass. The Spearman correlation coefficient is 0.39 ({\it{exact}} $N_{mm}$ vs. $M_{clump}$, 0.38 of {\it{rounding}} $N_{mm}$ vs. $M_{clump}$, p-value$<$0.001), indicating a moderate correlation. 
We hereby define the specific fragmentation level as a number of fragments per unit clump size, as $N_{mm}$/R$_{clump}$. The normalisation by clump size is necessary given our limited angular resolution improvement in identifying fragments, such that one tends to pick up more fragments when the clump size is larger.
The middle panel shows the specific fragmentation level as a function of the clumps' luminosity-to-mass ratio, a proxy of the evolutionary stage (e.g. \citealt{Molinari08}; \citealt{Csengeri16a}; \citealt{Urquhart18}). There is a trend of subtle increase in the specific fragmentation as a function of $L/M$, with a Spearman correlation coefficient of 0.24 ({\it{exact}} $N_{mm}$/$R_{clump}$ vs. $L_{clump}$, 0.29 of {\it{rounding}} $N_{mm}$/R$_{clump}$ vs. $L_{clump}$, p-value$\sim$0.015). We caution that a higher $L_{\rm bol}/M$ ratio could also be due to that the most massive (proto)stars are forming within the clump (\citealt{Ma13}) rather than an evolved stage characterised by gas dispersal. 
Finally, comparing the number of fragments with the thermal Jeans fragmentation prediction, we examine the correlation between specific fragment level with clump gas density (middle right panel of Fig.\,\ref{fig:fmtt_level}), and find a Spearman correlation coefficient of 0.41 ({\it{exact}} $N_{mm}/R_{clump}$ vs. $\rho_{clump}$, 0.45 of {\it{rounding}} $N_{mm}/R_{clump}$ vs. $\rho_{clump}$, p-value$<$0.001). Based on the clump's temperature and gas density, we also find a moderate correlation between the fragment level with the predicted Jeans number ($N_{Jeans}$ = $M_{clump}/M_{Jeans}$), with a Spearman correlation coefficient of 0.31 ({\it{exact}} $N_{mm}$ vs. $N_{Jeans}$, 0.32 of {\it{rounding}} $N_{mm}$ vs. $N_{Jeans}$, p-value$\sim$0.001). We also notice that the clumps considered here have a positive correlation of their gas volume density and temperature ($T$$\,\propto\,$$\rho^{0.22}$), corresponding to the gas equation of state (EOS) having a polytropic exponent of $\gamma$$\sim$1.22, close to the upper limit relevant for Galactic molecular clouds (\citealt{SpaansSilk2000}). The increase of temperature with density means that the Jeans mass decreases slowly with density than isothermal gas, favouring less fragmentation. If compared with an assumed initial temperature of 20 K, i.e. the clump temperature when fragmentation happened (e.g. \citealt{Palau15}), a stronger correlation between observed fragmentation level with predicted Jeans number is found, with a Spearman correlation coefficient of 0.40 ({\it{exact}} $N_{mm}$ vs. $N_{Jeans}$, 0.42 of {\it{rounding}} $N_{mm}$ vs. $N_{Jeans}$, p-value$<$0.001). 
On smaller scales towards more evolved massive clumps, it seems Jeans fragmentation is at work (e.g. \citealt{Palau15}, \citealt{Beuther18}, \citealt{Liu18}). It is possible that turbulence dominates the fragmentation process at larger scale (e.g. \citealt{Zhang09}, \citealt{Wang14}) to induce a larger critical mass. On the other hand, the time delay of fragments having Jeans mass may be explained within the global hierarchical collapse scenario (\citealt{VS19}), since the density of the fragments must grow to sufficiently high contrast, and while this happens, the Jeans mass in the contracting clump will have decreased. This implies that the fragment masses, especially in early stage clumps, will in general be larger than the average Jeans mass measured for the clump.

Since {\it{Gaussclump}} is a non-hierarchical source extraction method, the association of child structures to parental structures suffer from ambiguity, which introduces a source of uncertainty in the fragmentation level and the correlations with properties of the parental clumps. Here we provide two set of fragmentation levels as a first attempt to deal with the association uncertainties. As a further benchmark, we adopt a completely independent method of source extraction ({\tt{dendrogram}}, \citealt{Rosolowsky08}) which is not hampered by the association issue, and present the same correlations between fragmentation level with clump properties in Appendix C. We find that the moderate correlations we derive here are also present in the dendrogram analysis, with some of them showing slightly tighter correlations, i.e. $N_{mm}$ vs. $M_{clump}$ (Spearman correlation coefficient of 0.42), $N_{mm}$ vs. $N_{Jeans}$ (Spearman correlation coefficient of 0.48). These results help to solidify the analysis in this section.


Studies of smaller, but more homogeneous samples of massive cores and high-mass protostars based on higher angular resolution observations, e.g. thousands of au, did not find a clear correlation between the core properties and fragmentation level, except that a larger central volume density, a larger clump mass may lead to a larger number of fragments (\citealt{Palau15}). Although we discuss here the core scale and a considerably larger sample, we can only conclude that there is moderate indication of Jeans fragmentation at work, and there is no trend of fragmentation level with clump evolution. It is, however, possible, that the overall sample covers a too wide range of Galactic clumps to reveal which clump properties could influence the fragmentation level. In addition, our source selection may not be homogeneous for massive clumps with luminosity-to-mass ratio range larger than 10 $L_{\odot}/M_{\odot}$.

In Fig.\,\ref{fig:mmf} we illustrate the relationship between the most massive fragment and clump properties for sources in 1-2 kpc (triangles) and 2-4 kpc (circles) distance bins. For these plots, we again only include cases in which the parental clump has at least one fragment lying entirely within the clump. Comparing the mass of the most massive fragment with the clump $L_{\rm bol}/M$ ratio, it seems that more evolved clumps with a larger $L_{\rm bol}/M$ ratio (Fig.\,\ref{fig:mmf}, left panel) only host core masses above $\sim$20\,M$_{\odot}$.  We hereby derive the correlations and regressions only for sources in 2-4 kpc distance range, and sources in 1-2 kpc distance range are shown in the figure for comparison purpose.  We find a strong correlation between the clump mass and the mass of the most massive core. This correlation is statistically significant and has a Spearman coefficient of 0.92 and $p$-value less than 0.001. This clearly indicates more massive cores tend to form in more massive clumps. A linear fit in the logarithmic scale yields a relation of log$M_{\rm mmf}$ = 0.96log $M_{\rm clump}$ - 1.18, which corresponds to $M_{\rm mmf}$ = 0.31 $M_{\rm clump}^{0.96}$. This relation is close to a fraction of 30$\%$ of the clump mass is assembled in the most massive fragment. In App. \ref{sec:dendro} we demonstrate this strong correlation is present when {\tt{dendrogram}} source extraction method is applied, regardless of the distinct concepts behind these algorithms in recognising cloud structures. As a comparison, \citet{Merello15} find that 19\% of the clump mass resides within the total dense substructures seen at 350\,$\mu$m, when going from 30$''$ to 8$\as$5 resolution maps. We note that the exact values of this fraction depend primarily on the relative resolution differences in defining the parental structure and substructures. We also find that the surface density of the most massive fragment correlates well with the clump surface density, and has a fitted relation of log$\Sigma_{\rm mmf}$ = 1.02log $\Sigma_{\rm clump}$ + 0.42, close to $\Sigma_{\rm mmf}\sim$1.5$\Sigma_{\rm clump}$. This indicates that the gas in the cores is denser compared to their parental clumps. 

With the approximate relations between clump and its most massive fragment $\Sigma_{\rm mmf}$$\sim$1.5$\Sigma_{\rm clump}$ and $M_{\rm mmf}$$\sim$0.31$M_{\rm clump}$, assuming the clump density profile follows a power-law form as a function of radius n(r)$\,\propto\,$r$^{-k_{\rho}}$, a density structure of n(r)$\,\propto\,$r$^{-1.5}$ is obtained, equivalent to that the clump enclosed mass profile is M(r)$\,\propto\,$r$^{1.5}$. 
The self-similar density profile of n(r)$\,\propto\,$r$^{-k_{\rho}}$ with $k_{\rho}$ in range of 1.6-1.9 is obtained in gas spheres of high density from simulations of self-gravitating supersonic turbulence (\citealt{Collins12}, \citealt{MurrayChang15}). 
In particular, \citet{MurrayChang15} suggests that the n(r)$\,\propto\,$r$^{-1.5}$ is an attractor solution at clump inner radii that is reached over clumps' self-similar gravitational collapse, when the density is then independent of time, i.e. n(r, t)$\,\rightarrow\,$n(r). 
Furthermore, the clump density power-law forms can be directly linked with the power-law tails seen in molecular cloud column density probability distribution functions (N-PDF), lnN $\sim$ N$^{-s}$, with s = 2/($k_{\rho}$-1) (\citealt{FK13}). k$_{\rho}$$\sim$1.5 thus corresponds to an N-PDF slope of -4.  \citet{Lin17} proposed a tentative critical value of the N-PDF power-law slope of -4 to separate massive star-forming cloud intermediate evolutionary stages, which have a shallower slope, with early or late stages, by analysis of 10 massive star-forming clouds from infrared dark stage to evolved stage with embedded luminous OB clusters. Since most of the clumps in consideration here are of L/M $<$ 10,  which are in relatively early evolutionary stage (\citealt{Molinari08}), this small scale clump density distribution is in accord with the evolution trend of cloud N-PDF, which are compatible indirect observational evidence of cloud multi-scale collapse, as in global hierarchical collapse scenario (\citealt{VS19}).

In middle panel of Fig.\,\ref{fig:mmf} we also plot the mass of the most massive star in a cluster ($M_{\rm max}$) as a function of the cluster mass ($M_{\rm ecl}$) (\citealt{Weidner13}) scaled up by 3.3 times, together with the relation of masses between the most massive fragments and parental clumps. Here we make the simple assumption that the mass of most massive stellar member formed in the most massive fragment is proportional to the fragment mass with a constant SFE of 30$\%$ regardless of the mass-scale. The relation between $m_{\rm max}$ and $M_{\rm ecl}$ indicates that the mass of the most massive star systematically changes with the cluster mass in stars, and is incompatible with a scale-free IMF (\citealt{Weidner13}). The sampling procedure that introduces such a relation also suggests star formation is highly regulated,  with low-mass stars forming until feedback from the massive stars is able to prevail against the gravity dominated formation process (\citealt{weidnerkroupa06}). The fact that the formation of low-mass stars starts early and continues until massive stars are formed is seen in the numerical simulations of cloud formation and collapse by converging flows, over a time span of a few Myr (\citealt{VS17}). The relation of the most massive fragment and clump mass is compatible with the $m_{\rm max}-M_{\rm ecl}$ relation, with the most massive core likely to be the progenitor of the most massive stellar member and perhaps fragmenting into somewhat less massive cores which may also remain unresolved at present, indicating the reservoir clump mass to be essential in regulating the mass distribution at smaller scales. The larger deviation of our observed most massive core-clump mass relation to $m_{\rm max}-M_{\rm ecl}$ in larger mass range is understandable since we do find more massive objects tend to have more fragments as discussed before. With higher angular resolution observations we might resolve the bending of the most massive core-clump mass relation. Numerical simulations of rotating massive cloud cores with radiative feedback have indicated that fragmentation-induced starvation could be a possible scenario that set the limit of the mass of the most massive stars (\citealt{Peters10}).

\subsection{Massive quiescent cores: candidate high-mass pre-stellar cores?}\label{sec:irquietmassive}

\begin{table*}

\begin{threeparttable}
\caption{Candidate massive pre-stellar cores with $>$100 M$_{\odot}$ at < 5 kpc.}
 \label{tab:pre_cores}

\begin{tabular}{llllllll}
\toprule
                name &   Distance & v$_{LSR}$& SiO ($J$=2--1)& HCO+/H$^{13}$CO+ lines & Outflow?$^+$ & Infall? & \\
                &(kpc)&(km/s)& detection & \\
\midrule      
 GS327.2954$-$0.5787 &   2.8 &  -44.2$\pm$1.0 &   det.$^a$ & det.  &  -- & y \\
 GS327.2990$-$0.5735 &   2.8 &      -44.0$\pm$1.3&det.$^a$$^d$ &  det. & -- & y & \\
 GS333.1274$-$0.5599 &   3.5 &        -54.9$\pm$1.7&det. $^{a,d}$ &   det. & -- & y & \\
 GS333.1294$-$0.5554 &   3.5 &    -57.9$\pm$2.1&no det.$^{a,d}$ &  det. & -- &  y & \\
 GS333.1353$-$0.5645 &   3.5 &    -57.8$\pm$1.9&   det.$^{a,d}$ &  det.  & -- & y   &\\
 GS333.1241$-$0.5594 &   3.5 &        -57.3$\pm$2.2& det.$^{a,e}$ &  det.  & -- &  y   &\\
 GS333.1280$-$0.5520 &   3.5 &       -57.7$\pm$1.9 &no det.$^{a,e}$ &  det.  & -- &  y & \\
 GS333.4641$-$0.1614 &   2.9 &    -43.1$\pm$1.4&no det.$^a$ &   det. & -- & -- &  \\ 
 GS333.6570$+$0.0587 &   5.0 &       $-$84.6$\pm$2.3&no det.$^a$ &  det. & -- &--  & \\
 GS334.7438$+$0.5049 &   3.9 &       -62.2$\pm$5.8&no det.$^a$ & det. & -- &--  & \\
 GS336.8802$-$0.0079 &   4.6 &      -73.8$\pm$3.0&no det.$^a$ & det. & -- & -- & \\
 GS340.2403$-$0.3735 &   3.6 &       -50.1$\pm$0.9&det. $^a$ & det.  &  -- &  y  &\\
 GS340.2504$-$0.3804 &   3.6 &       -50.3$\pm$0.8& det.$^{a,d}$ &   det.  &   -- & -- & \\
 GS341.9324$-$0.1741 &   3.3 &       -43.6$\pm$0.7& no det. $^a$  &  only HCO$^+$ $^a$ & -- & -- & \\
 GS344.5771$+$0.2166 &   4.8 &       -62.3$\pm$0.8&no det & no det.$^a$ & -- & -- &  \\
 GS351.4378$+$0.6537 &   1.4 &       -4.2$\pm$0.4&det. &  det.$^{a,c}$ & y & y & \\
 GS351.4416$+$0.6534 &   1.4 &       -4.2$\pm$0.4& det. & det.$^{a,c}$ & y &  y & \\
   GS008.4009$-$0.2888 &   4.4 &       37.2$\pm$1.1&no det &  det.$^a$ & -- &  y  & \\
   GS008.4068$-$0.2995 &   4.4 &       37.0$\pm$1.4&no det &  det.$^{a,d}$ & -- & y & \\
   GS009.2098$-$0.2012 &   4.8 &       42.6$\pm$2.1&no det & det.$^a$   & -- &  -- & \\
  GS019.0131$-$0.0268 &   4.2 &  59.8$\pm$0.5&     -- & --& & &  \\
  GS012.8796$-$0.2871 &   3.0 &  34.2$\pm$2.7&     det. $^{b,c}$ &  det.$^a$  & -- & y &  \\
  GS028.5585$-$0.2398 &   4.7 &    85.5$\pm$0.07&det.$^{b,d}$ &   -- &  &  \\
  GS028.5643$-$0.2313 &   4.7 &    85.5$\pm$0.07&det.$^{b,d}$ &   -- &  &  \\
  GS028.5665$-$0.2376 &   4.7 &     85.5$\pm$0.07&det.$^{b,d}$&   -- &   &  \\
  GS035.5766$+$0.0068 &   3.2 &     54.2$\pm$0.18&det.$^b$  & -- &   &  & \\
  GS035.5816$+$0.0116 &   3.2 &  54.2$\pm$0.18&det.$^b$  & -- &  & &   \\
\bottomrule

\end{tabular}
    \begin{tablenotes}
      \item\small $^a$: MALT90  data, $^b$: results from \citet{Csengeri16a}, $^c$: possible contamination by a nearby bright source, $^d$: within the same beam as the core above. In the last two columns, sources without observations are left blank; ``y'' stands for confirmation and ``-'' otherwise.
   
      \end{tablenotes}
  \end{threeparttable}
\end{table*}

The higher resolution 350\,$\mu$m observations with SABOCA compared to the 870\,$\mu$m LABOCA maps of ATLASGAL offer the possibility to better distinguish between star-forming and quiescent cores. We take this opportunity to investigate in more detail the brightest SABOCA sources that are quiescent and lack embedded compact sources at 24 and 70 $\mu$m. To select the most massive cores, we use our mass estimates measured in a radius of 1$\sigma$, corresponding typically to a mass measurement within an $FWHM$ diameter of 0.17\,pc for a 0.2\,pc core. We select cores that are located at $d$$<$5 kpc and have $M_{\rm{1}\sigma}$$>$100 $M_{\odot}$, assuming that they are capable of forming at least one early B or O-type star. Among all SABOCA cores, we find 87 sources that fulfil this distance and mass criteria, of which 34 (37\%) are not associated with any embedded compact sources at 24 and 70\,$\mu$m. The fraction of star-forming versus quiescent cores is higher in this high-mass regime than towards all sources within 5\,kpc where 58\% are found to be quiescent. Considering only the most massive cores the relative fraction of star-forming versus quiescent cores is significantly higher, which is consistent with other studies (e.g. \citealt{Csengeri14, Urquhart18}) suggesting shorter formation timescales for the most massive clumps.

A quiescent massive core with $>100$\,M$_{\odot}$ within a typical radius of 0.085\,pc has a volume density $>7.5\,\times10^5$\,cm$^{-3}$, and a mass surface density of 0.92\,g\,cm$^{-2}$ assuming a uniform density distribution. However, most of these sources typically have non-flat flux distribution, suggesting a steeper density profile, as it is the case for their embedding clump. The bulk of the gas is cold with dust temperatures between 14 and 25\,K, and on average of 18\,K. Like the overall population of quiescent cores, the massive quiescent cores also exhibit lower bolometric luminosities between $60$ and $2300$\,L$_{\odot}$.

To investigate whether these massive quiescent cores could qualify as candidates for extremely elusive and rare high-mass pre-stellar cores (\citealt{Csengeri17a}; \citealt{Motte18}), we visually inspected these candidates, and removed 7 cores from the final list that are too close to and blended with a bright source at 24 \,$\mu$m. This left us with a final list of 27 sources (Table \ref{tab:pre_cores}). Being embedded in a larger scale clump, these sources are typically surrounded by other cores, some of which are lower mass and/or star-forming. We do not find any isolated or single candidate high-mass pre-stellar core above the $>100$\,M$_{\odot}$ mass limit. An example of the massive quiescent core embedded in clump structure with lower-mass counterparts is shown in Fig.\,\ref{fig:mm_dark_frag}

We searched for available ancillary higher angular resolution archival data, that would help us to investigate the nature of these cores further, and found that G333.1298-0.5602 and G333.4659$-$0.1641 have been observed at high angular resolution as part of the SPARKS project using ALMA (\citealp{Csengeri17b,Csengeri18}). These studies suggest, however, that the brightest fragments on a 2000\,au scale in these sources are high-mass protostellar objects rather than pre-stellar cores.

To further probe the star formation activity of these cores we examined ancillary data on molecular tracers obtained with the IRAM 30m telescope described in \citet{Csengeri16a}, and the MALT90 survey observed with the MOPRA telescope \citep{Foster13} which are available for all but one of the sources. These observations, however, have a coarser angular resolution than the SABOCA maps are, therefore, only indicative as an independent probe of their star formation activity.

Similarly to \citet{Motte07}, we first investigate here the velocity profiles of the SiO (2--1) line. SiO is a widely used tracer for shocks, in particular, it becomes abundant in fast shocks associated with jet activity (e.g. \citealt{Schilke97}; \citealt{Codella99}). Therefore, high-velocity line wings could be indicative of the presence of jets and outflow activity related to embedded protostars, however, it might also originate from a nearby more evolved core if the resolution is lower than our SABOCA maps. We use here the pointed observations of \citet{Csengeri16a} and the MALT90 maps \citep{Foster13}, where we extracted and averaged the spectra towards the core position within the MOPRA beam ($\sim$38$''$). For the MOPRA data, these figures are shown in App. C. Except for the cores GS351.4379+0.6537 and G351.4416+0.6534 we find no clear evidence of a broad velocity component in this line towards our candidate high-mass pre-stellar cores, however, the lack of SiO line-wings could also be due to sensitivity limitations. Where available (6 sources), we checked the IRAM 30m observations for the CO (1--0) line covered by our spectral survey described in \citet{Csengeri16a} that could be a more sensitive probe for high-velocity outflowing gas. Although the line shape is strongly affected by absorption from the reference position, there is no clear evidence for high-velocity emission that could be related to outflowing gas in this tracer either.  The observed line wings of the SiO (2-1) transitions towards GS351.4379+0.6537 and G351.4416+0.6534 are likely due to blending with a star-forming core within the beam of the MOPRA observations.  From the remaining 24 sources, 13 (54\%) have detections in the SiO (2--1) line showing a narrow SiO component and altogether 11 sources show no detection in the SiO (2--1) line at all. To conclude, the SiO (2--1) line does not provide an indication for deeply embedded ongoing star formation activity.

To make use of the available spectroscopic data, we also investigated whether infall signatures on the clump scale could be revealed towards our sample. While \citet{Csengeri16b} suggests that most of the massive clumps are gravitationally unstable, spectroscopic signatures of infall (e.g.\,\citealt{Myers1996}; \citealt{Mardones97}; \citealt{WuEvans03}) could provide an indication that the clump itself is collapsing (see, however, \citealt{Smith2013}). To search for an infall signature, i.e. red-shifted self-absorption of optically thick lines, we use the HCO$^+$ and H$^{13}$CO$^+$ ($J$=1--0) lines, and where detected, we used a single Gaussian fit to the H$^{13}$CO$^+$ line to extract the $v_{\rm lsr}$ of the clump. Although a proper interpretation of these line profiles would require dedicated modelling, visual inspection of the spectra suggests possible infall motions towards 16 sources (60\%). The information obtained from molecular line data is listed in Table\,\ref{tab:pre_cores}.

Considering a typical star formation efficiency of 20-40$\%$ (\citealt{Tanaka17}), the SABOCA cores with $>$100\,M$_{\odot}$ are good candidates to form at least one massive star. The absence of 24-70\,$\mu$m emission, together with the indications from SiO/HCO$^+$/H$^{13}$CO$^+$ tracers suggest that except for two cores, there is no further indication for star formation activity related to jets and outflows in the sample based on the SiO (2--1) line, while more than half of the clumps hosting half of the cores shows indications of the presence of global infall motions. We identify 7 sources where the SiO (2--1) line is not detected and there is no indication for infall. On the clump scale, these sources could be considered as starless massive sources. However, we cannot exclude on-going low-mass star-formation when the core would fragment and form only a low-mass cluster. What physical processes determine the possibility for a massive core to form high-mass stars is not yet clear.

 We conclude that considering all of the available low angular-resolution data, our sample of candidate high-mass pre-stellar cores is robust, although deeply embedded high-mass protostars have been detected towards mid-infrared quiet massive cores and clumps (\citealt{Bontemps10}; \citealt{Wang11}; \citealt{Ohashi16}; \citealt{Csengeri17b}). High angular resolution observations are needed to confirm their nature.

\begin{figure}
\begin{tabular}{p{0.75\linewidth}}
\hspace{-0.1cm}\includegraphics[scale=0.45]{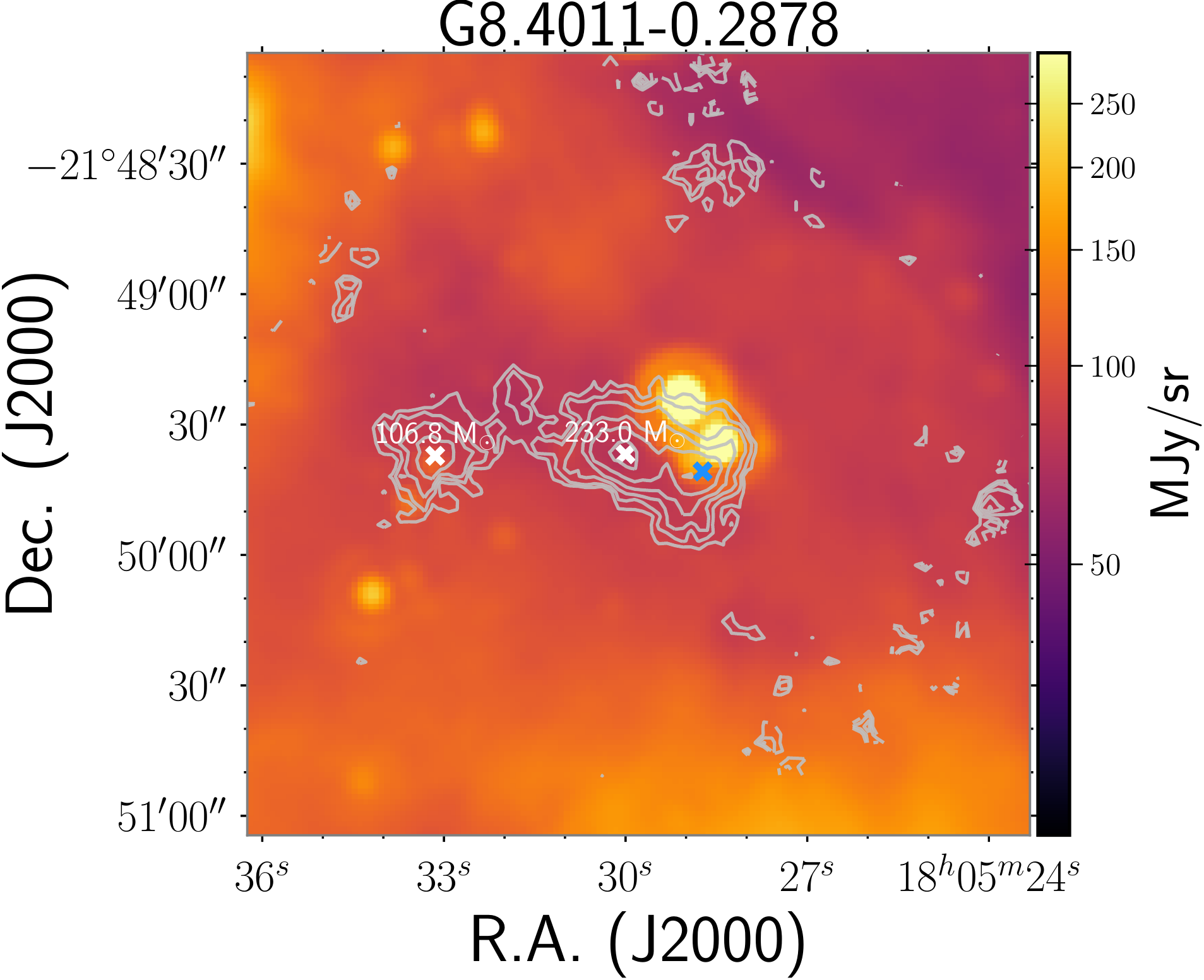}\\
 \end{tabular}
 \caption{Example of a clump hosting massive quiescent cores. White crosses mark the $>$100\,M$_{\odot}$ quiescent cores and blue cross marks the star-forming cores. The masses for quiescent cores are indicated in the figure. Contour levels are from 7$\sigma$ to the peak flux with 6 logarithmic linear levels in-between.}
 \label{fig:mm_dark_frag}
\end{figure}

\section{Conclusions}

We present the largest sample of Galactic clumps studied at a moderate resolution of $8\as5$ at 350\,$\mu$m observed with APEX/{\it{SABOCA}}. By combining with available {\it{Herschel}} and {\it{LABOCA}} observations from ATLASGAL, we analyze the core properties derived from SED fits and investigate the fragmentation as a function of the physical properties of clumps. Our major findings are as follows:

\begin{enumerate}

\item Our {\it{SABOCA}} observations detect the 350\,$\mu$m emission towards all targeted ATLASGAL clumps and reveal a variety of morphology from compact, relatively isolated sources to complex filamentary structures with branches. We find that the majority of the targeted sources exhibit a filamentary structure emission. Fields dominated by a single, bright source at 350\,$\mu$m represent only a minor fraction of the sample.\\

\item We identify 1120 compact sources at 350\,$\mu$m using {\it{Gaussclumps}}. We estimate the physical properties towards 971 SABOCA sources from SED fits. The average $FWHM$ size is 0.32\,pc, dust temperature is 21.6\,K, average mass is 198\,M$_{\odot}$ and average surface density is 0.40\,g\,cm$^{-2}$. Among them, 405 cores are located in a distance range of 2-4\,kpc; these have an average $FWHM$ size of 0.19\,pc, an average dust temperature of 21.8\,K, average mass of 52\,M$_{\odot}$ and average surface density of 0.40\,g\,cm$^{-2}$. We find a systematic shallowing of slopes of the cores' mass-radius relation as a function of distance, from $\sim$3 to $\sim$2. \\

\item Using the presence of 22-24 $\mu$m and 70\,$\mu$m point sources, we distinguish between star-forming and quiescent SABOCA sources. The majority of the sources lack mid- and far-infrared counterparts (56.1$\%$) and are classified as quiescent cores. For the distance-limited sample (2-4 kpc), we find that the quiescent cores have a similar mass range as the star-forming cores but are somewhat more extended and hence less dense, and of lower dust temperature (0.16 pc, 0.40\,g\,cm$^{-2}$,18.9 K) than the star-forming cores (0.15 pc, 0.50\,g\,cm$^{-2}$, 24.0 K). \\

\item  We find on average each clump hosts two cores, which is broadly consistent with the length scale of thermal Jeans fragmentation. The core masses are, however, a factor of 10-100 times larger than what the thermal fragmentation would predict. The observed number of fragments is moderately correlated with the clump density and the number of fragments predicted by pure thermal Jeans fragmentation inferred from the clump properties (correlation coefficient of $\gtrsim$0.4, p-value$\sim$0.001). We do not find strong indication that a larger number of cores would form as the clumps evolve as traced by their increasing $L_{\rm bol}/M$ ratio. \\

\item We find a strong correlation between the clump mass and the mass of the most massive core, close to that $\sim$30\% of the clump mass is assembled in the most massive core. The surface density of the most massive fragment is increased compared to the parental clumps. \\

\item We identify 27 massive ($>100$\,M$_{\odot}$) quiescent cores at $d<5$\,kpc that represent promising candidates of massive star-forming progenitors at earliest stages.  Further high-resolution observations are necessary to confirm the nature of these cores and settle whether they could host genuine high-mass pre-stellar cores.\\


\end{enumerate}




\begin{acknowledgements}
     Y. Lin thanks Pavel Kroupa for helpful discussions, and Dario Colombo for his suggestions on {\tt{Dendrogram}} analysis.
     T.Cs. acknowledges support from the \emph{Deut\-sche For\-schungs\-ge\-mein\-schaft, DFG\/}  via the SPP (priority programme) 1573 'Physics of the ISM'. 
     \\
     This work have made use of the following Python libraries/packages: {\tt{Astropy}} (\citealt{astropy}), {\tt{Matplotlib}} (\citealt{matplotlib}), {\tt{Numpy}} (\citealt{numpy}), {\tt{Scipy}} (\citealt{scipy}), {\tt{Scikit-image}} (\citealt{scikit-image}), {\tt{Pandas}} (\citealt{pandas}), {\tt{PySpecKit}} (\citealt{pyspeckit}), and {\tt{Aplpy}} (\citealt{aplpy}). 
\end{acknowledgements}

\bibliography{all}

\clearpage
\begin{appendix} 
\section{SABOCA maps}\label{app:sab_maps_all}
\begin{python}%
import os
directory = r"./appendix1_sm/"
extension = ".pdf"
files = [file for file in os.listdir(directory) if file.lower().endswith(extension)]
for i, file in enumerate(files[::4]):
       if i!=102:
        print r"\begin{figure*}[h!]"
        print r"\hspace{-0.8cm}"
        print r"\vspace{-0.2cm}"
        print r"\begin{tabular}{p{5.2cm}p{4.2cm}p{5.2cm}p{4.2cm}}"

        file1 = files[i*4:i*4+1][0]
        file1 = file1[:-4]
        file2 = files[i*4+1:i*4+2][0]
        file2 = file2[:-4]
        file3 = files[i*4+2:i*4+3][0]
        file3 = file3[:-4]
        file4 = files[i*4+3:i*4+4][0]
        file4 = file4[:-4]
        if file1 !=[]:
           print r"\hspace{0.55cm}\includegraphics[width=5.cm,keepaspectratio]{./appendix1_sm/{
        else:
           print r"&"
        if file2 !=[]:
           print r"\hspace{-0.0cm}\includegraphics[width=4.cm,keepaspectratio]{./appendix1_sm/{
        else:
           print r"&"
        if file3 !=[]:
           print r"\hspace{-0.3cm}\includegraphics[width=5.cm,keepaspectratio]{./appendix1_sm/{
        else:
           print r"&"
        if file4 !=[]:
           print r"\hspace{-0.75cm}\includegraphics[width=4.cm,keepaspectratio]{./appendix1_sm/{
           
        else:
           print r"\\"

        print r"\end{tabular}"
        if (i+1)
        	   print r"\caption{Continued.}"
        if i==4:
            print r"\caption{SABOCA 350\,$\mu$m and 24 $\mu$m emission from Spitzer/MIPS emission maps of target sources. Contour levels start at 5$\times \sigma$ and show 5 uniformly spaced intervals on a logarithmic scale up to peak flux density of each field of 350\,$\mu$m flux. The green crosses show the position of ATLASGAL sources. Coloured images show the 24 $\mu$m emission from Spitzer/MIPS on a logarithmic scale, contours are the same as in the top panel of the 350\,$\mu$m emission. red crosses mark the positions of the compact sources identified in the SABOCA maps. The target name from the ATLASGAL Gaussclumps catalog is given in each plot. }"

        print r"\end{figure*}"
        if (i+1)
           print r"\clearpage"
\end{python}%

\section{The 70 $\mu$m scaling factor and the comparison between the 25$''$ and 10$''$ $N(H_{\rm2}$) and $T_{\rm d}$ maps}
The method we adopt for the 10$''$ resolution SED fit is based on the assumption that the SED at longer wavelengths 160, 250, 350 and 870 $\mu$m fluxes represent the cold component of gas which dominate the bulk of mass (envelope of cores) at the scale we are probing.  
Thus the predicted 70 $\mu$m fluxes from this SED compared to the observed fluxes provide a correction factor to account for the hot gas component, which otherwise would bias the derived $T_{\rm d}$ to a higher value and hence underestimate the gas column density. 

\begin{figure*}
 \begin{tabular}{p{0.5\linewidth}p{0.5\linewidth}}

\hspace{-0.2cm}\includegraphics[scale=0.31]{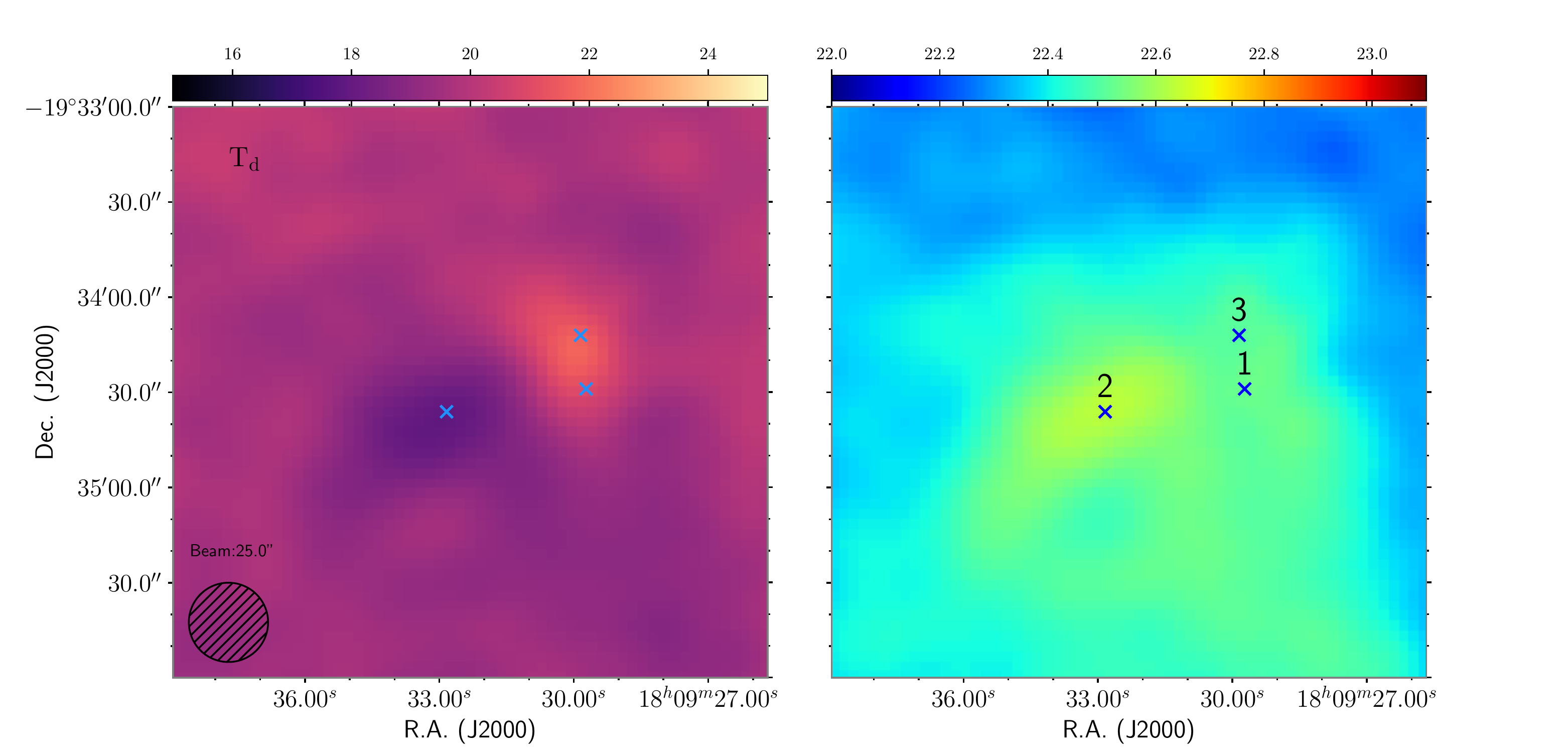}&
\includegraphics[scale=0.31]{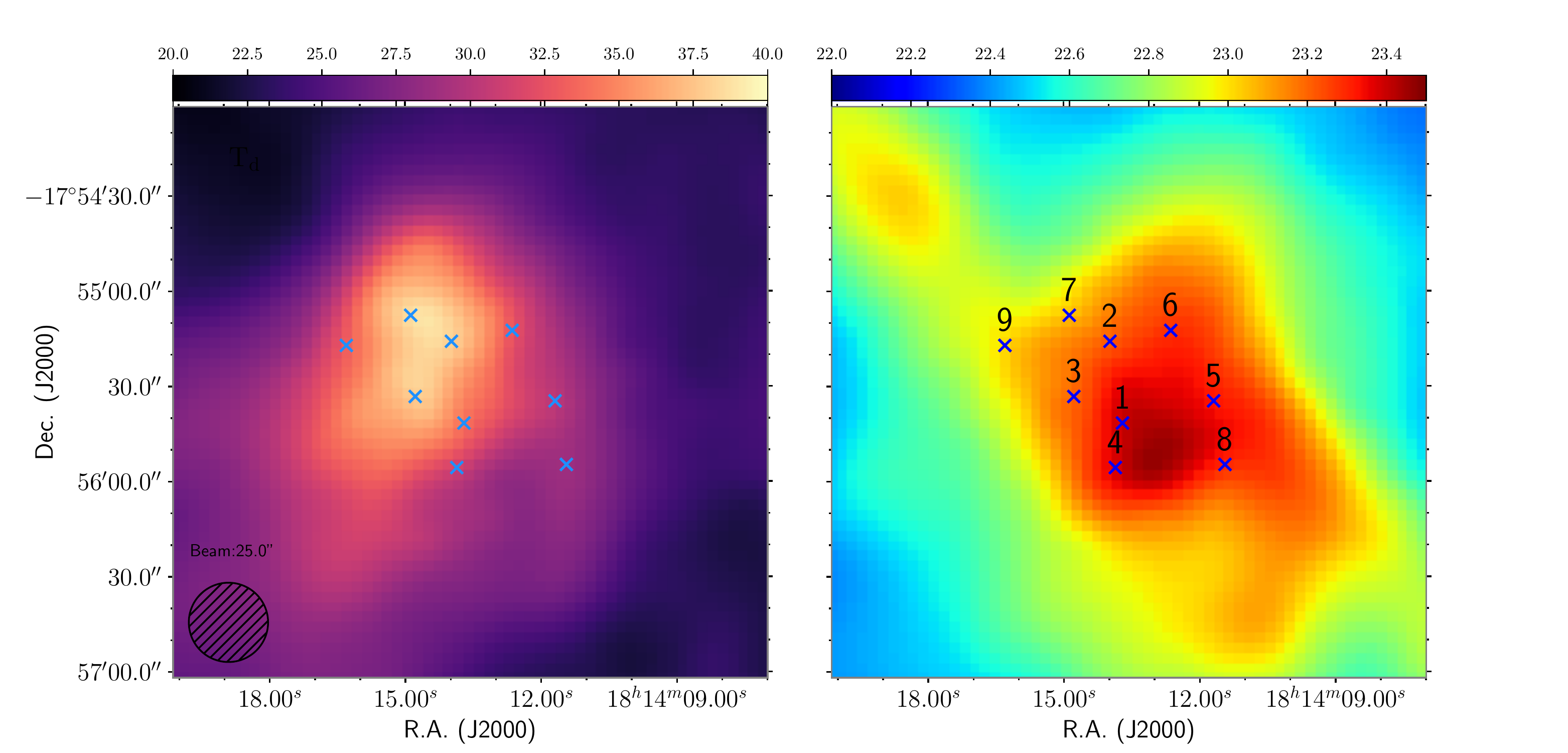}\\
\hspace{-0.8cm}\includegraphics[scale=0.33]{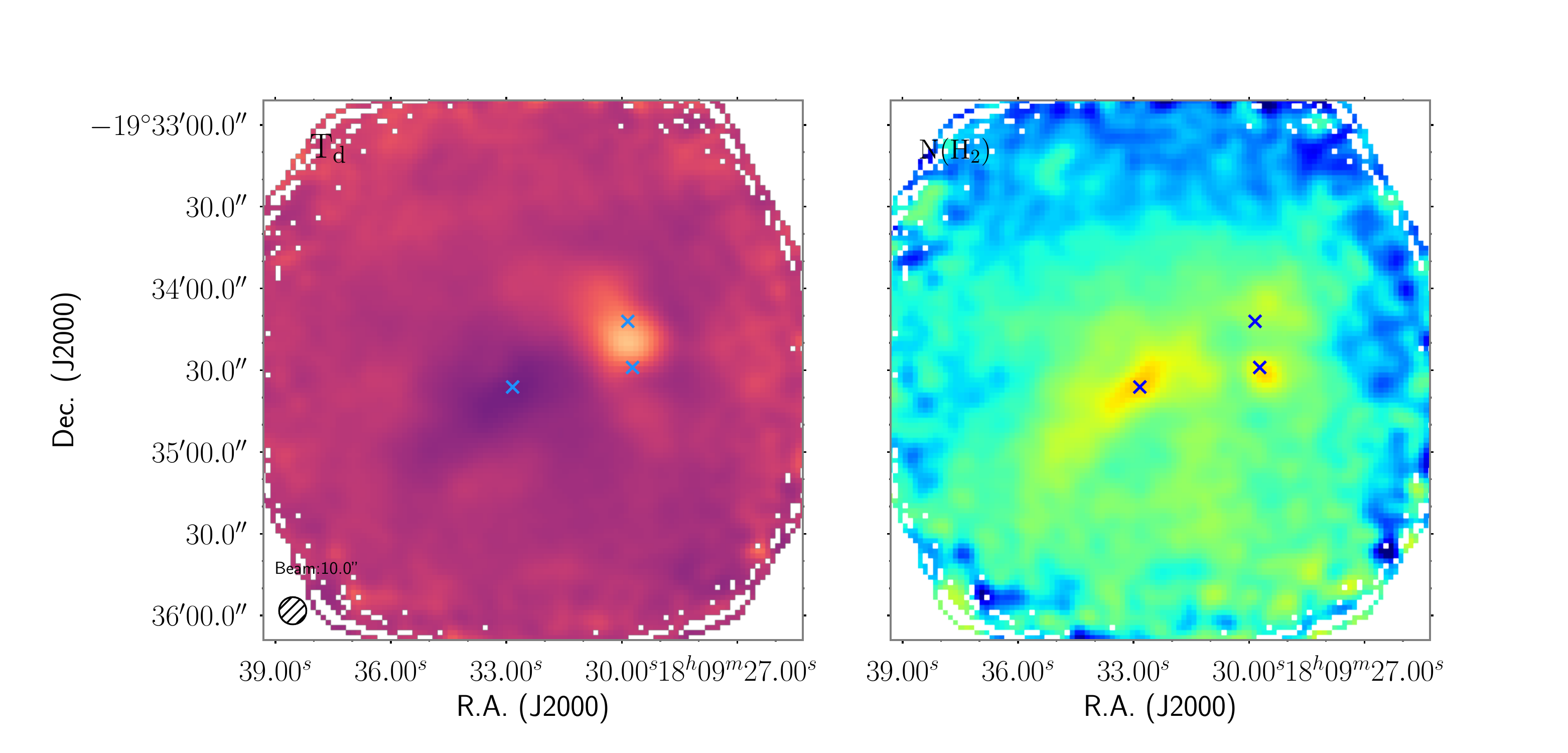}&
\hspace{-0.5cm}\includegraphics[scale=0.33]{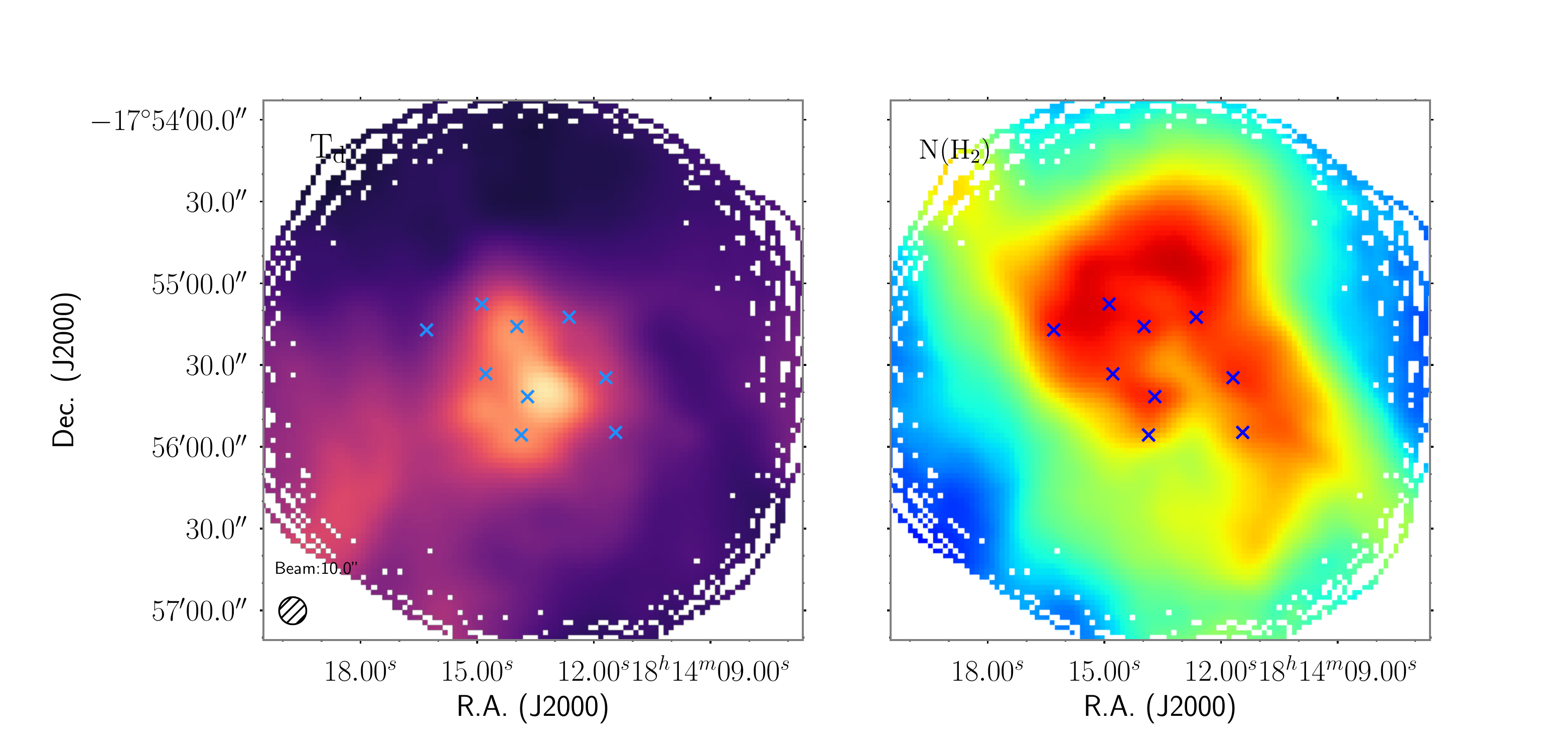}\\
\end{tabular}
\caption{25$''$ and 10$''$ column density and dust temperature maps for source G10.8278-0.0184 and G12.7914-0.1958. The corresponding SED curves for the positions marked by red crosses with number labels are given in the subplots of Fig. \ref{fig:sed_curves} with same number labels in red.}
\label{fig:sed_curves_e}
\end{figure*}

\begin{figure*}
 \begin{tabular}{p{0.5\linewidth}p{0.5\linewidth}}
\hspace{-0.2cm}\includegraphics[scale=0.4]{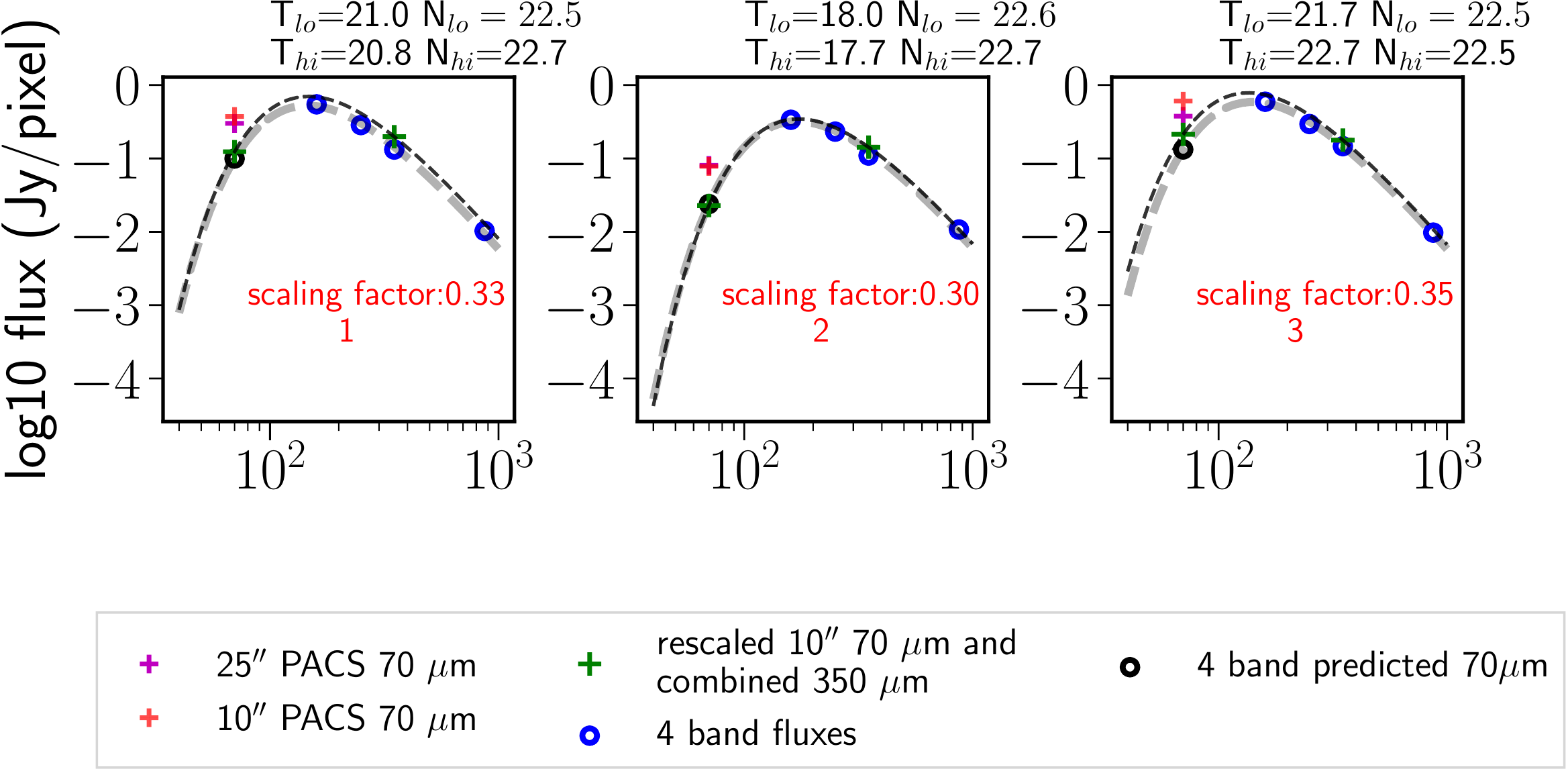}&
\hspace{0.2cm}\includegraphics[scale=0.4]{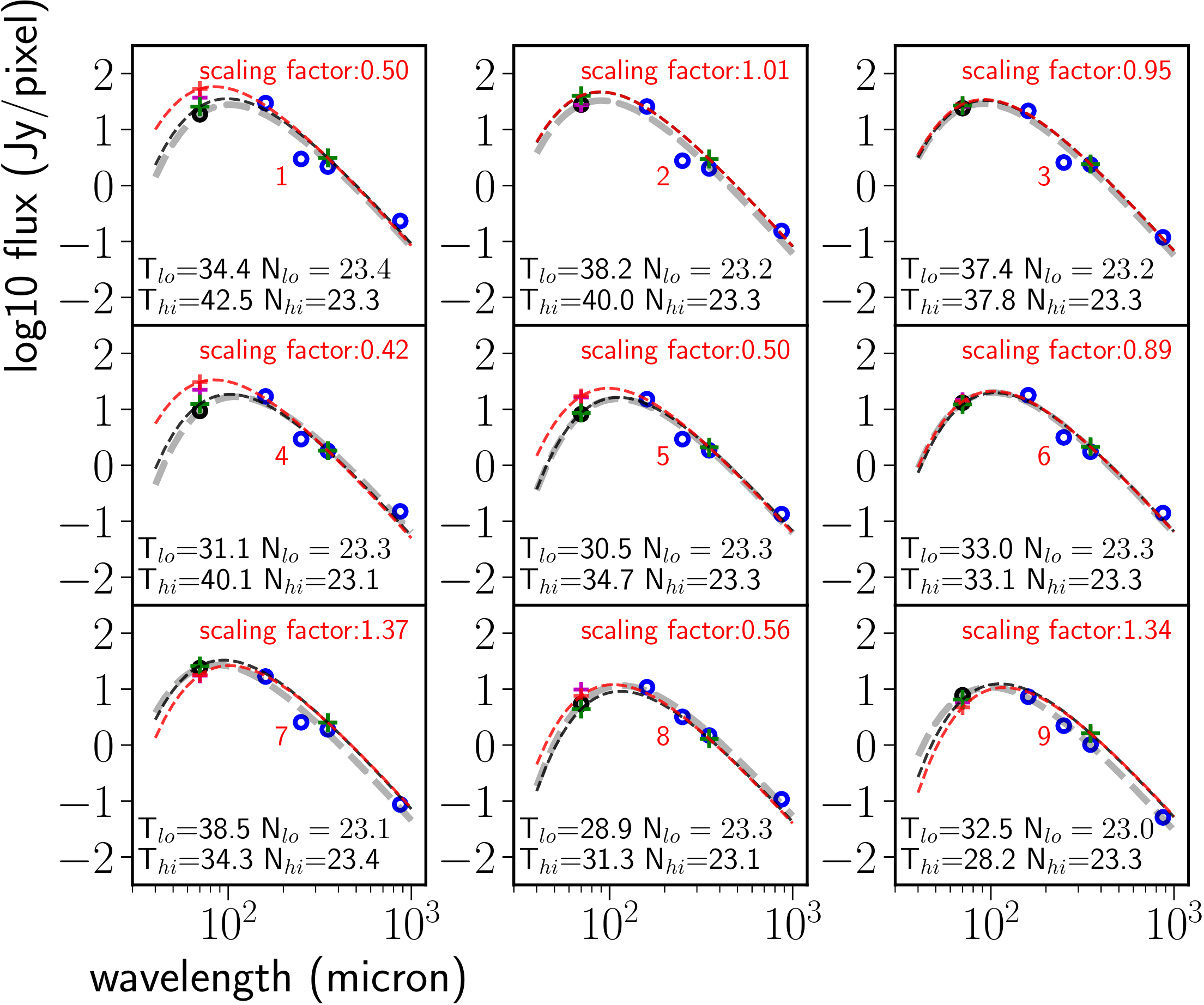}\\
\end{tabular}
\caption{Example SED curves for pixels in Fig. \ref{fig:sed_curves_e}. The gray dashed line represents the 4 band fitted SED curve for the 25$''$ resolution maps and the black dashed line represents the 2 band fitted SED curve for the 10$''$ resolution maps after rescaling 70 $\mu$m fluxes. Red and magenta crosses mark the fluxes from smoothed 10$''$ and 25$''$ PACS 70 $\mu$m maps, respectively. Red circles represent 25$''$ fluxes for 4 band fits, and black circles extrapolated 25$''$ 70 $\mu$m flux from the 4 band fits.
In the right panel, the additional red dashed line in each subplot represents the 2 band fitted SED curve for the 10$''$ resolution maps from unscaled 70 $\mu$m fluxes.}
\label{fig:sed_curves}
\end{figure*}

\begin{figure}

\hspace{0.2cm}\includegraphics[scale=0.35]{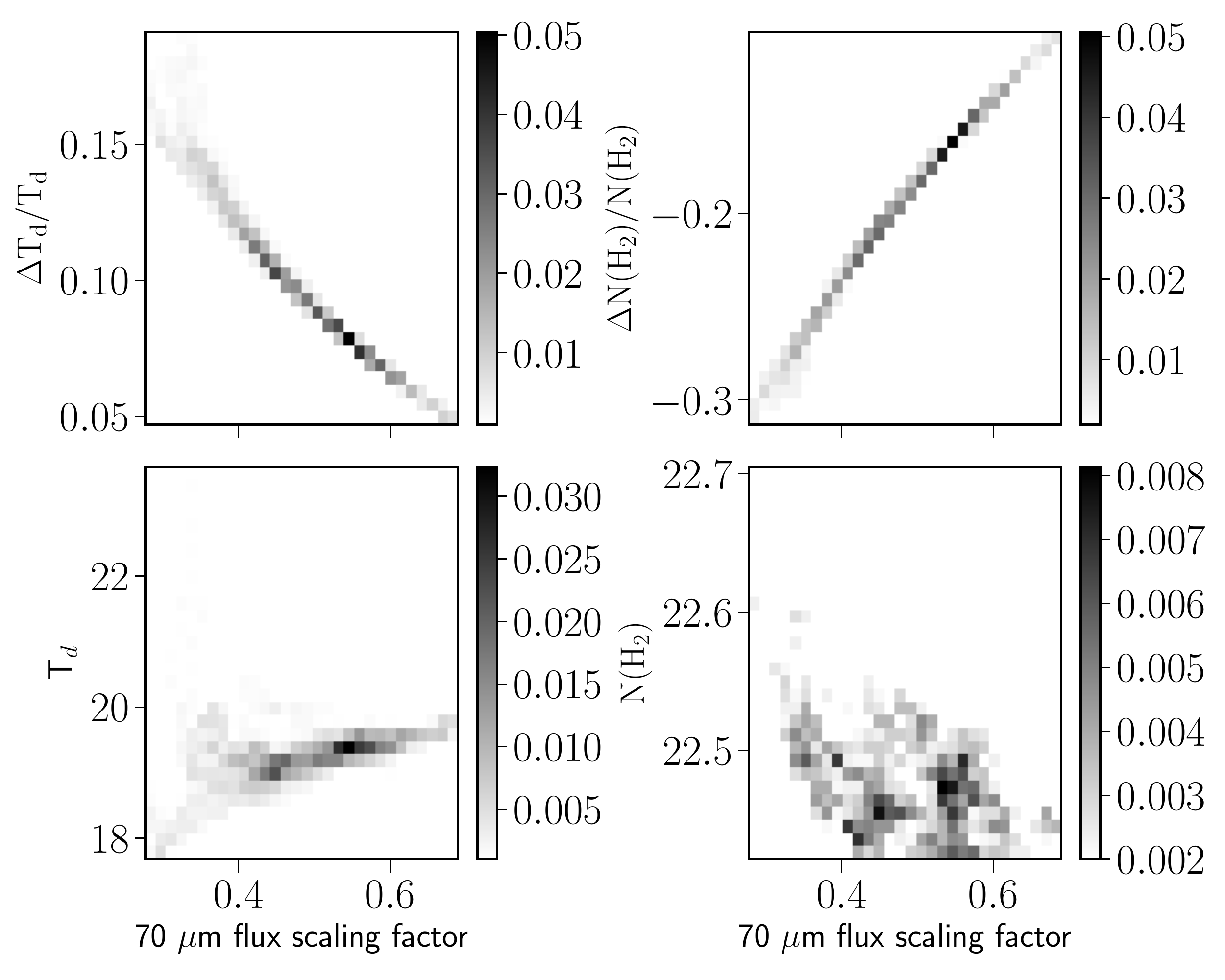}
\caption{2-dimensional distribution of 70 $\mu$m flux scaling factors with derived T$_{d}$, N(H$_{2}$) and $\Delta$N(H$_{2}$)/N(H$_{2}$), $\Delta$T$_{d}$ /T$_{d}$ for pixels in 10$''$ derived maps, of source G10.8278-0.0184.}
\label{fig:70_sf_dist}
\end{figure}

In Fig. \ref{fig:sed_curves_e},  we present 25$''$ and 10$''$ resolution column density and dust temperature maps for two fields, for source G10.8278-0.0184 (left panel) and G12.7914-0.1958 (right panel).  
In the column density and dust temperature maps, the identified SABOCA cores are marked and labeled with number. 
The corresponding SED curves for the pixels of these core positions are shown in Fig. \ref{fig:sed_curves}. 
We denote the difference between the derived column density and temperature, $N(H_{\rm 2})$ and $T_{\rm d}$ from the unscaled 70 $\mu$m flux and that from the scaled 70 $\mu$m as $\Delta N(H_{\rm 2})$
and $\Delta T_{\rm d}$.
 In Fig. \ref{fig:70_sf_dist} we present the 2d histogram of the pixels in the 10$''$ derived maps of G10.8278-0.0184, in terms of their 70 $\mu$m scaling factor, f with $\Delta N(H_{\rm 2})$/$N(H_{\rm 2})$ and $\Delta T_{\rm d}$/$T_{\rm d}$, with the color scale indicating the pixel fractions.
The anti-correlation between scaling factor with $\Delta T_{\rm d}$/$T_{\rm d}$ reflects the increase of the dust temperature of these relatively cold sources ($<$20 K), i.e. core 2 in source G10.8278-0.0184 (left panel of Fig. \ref{fig:sed_curves_e}), is correlated with the increase of the 70 $\mu$m flux fraction from the bulk gas component, representing a gradual warm-up process of the envelope. 
The values of $\Delta T_{\rm d}$/$T_{\rm d}$ are within 30$\%$, depending on the absolute value of the rescaling factor $f$. $\Delta N(H_{\rm 2})$/$N(H_{\rm 2})$ has a more tight relation to the rescaling factor $f$, close to a linear relation of $\Delta N(H_{\rm 2})$/$N(H_{\rm 2})$ = 0.5$f$-0.45. 

Scaling down the 70 $\mu$m flux according to cold component extrapolated 70 $\mu$m flux is not equivalent to fixing temperature to the coarser resolution temperature map. 
As we can see from the left panel of Fig. \ref{fig:sed_curves_e}, the enhanced temperatures of larger than $\sim$25 K, i.e. for core 1 and 3 and their adjacent region, are better resolved in the 10$''$ temperature map.  
In addition, we can now compare the derived fragment mass to the clump mass on a common basis to avoid an overestimate of temperature due to the inclusion of 70~$\mu$m map in the higher resolution fits.

We also show cases in which the scaling factors are larger than 1, i.e. cores 2, 7, 9 in source G12.7914-0.1958 (right panel of Fig. \ref{fig:sed_curves_e}). 
This is due to external heating by the nearby mid-infrared bright object, therefore along the line-of-sight there is a mixture of hot and cold dust with the hot component not well-shielded due to its relatively low density cold component.  In this case the 4 band fits probe a temperature close to the hot dust and a column density close to the cold component,  thus giving a predicted 70 $\mu$m flux higher than the observations. 
These sources should be intrinsically hotter, for which the inclusion of 70 $\mu$m flux into the SED would better constrain the $T_{\rm d}$ and $N(H_{2})$.
Thus for these sources, we use directly the unscaled PACS 70 $\mu$m fluxes in the high resolution fit. 


\clearpage
%
%
\section{Fragmentation analysis: a comparison with Dendrogram results}\label{sec:dendro}

\begin{figure}[htb]
\includegraphics[scale=0.33]{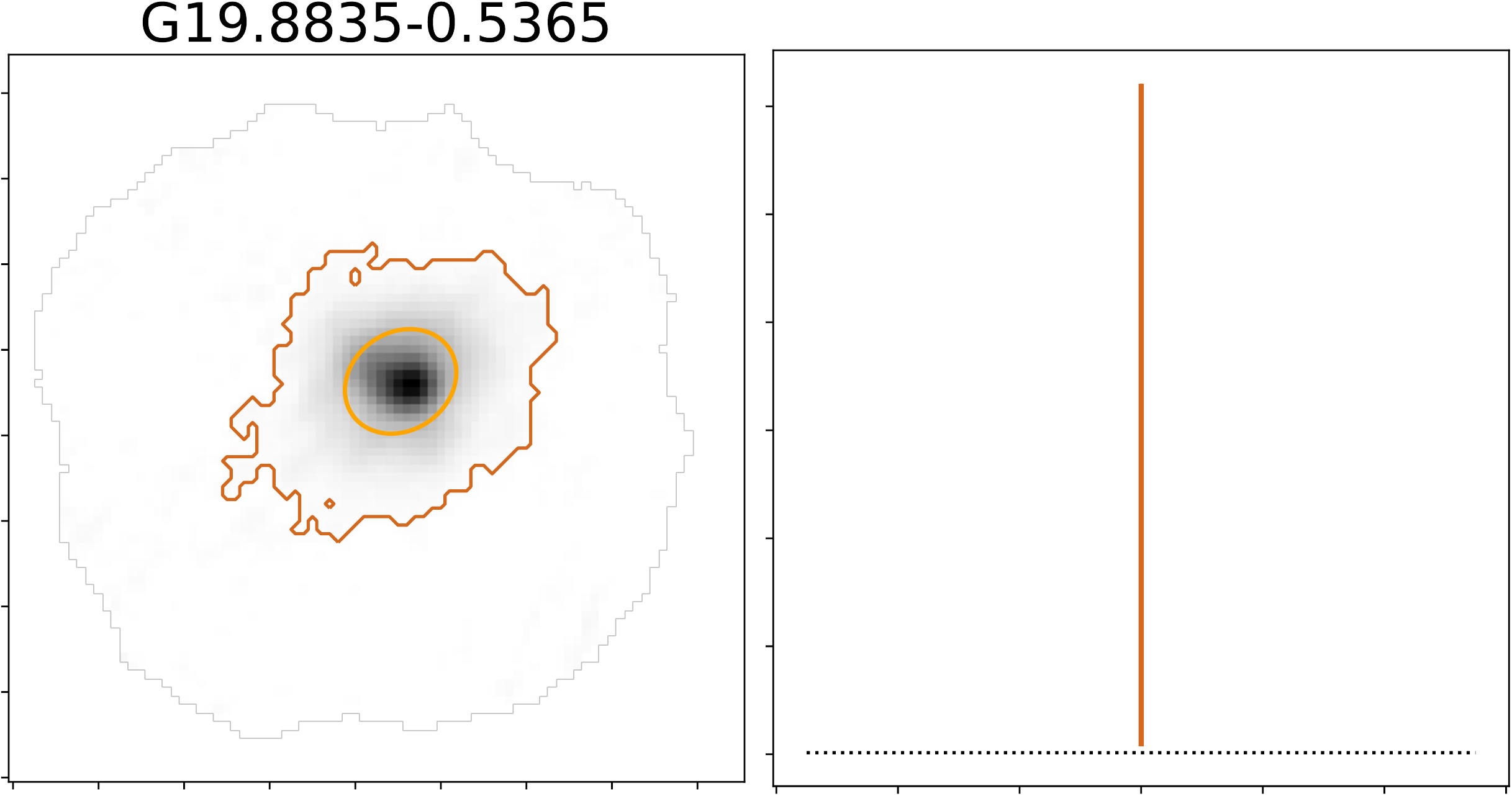}\\
\includegraphics[scale=0.33]{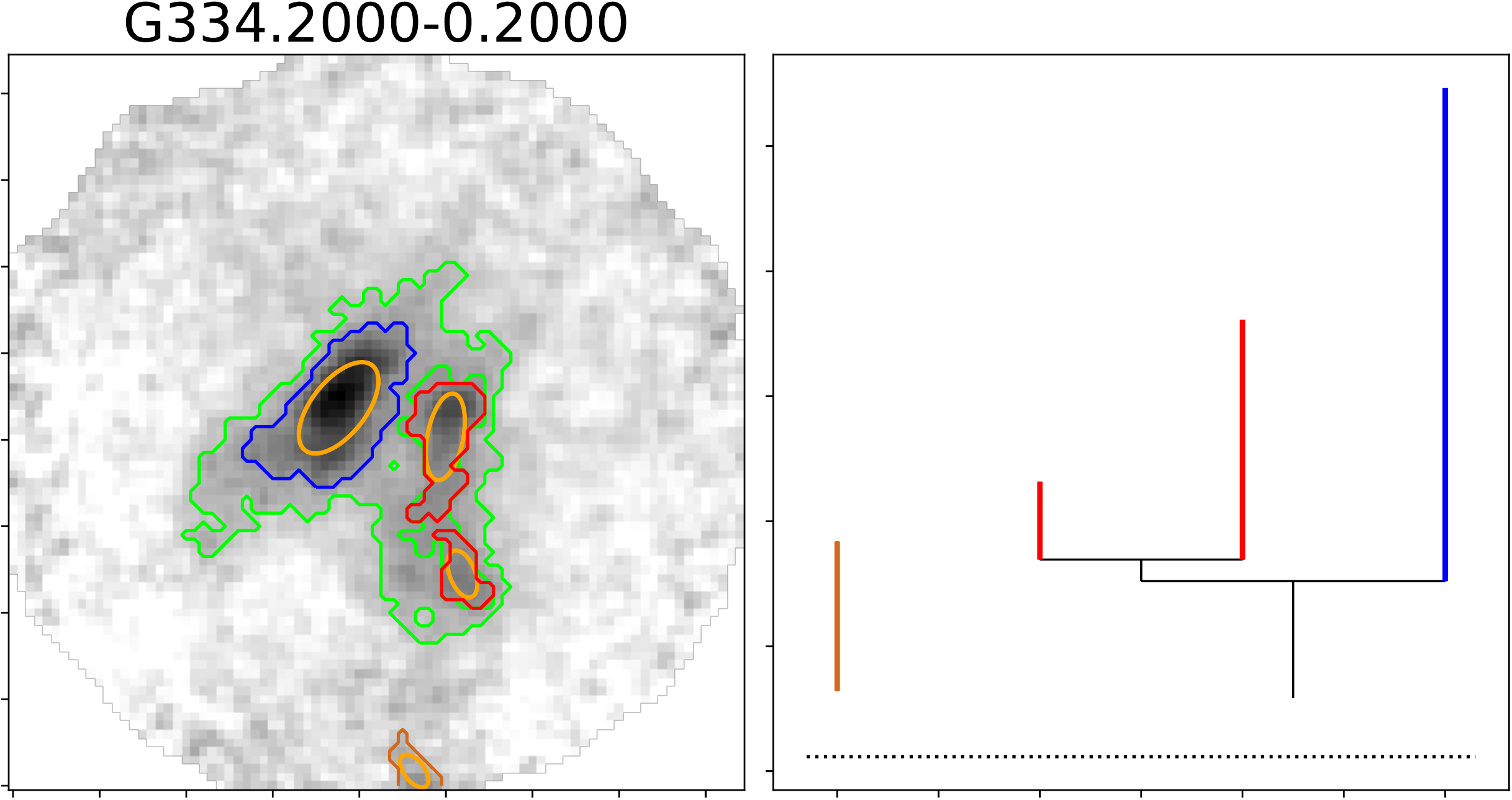}\\
\includegraphics[scale=0.33]{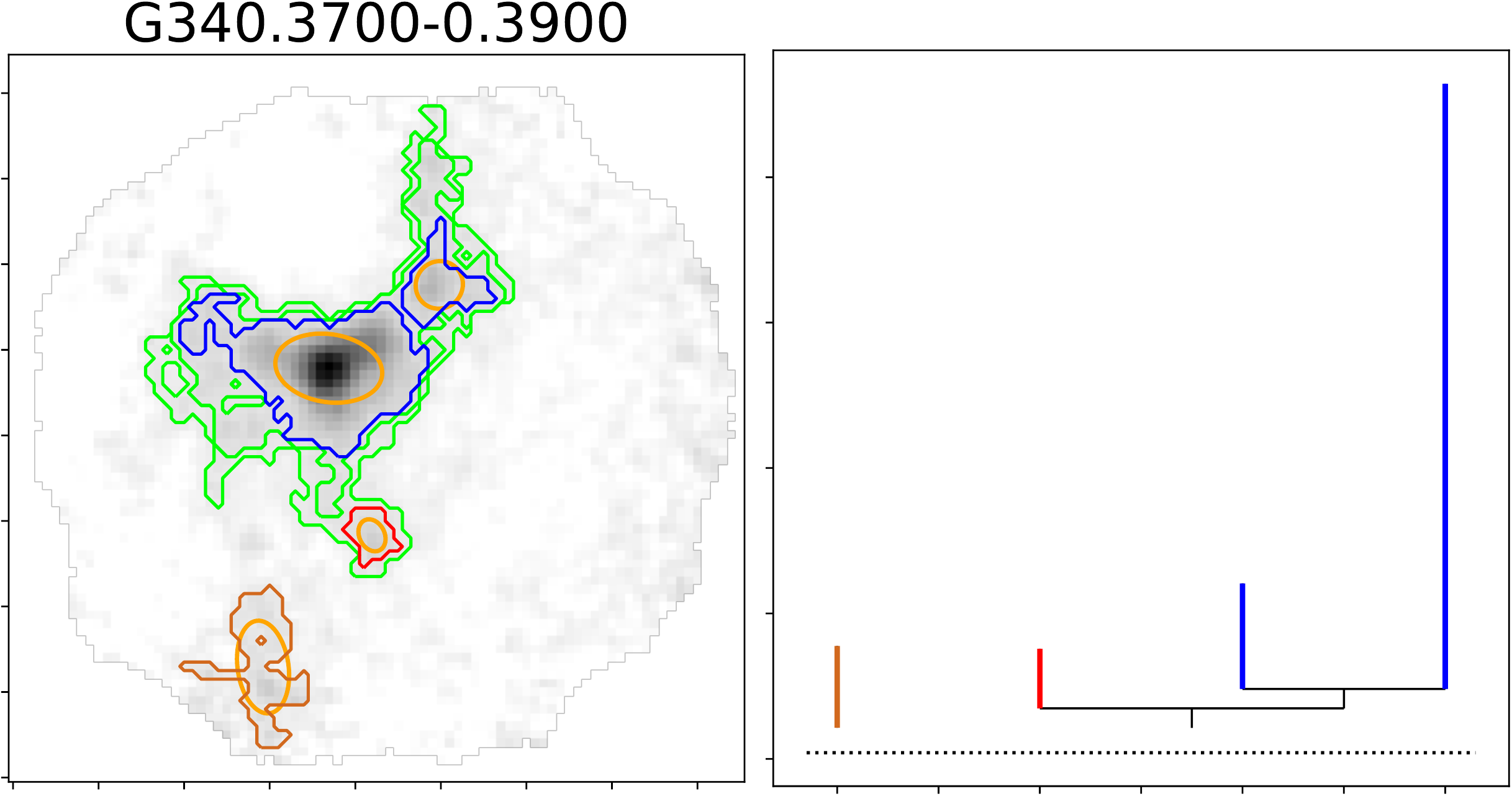}\\
\includegraphics[scale=0.33]{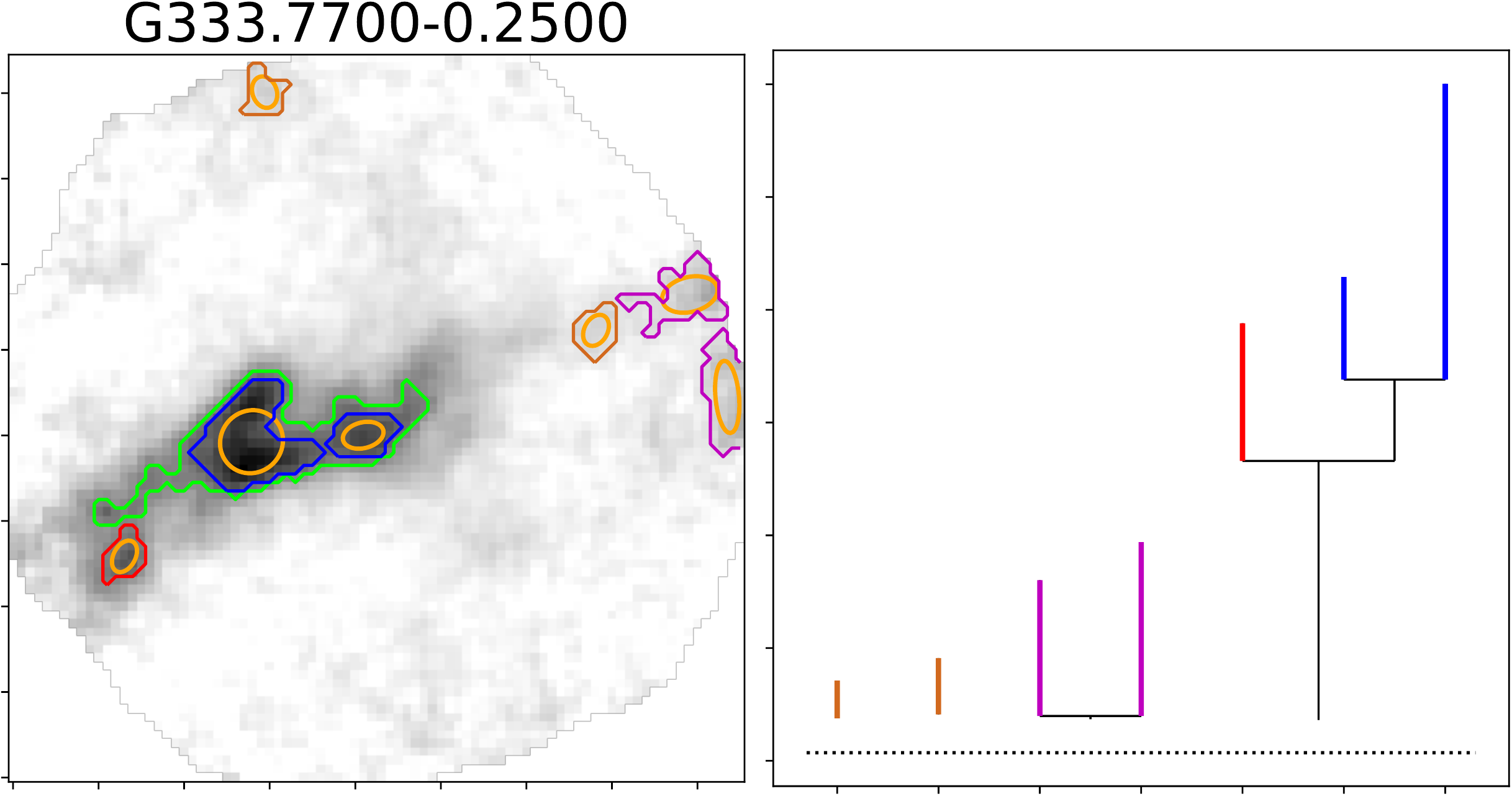}
\caption{Dendrogram source extraction results of four sources. {\it{Left}}: Dendrogram identified multi-level structures overlaid on the SABOCA emission map. Leaves originating from same top branch structures are marked with same color (in red, blue, cyan or magenta) as contour lines. Leaves with brown color are isolated leaf structures, i.e. with no parental structure present in the emission map. Orange ellipses are the corresponding schematic ellipse calculated based on the weight of intensity. Branches that do not intersect with the map edge, i.e. with closed contours as defined in the texts, are marked with lime contour lines. {\it{Right}}: Dendrogram tree plot representation of the left panel. Similarly, leaves from same top branch are marked with same color as in left panel. Horizontal dotted line indicates the 3$\sigma$ threshold we use for minimum level as input parameter of dendrogram identification.}
\label{fig:dendro_output}
\end{figure}

In this paper we use {\it{Gaussclumps}} to identify the compact sources in the SABOCA emission maps. 
{\it{Gaussclumps}} is a non-hierarchical source extraction method, which assumes that cloud structures are composed of multiple Gaussian sources which can overlap. 
In contrast, another widely-used source extraction method, {\tt{dendrogram}}, is a hierarchical source identification method, which is discreet in the sense that a given emission area (or pixel) can only belong to a single structure, no overlap is allowed among the structures (of same level).
We explore here the dendrogram representation of the cloud structure seen in the SABOCA maps, and compare these results with {\it{Gaussclumps}}.

There are several parameters to control the dendrogram source extraction,  {\tt{min$\_$value, min$\_$delta and min$\_$npix}}, representing the minimum threshold, the minimum significance of a structure compared to its merging level, and the minimum size of a structure. Similarly as we did with {\it{Gaussclump}}, we set the {\tt{min$\_$value}} to 3 times the rms of the SABOCA map, and {\tt{min$\_$npix}} as the number of pixels corresponding to beam size. For {\tt{min$\_$delta}}, we set this to zero, as an attempt to pick up the maximum number of sources as well as to be consistent with {\it{Gaussclump}}, which does not have a similar control parameter. 

We run {\tt{dendrogram}} on the SABOCA emission maps and apply the extracted structures to the 10$''$ column density maps in order to derive the physical properties of these structures.
We define the size scale of the structures extracted by {\tt{dendrogram}} as the geometrical mean of the area, 
similarly, assuming this size scale is representative of the scale of the line-of-sight dimension, the average gas density ($\rho_{clump}$) is estimated.
For the analysis, we only include structures that have closed contours within the map. 
We find that the relatively small FOV of the SABOCA maps provide an incomplete view of the structures at the lowest contours, especially for those showing a filamentary structure. Therefore, requiring a closed contour for the first structure is a condition that strongly limits our statistics. 

With {\tt{dendrogram}}, we define the fragmentation level as the number of leaves residing within the lowest parental structure (branch or trunk) on the map. Isolated leaves without any parental structures are omitted. 
For example in Fig. \ref{fig:dendro_output},  source G19.8835$-$0.5363 appears to be an isolated structure,  while source G334.2000$-$0.2000 has 3 leaves (1 blue, 2 red) residing in the lowest branch (outer green contour) which gives one statistic. In general, compared to {\it{Gaussclump}}, {\tt{dendrogram}} picks up in addition numerous  structures with irregular morphology, i.e.  the magenta leaves at the edge of G333.7700$-$0.2500; the number of leaves is smaller than number of compact sources identified by {\it{Gaussclump}} with the average size appears larger.  The differences are intrinsically linked with their different ways of defining structures. Although a systematic comparison between these two methods is beyond the scope of this paper, here we benchmark the analysis in Sec. \ref{sec:fmtt} to see whether the choice of source extraction method causes systematic differences in the results.

\begin{figure*}[htb]
\begin{tabular} {p{0.25\linewidth}p{0.25\linewidth}p{0.25\linewidth}p{0.25\linewidth}}

\includegraphics[scale=0.3]{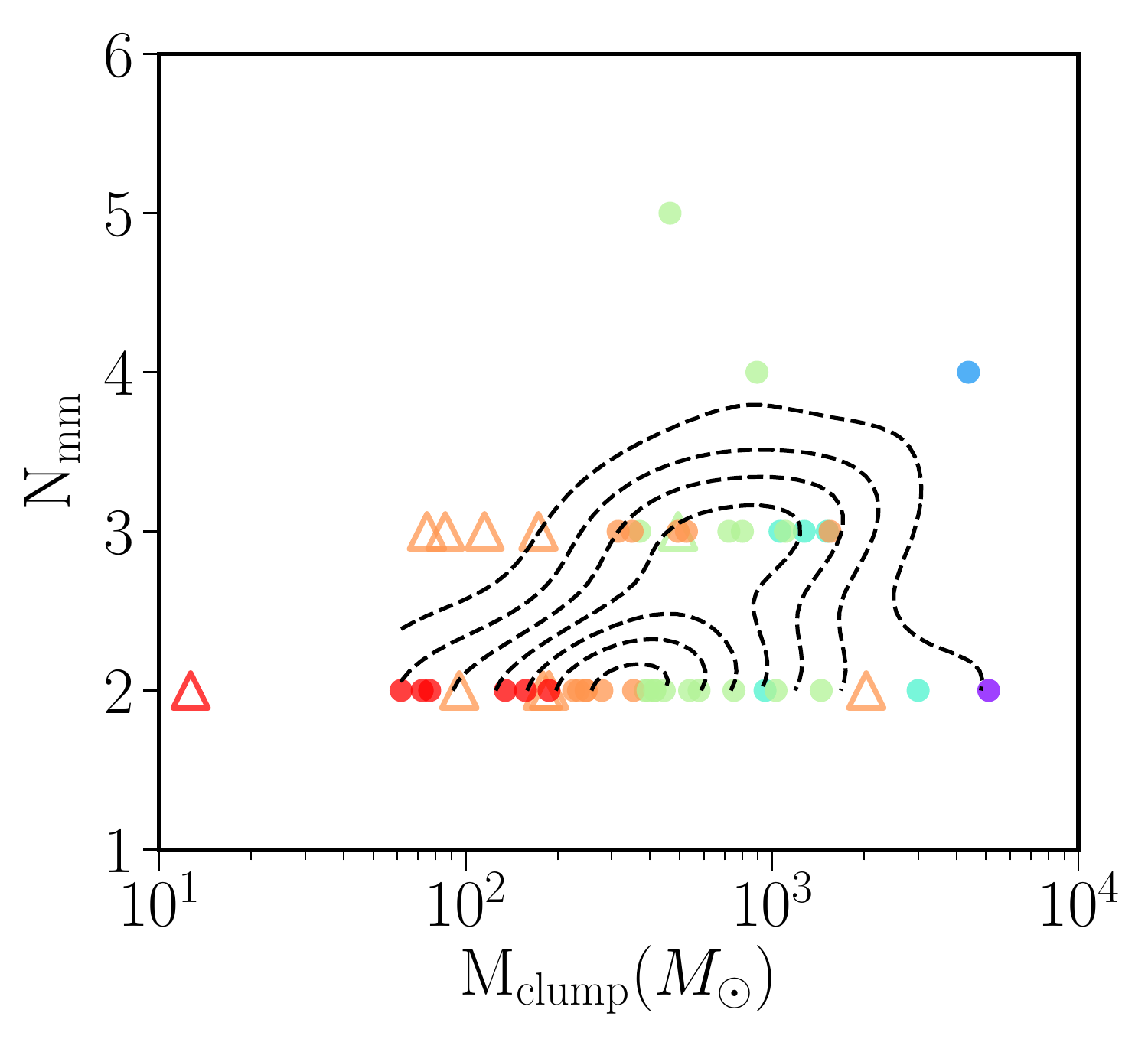}&
\hspace{-0.55cm}\includegraphics[scale=0.3]{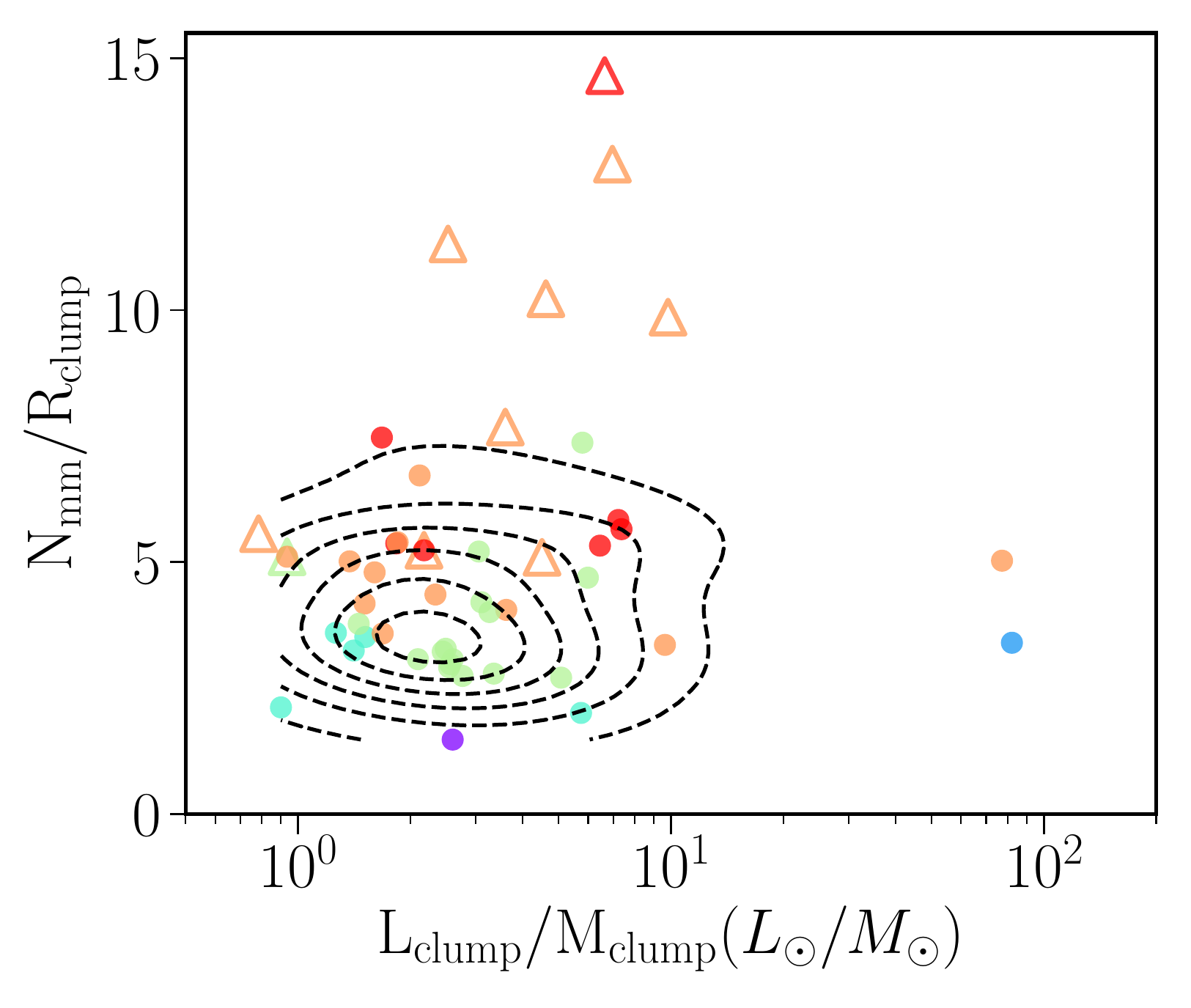}&
\hspace{-0.55cm}\includegraphics[scale=0.3]{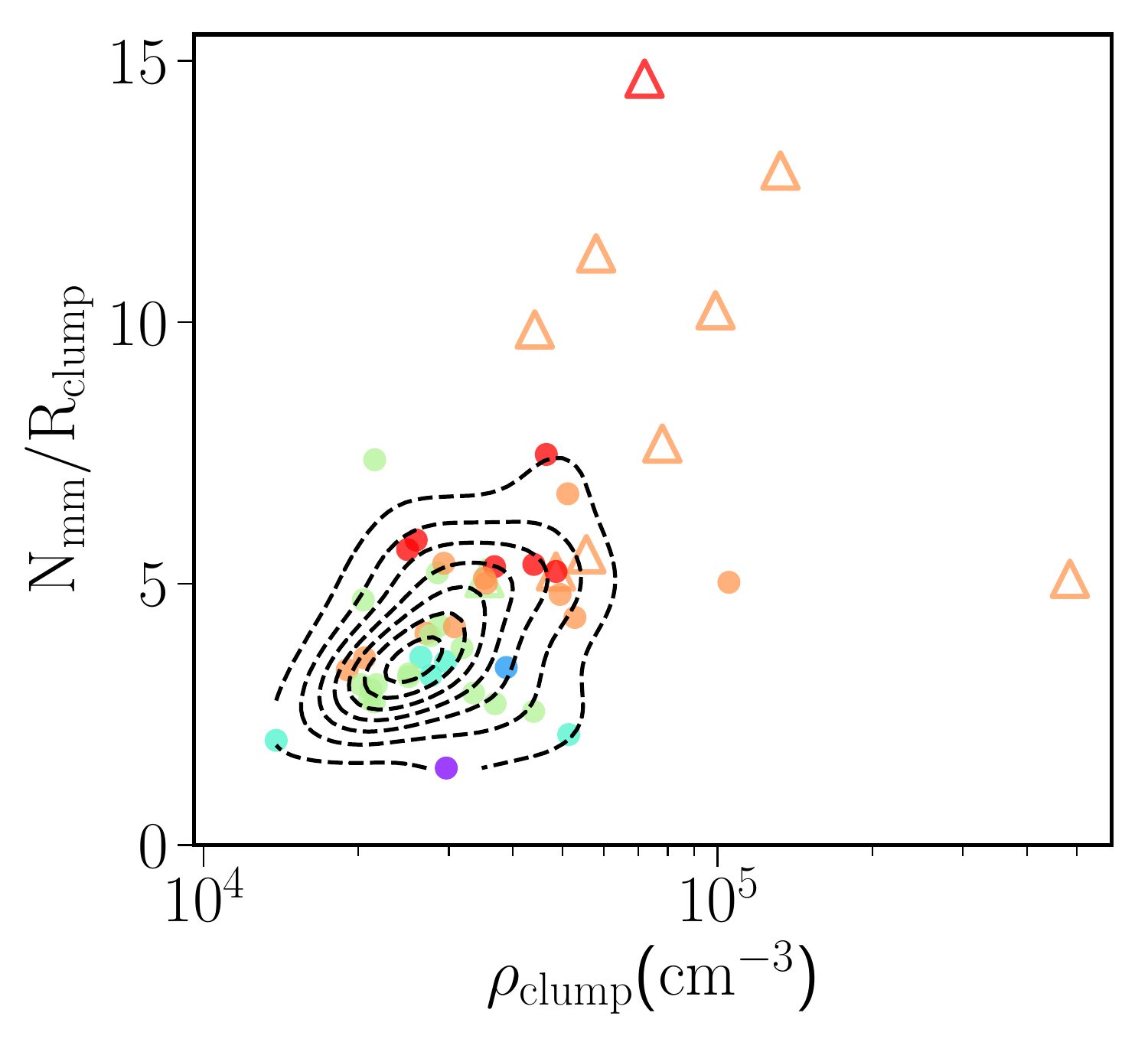}&
\hspace{-0.75cm}\includegraphics[scale=0.3]{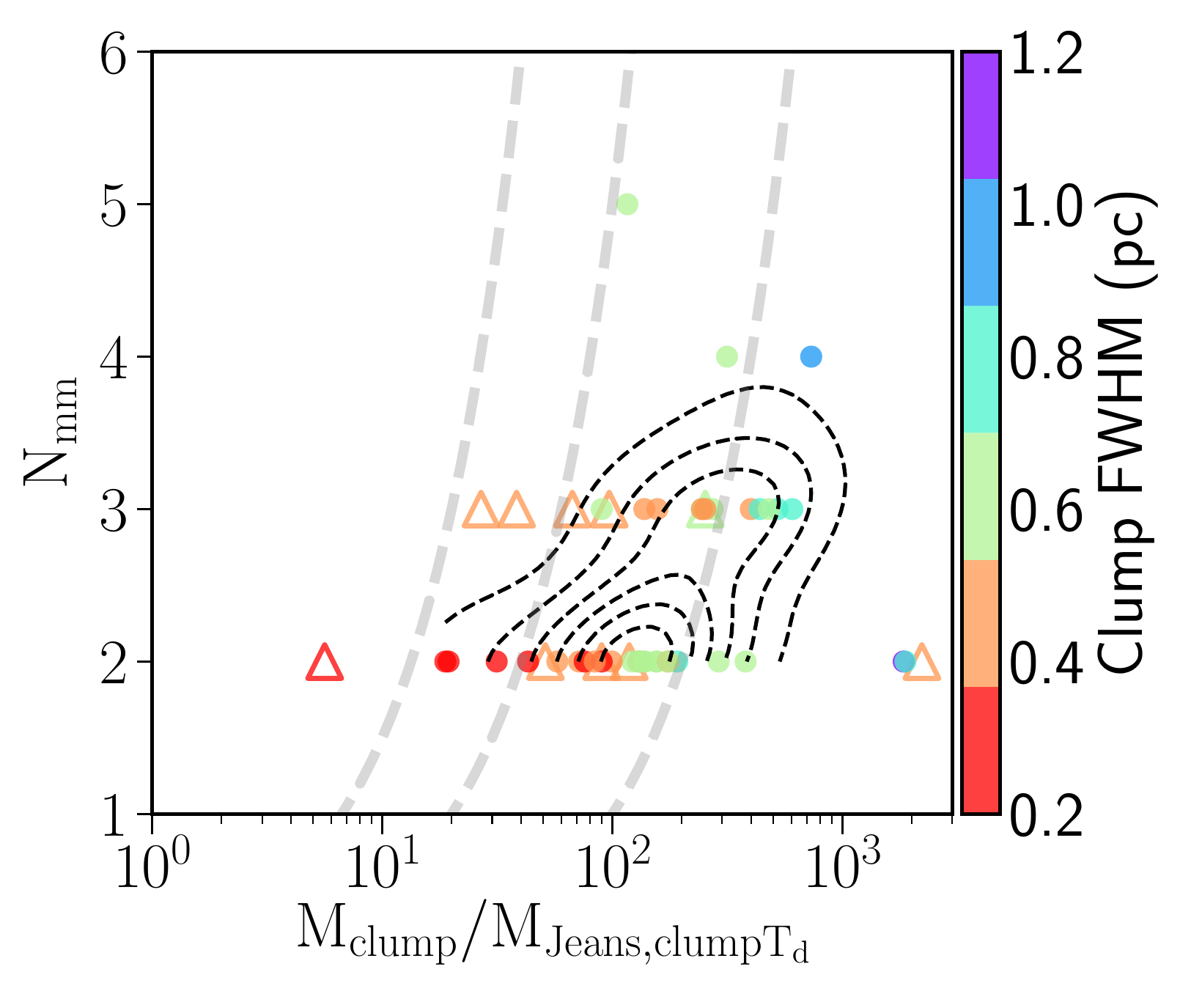}\\
\end{tabular}
\caption{Fragmentation level as a function of clump properties of sources located at a distance in 2-4\,kpc (dots) and 1-2\, kpc (triangles). In each plot, data points are color-coded according to the clump size.  Dashed contours show the distribution of sources at 2-4\ kpc from Gaussian kernel density estimation. \emph{Left}: Number of fragments as a function of clump mass. \emph{Middle left}:  Specific fragmentation level (normalised by parent structure size) as a function of the clump's luminosity-to-mass ratio. \emph{Middle right}:  Specific fragmentation level as a function of the clump's density. \emph{Right}:  Number of fragments as a function of predicted number of fragments based on Jeans fragmentation scenario. Gray dashed lines mark the lines of $N_{mm}$ = 0.15/0.05/0.01$\times$$M_{clump}/M_{Jeans,clumpT_{d}}$.}

\label{fig:fmtt_level_dendro}
\end{figure*}

\begin{figure*}[htb]
\begin{tabular} {p{0.33\linewidth}p{0.33\linewidth}p{0.33\linewidth}}
\hspace{-0.5cm}\includegraphics[scale=0.40]{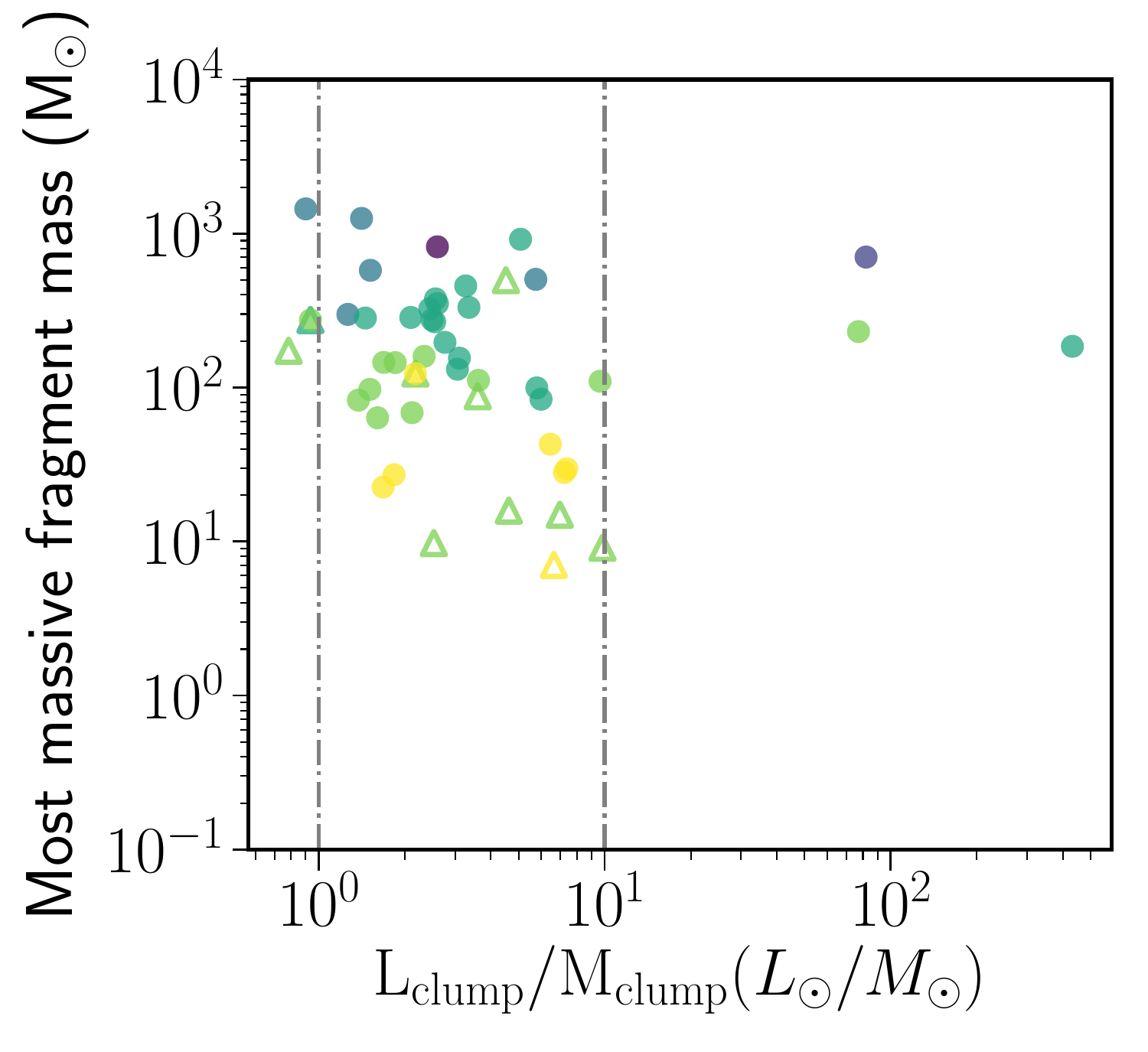}&
\hspace{-0.65cm}\includegraphics[scale=0.40]{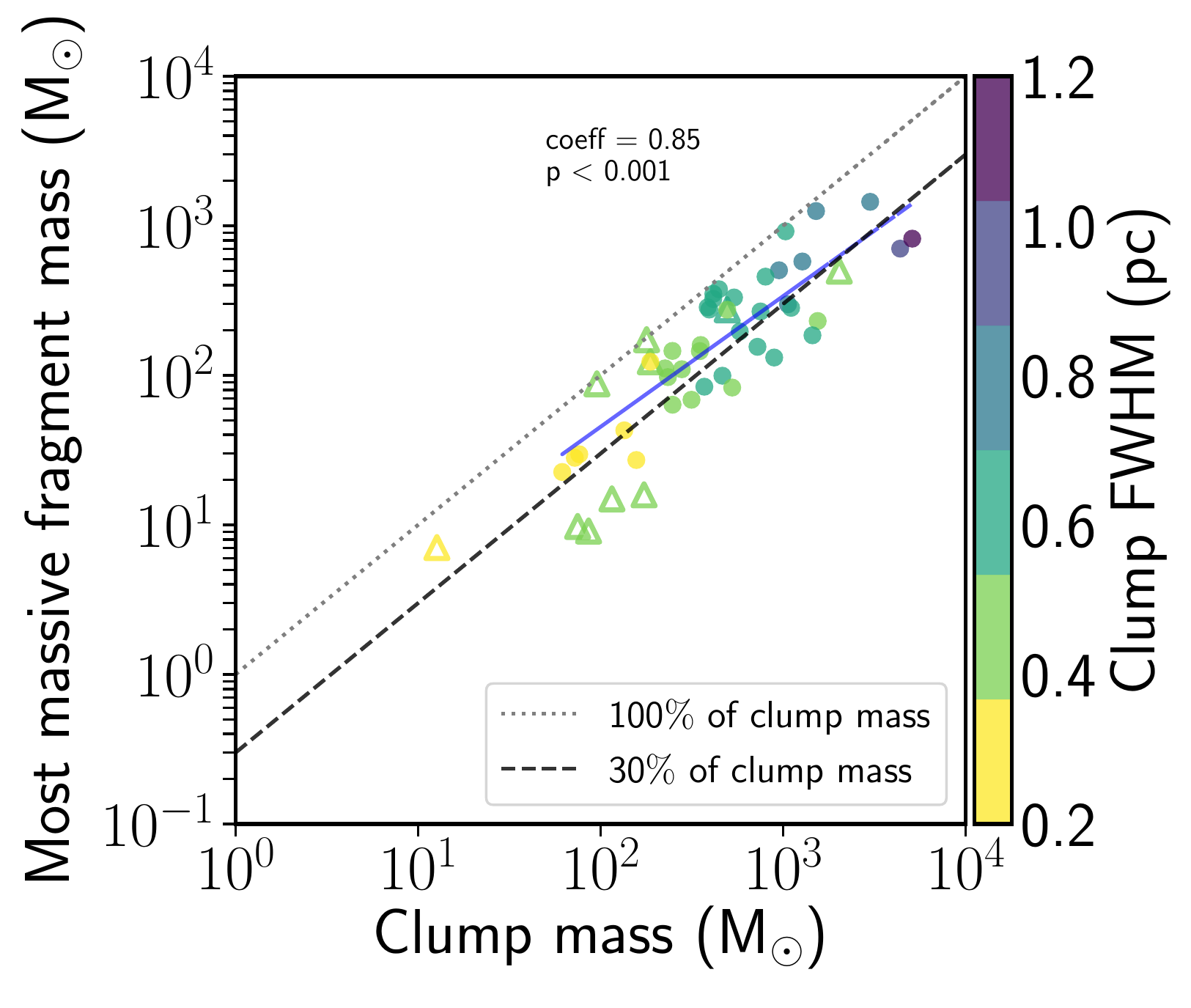}&
\hspace{-0.65cm}\includegraphics[scale=0.40]{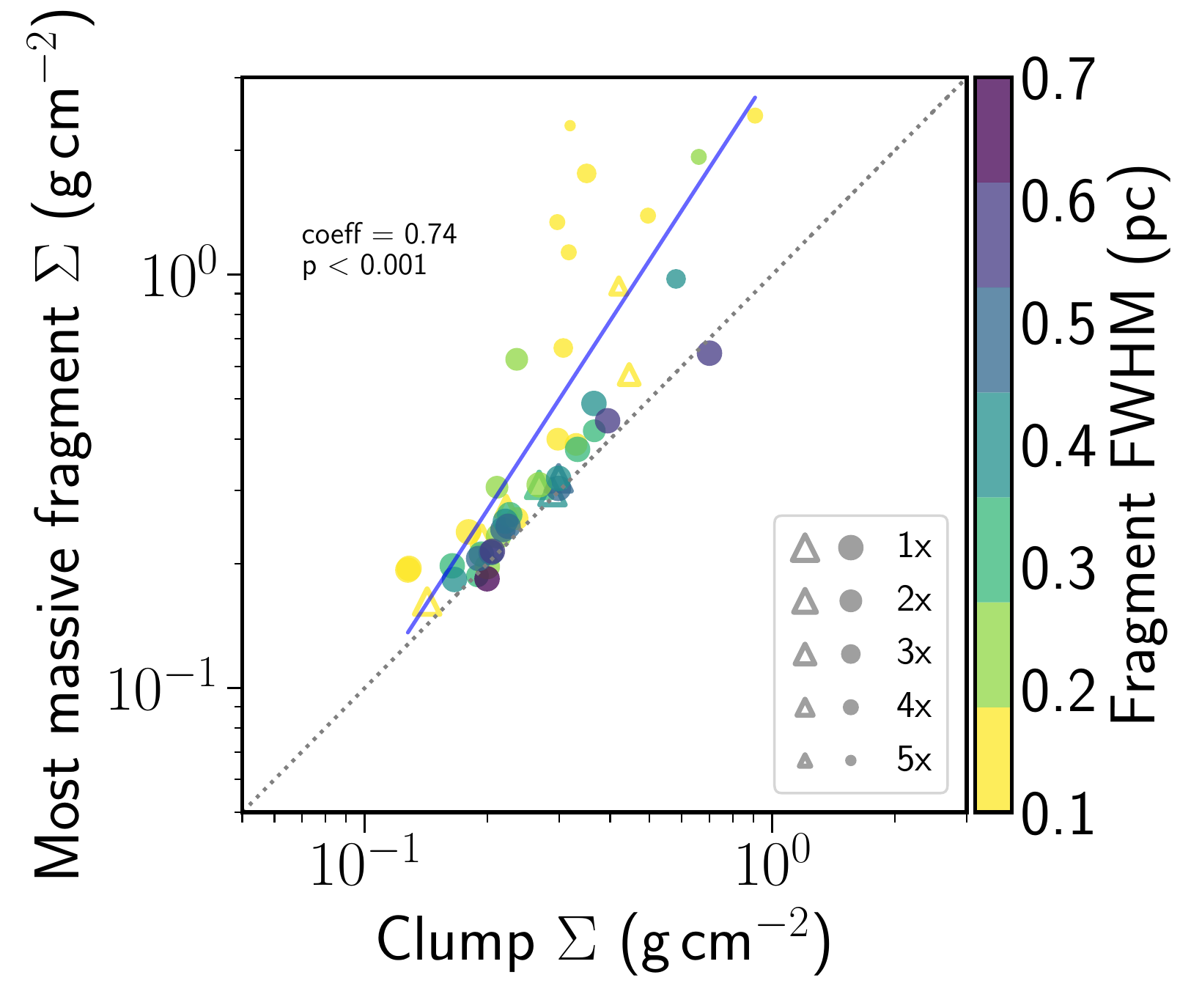}\\

\end{tabular}
\caption{\emph{Left}: Most massive fragment mass as a function of clump properties for sources at a distance 2-4\,kpc (dots) and 1-2\,kpc (triangles). \emph{Left}: Mass of the most massive fragment as a function of clump luminosity-to-mass ratio. Vertical lines mark the luminosity-to-mass ratio of 1 and 10. \emph{Middle}: Mass of the most massive fragment mass as a function of clump mass. Gray lines show the 30$\%$ and 100$\%$ proportion of clump mass; blue line shows the result of a linear fit on logarithmic scale to sources in 2-4\, kpc distance range of $M_{\mathrm{fragments}}$ = 0.87log$M_{\mathrm{clumps}}-0.21$. Left and middle plots share the same color bar, with clumps color coded according to their sizes. \emph{Right}: Surface density of the most massive fragment as a function of clump surface density. Gray line marks the line of equality; the blue line shows the result of a linear fit to sources in 2-4\, kpc distance range of log$\Sigma_{\mathrm{fragments}}$ = 1.51log$\Sigma_{\mathrm{clumps}}$+1.13. The different sizes of markers correspond to the size ratios between the parental clump size and most massive fragment size, as indicated in the legend, i.e. the smaller the size of a marker, the larger size difference between fragment and its parental clump, up to a factor of 5.}
\label{fig:mmf_dendro}
\end{figure*}

In Fig. \ref{fig:fmtt_level_dendro} the fragmentation level is compared with the clump properties, similarly as in Fig. \ref{fig:fmtt_level}. We again find a correlation between the fragmentation level with the parental clump mass, with a Spearman correlation coefficient of 0.42 (p-value<$0.001$), although we have smaller statistics here. We again define the specific fragmentation level as number of fragments normalised by clump size, as $N_{mm}/R_{clump}$. There is no correlation between specific fragmentation level with clump $L/M$ ratio, with Spearman correlation coefficient of 0.018. Finally, comparing the fragmentation level with the Jeans fragmentation prediction, we find a Spearman correlation coefficient of 0.30 of specific fragmentation level with clump gas density (p-value$\sim$0.05). As for the comparison of fragmentation level with number of Jeans mass predicted by clump dust temperature and gas density, a Spearman correlation coefficient of 0.48 is found (p-value$\sim$0.003). Assuming a fragmentation temperature of 20 K, the Spearman correlation coefficient is 0.44 (p-value$<$0.001). These results are broadly consistent with Fig. \ref{fig:fmtt_level} in Sec. \ref{sec:fmtt}.  The relation of mass of the most massive fragment with clump $L/M$ and clump mass are also compatible with Fig. \ref{fig:mmf}, despite the drastic difference in the contrast of surface densities between the most massive fragment and their parental clump as in the right panel of Fig. \ref{fig:mmf} and Fig. \ref{fig:mmf_dendro}. We size-coded the data points in the right panel of Fig. \ref{fig:mmf_dendro} according to the size difference between the most massive fragment and its parental clump. It is obvious that the large scatter of the two surface density values is related to the relative size difference defining the parent and child structures, which together lead again to the tight correlation between the mass of the most massive fragment and the parental clump mass. We therefore conclude that the source extraction method and the way we define the fragmentation level by allocating sources in a non-hierarchical way do not affect the analysis results within the scope of this work. 

\section{MALT90 spectrum towards the massive pre-stellar cores}
\let\cleardoublepage\clearpage

\begin{figure*}
\hspace{-0.8cm}
\vspace{-0.2cm}
\begin{tabular}{p{6.2cm}p{6.2cm}p{6.2cm}}
\hspace{0.55cm}\includegraphics[width=6.0cm]{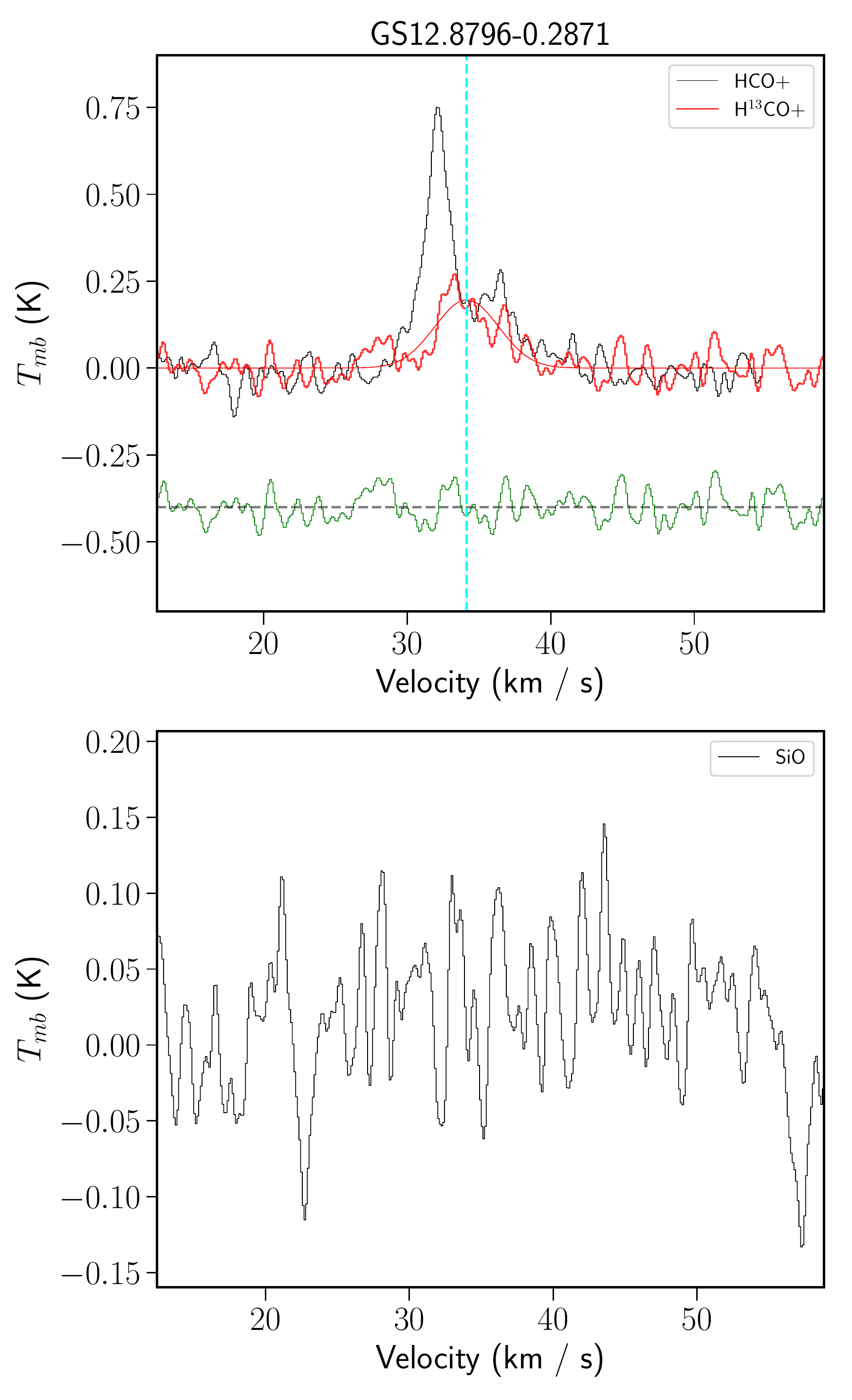}&
\hspace{-0.0cm}\includegraphics[width=6.0cm]{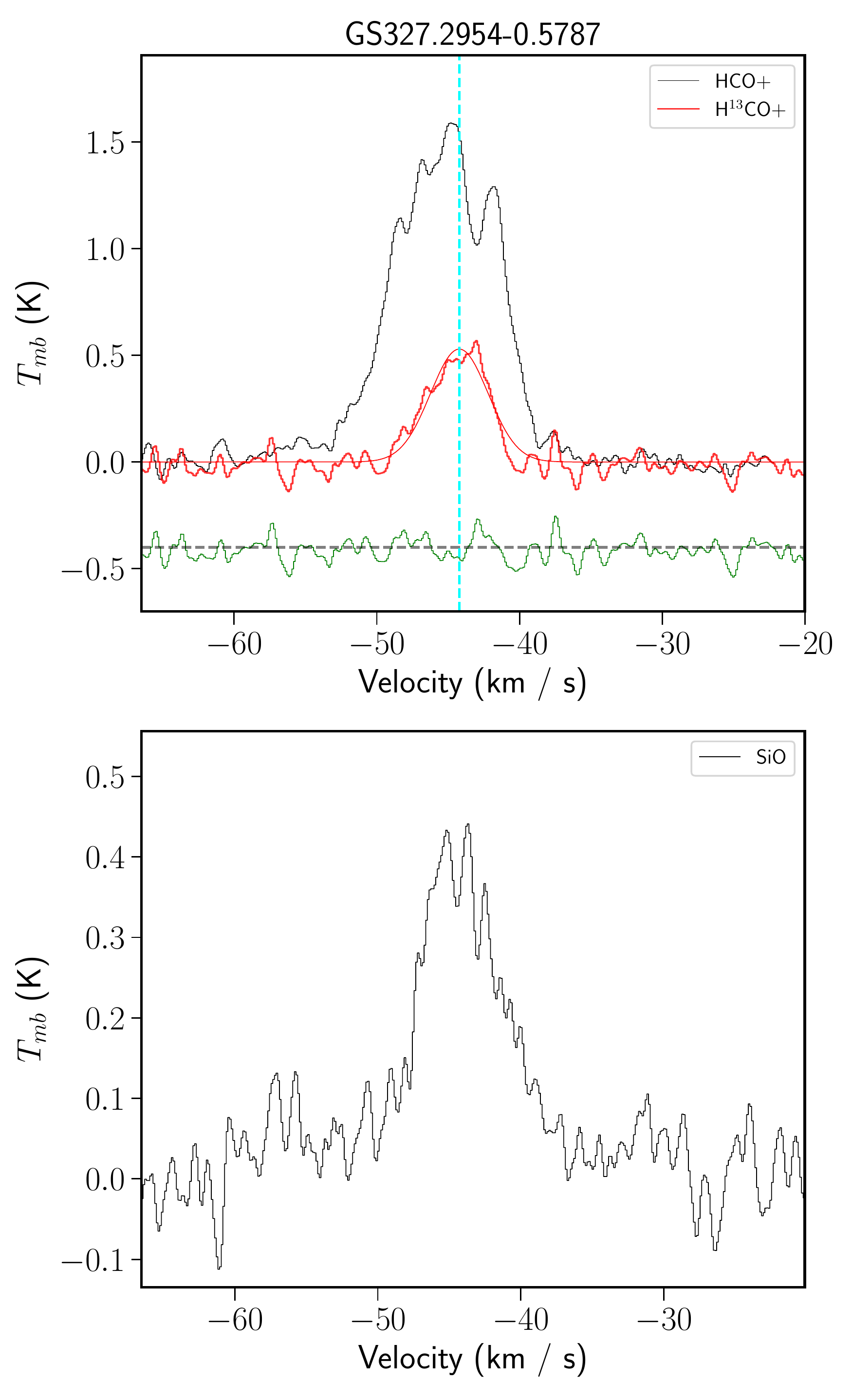}&
\hspace{-0.3cm}\includegraphics[width=6.0cm]{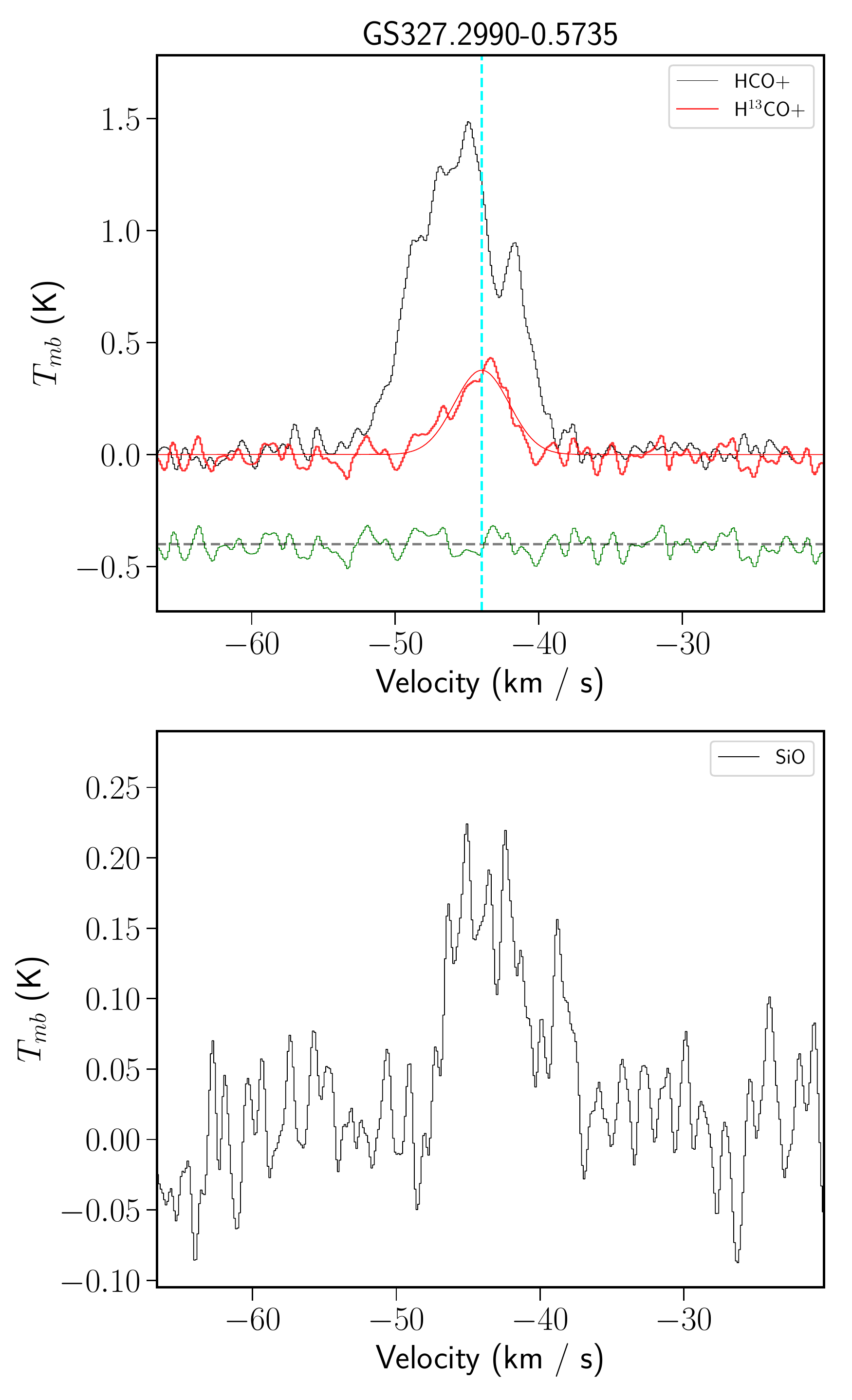}\\
\end{tabular}

\end{figure*}
\begin{figure*}
\hspace{-0.8cm}
\vspace{-0.2cm}
\begin{tabular}{p{6.2cm}p{6.2cm}p{6.2cm}}
\hspace{0.55cm}\includegraphics[width=6.cm]{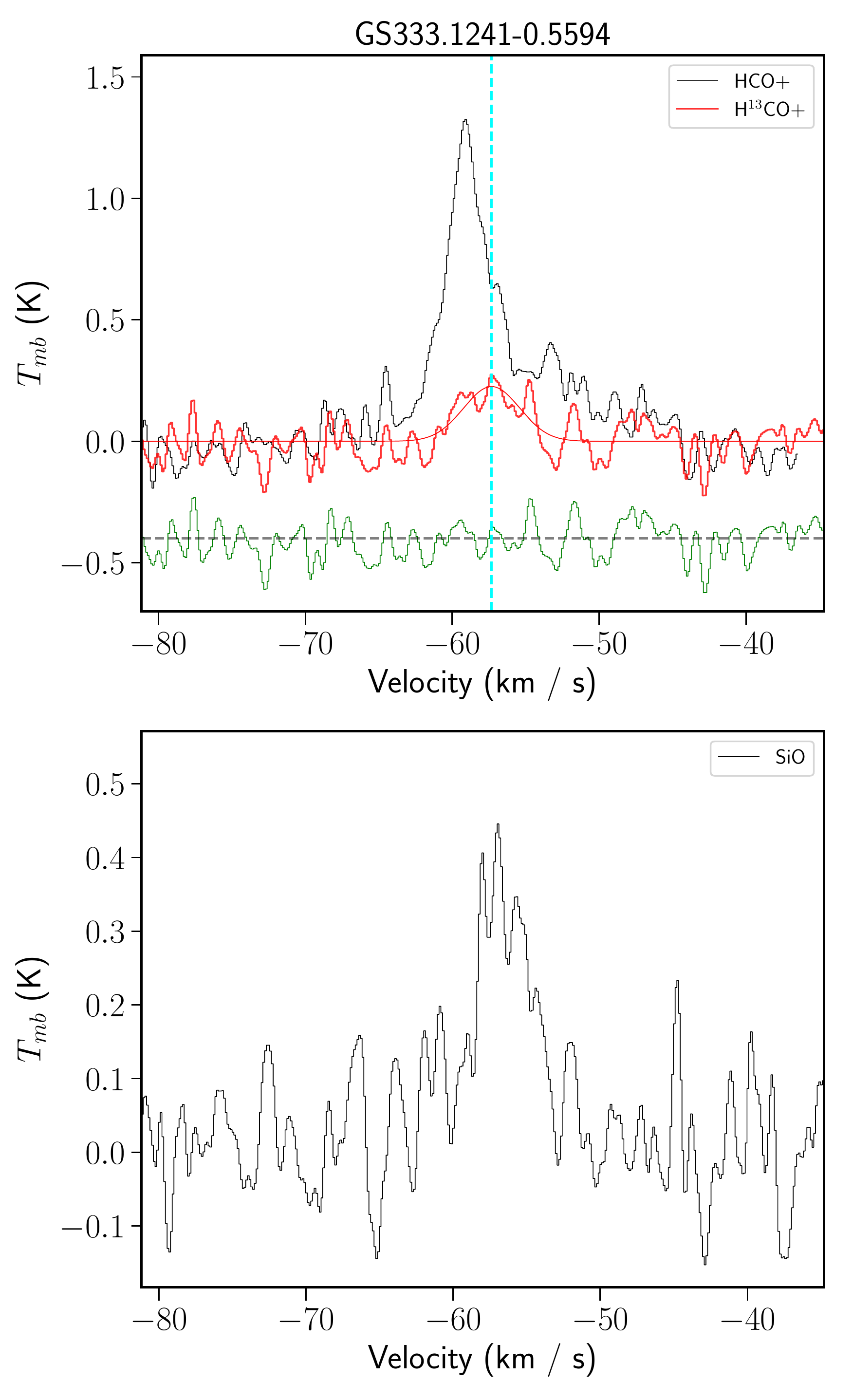}&
\hspace{-0.0cm}\includegraphics[width=6.cm]{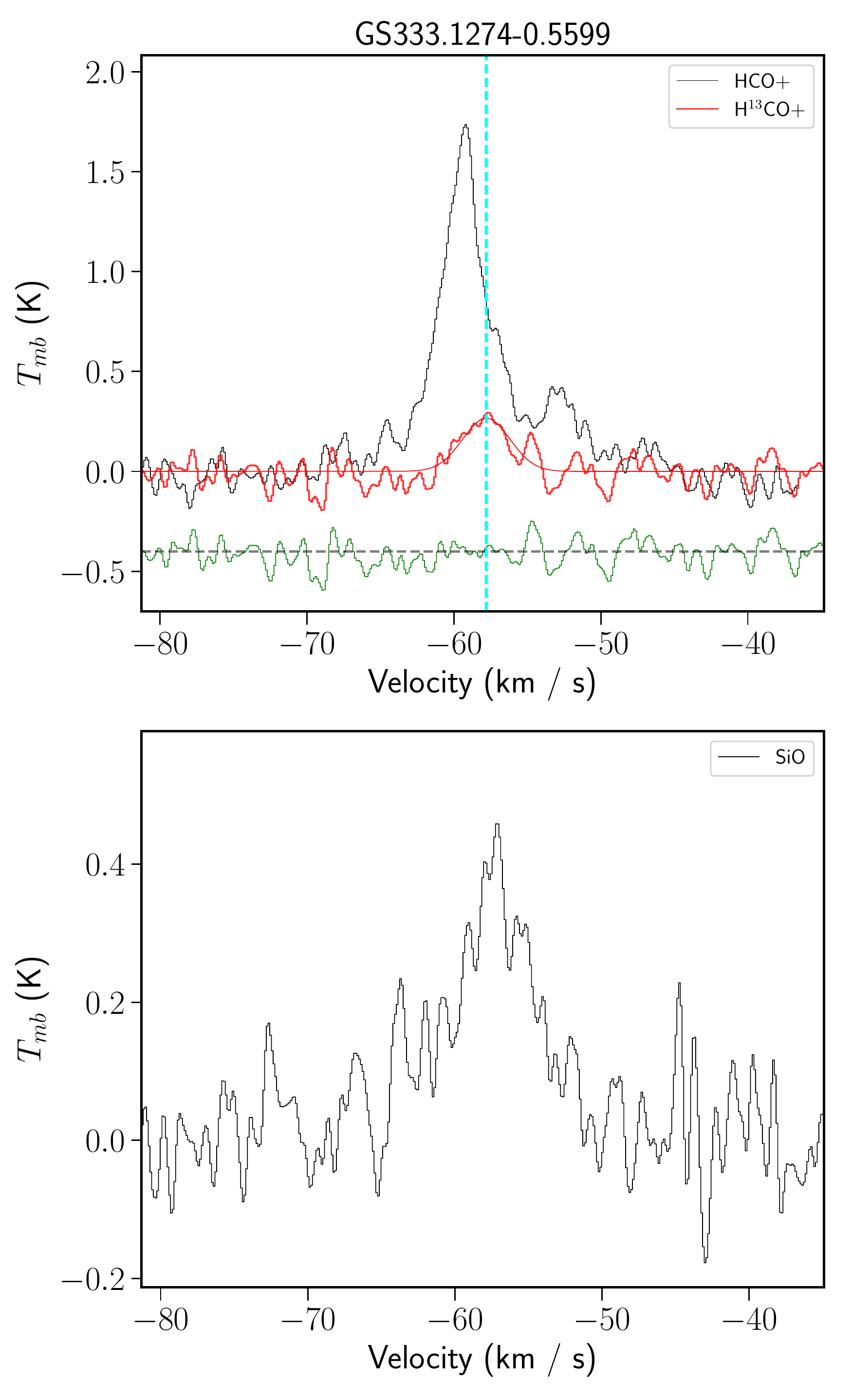}&
\hspace{-0.3cm}\includegraphics[width=6.cm]{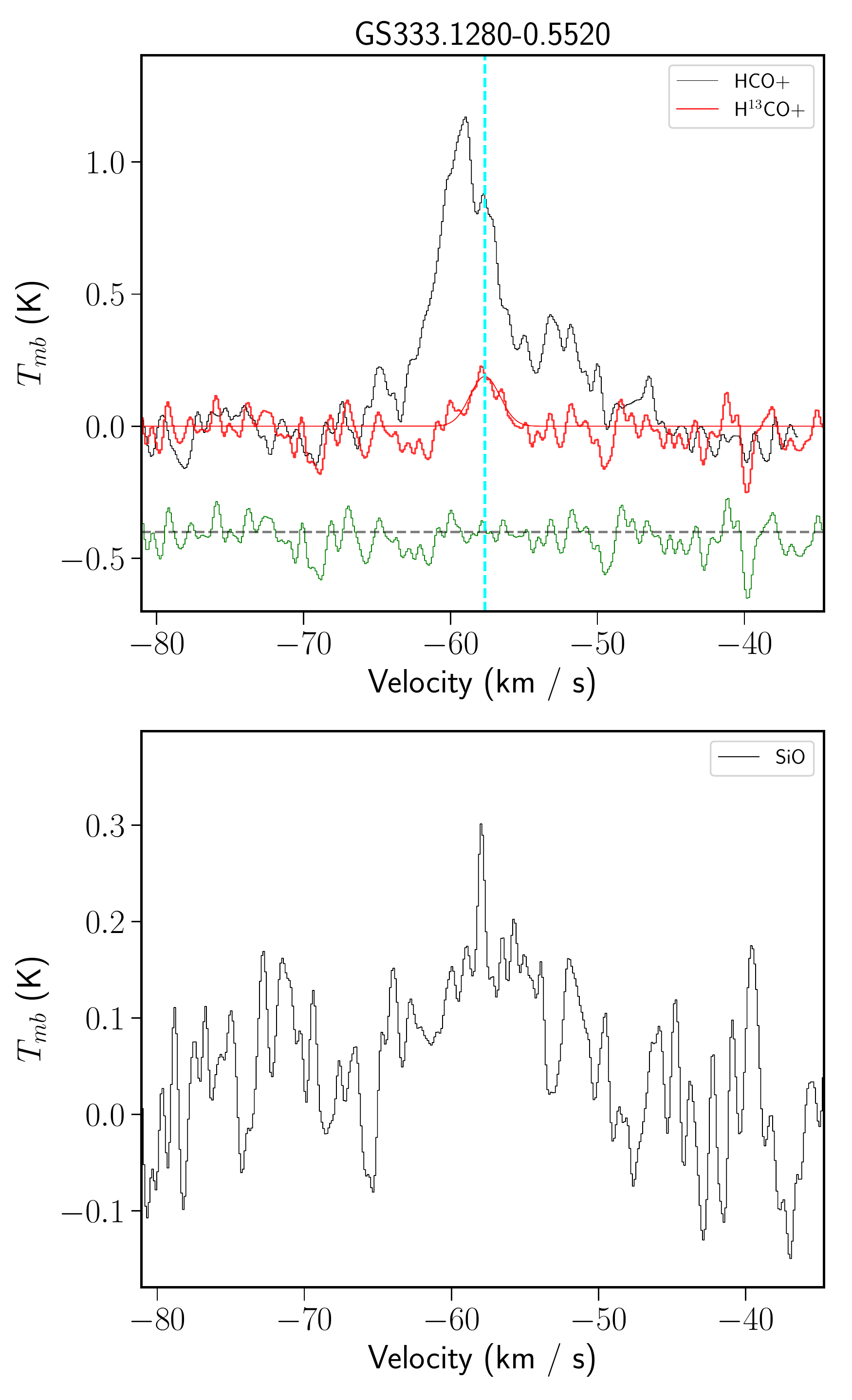}\\
\end{tabular}
\caption{MALT90 spectra towards the $>$100 M$_{\odot}$ quiescent cores at $<$ 5 kpc. The source names are shown in the title. HCO$^+$/H$^{13}$CO$^+$ and SiO lines are plotted in the top and bottom panels, respectively. The $V_{\rm LSR}$ is indicated with a vertical line in the HCO$^+$/H$^{13}$CO$^+$ plot. The residual of the H$^{13}$CO$^+$ fit is shown in green, offset by -0.4 K.}
\end{figure*}
\begin{figure*}
\hspace{-0.8cm}
\vspace{-0.2cm}
\begin{tabular}{p{6.2cm}p{6.2cm}p{6.2cm}}
\hspace{0.55cm}\includegraphics[width=6.cm]{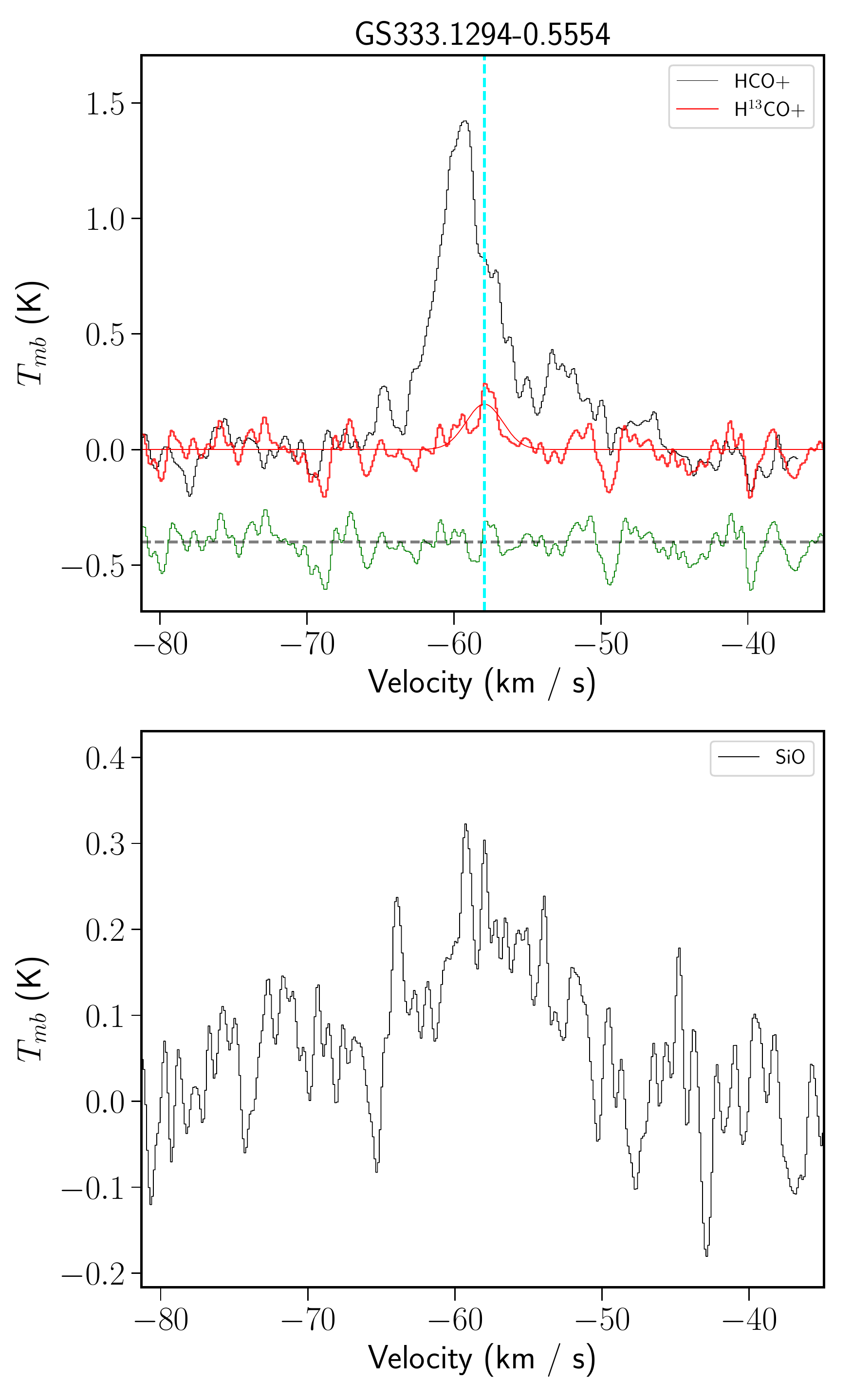}&
\hspace{-0.0cm}\includegraphics[width=6.cm]{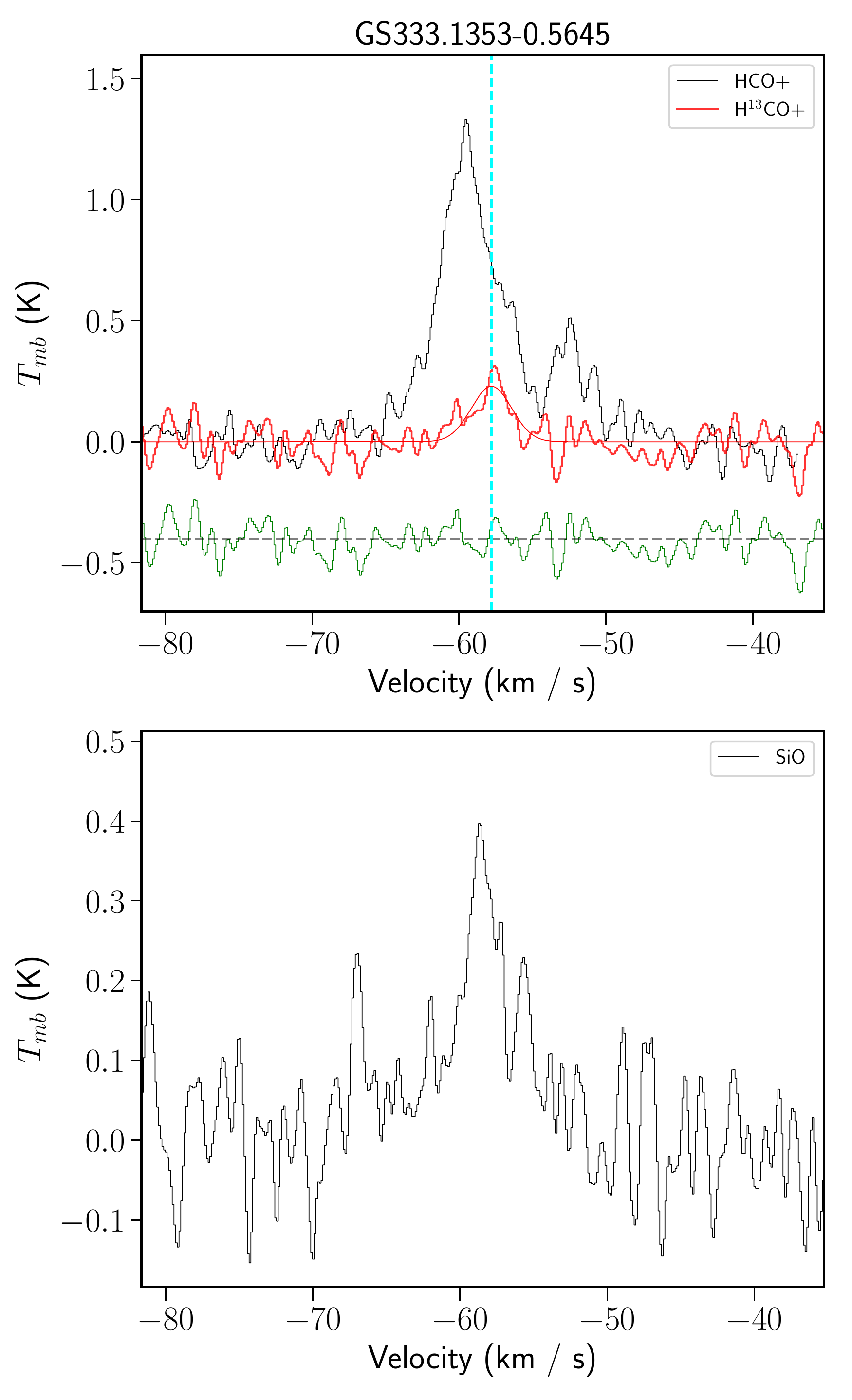}&
\hspace{-0.3cm}\includegraphics[width=6.cm]{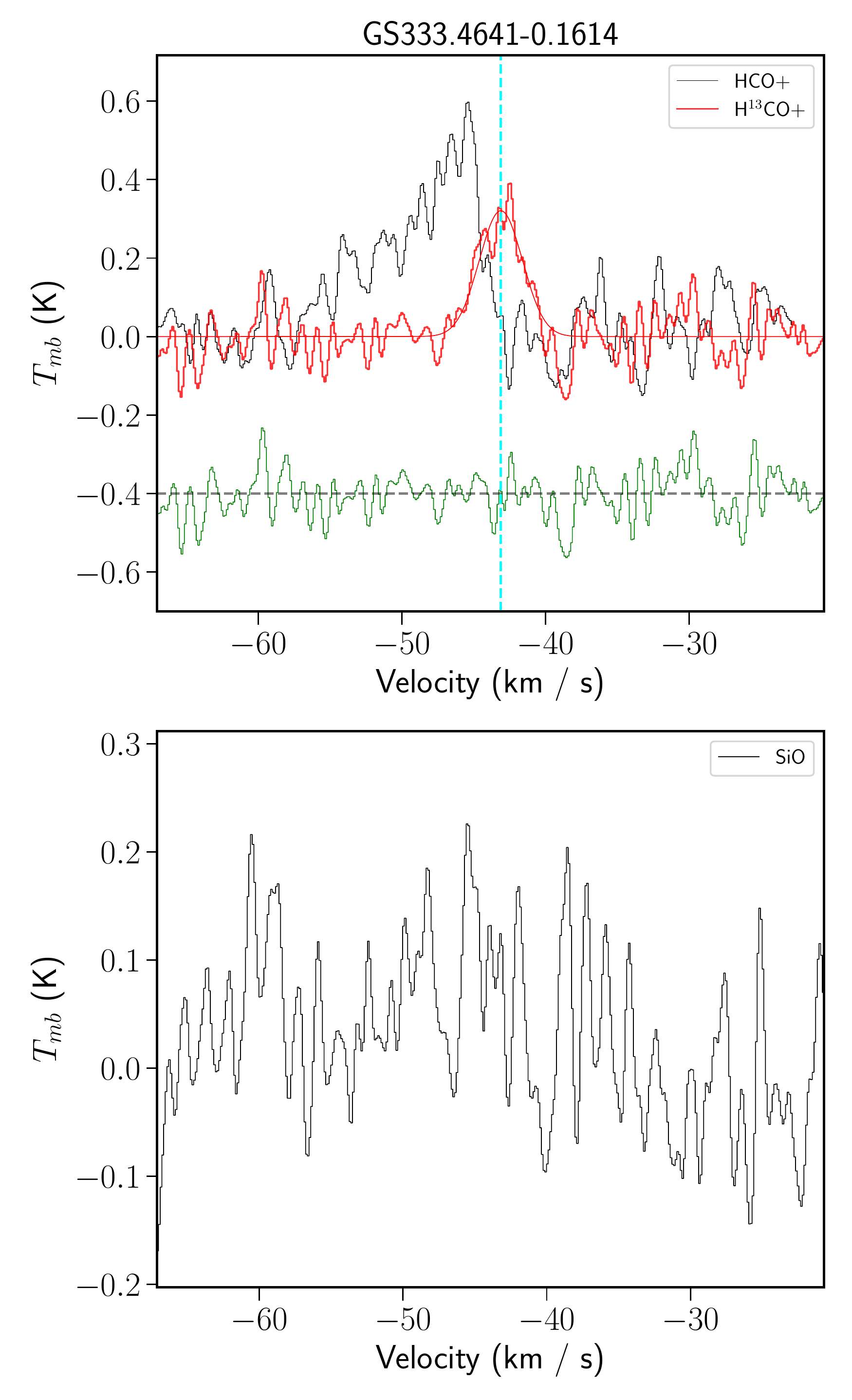}\\
\end{tabular}

\end{figure*}
\begin{figure*}
\hspace{-0.8cm}
\vspace{-0.2cm}
\begin{tabular}{p{6.2cm}p{6.2cm}p{6.2cm}}
\hspace{0.55cm}\includegraphics[width=6.cm]{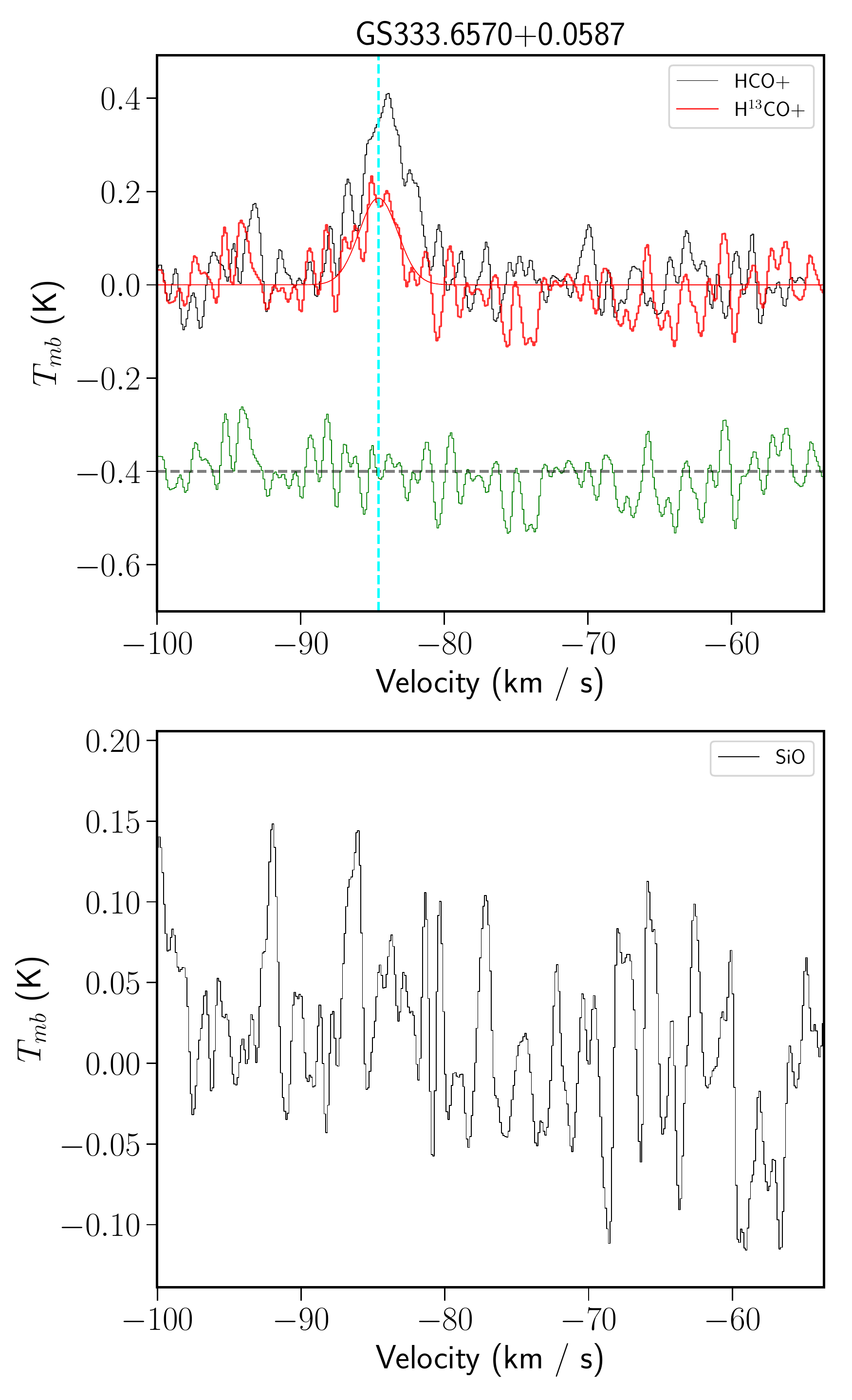}&
\hspace{-0.0cm}\includegraphics[width=6.cm]{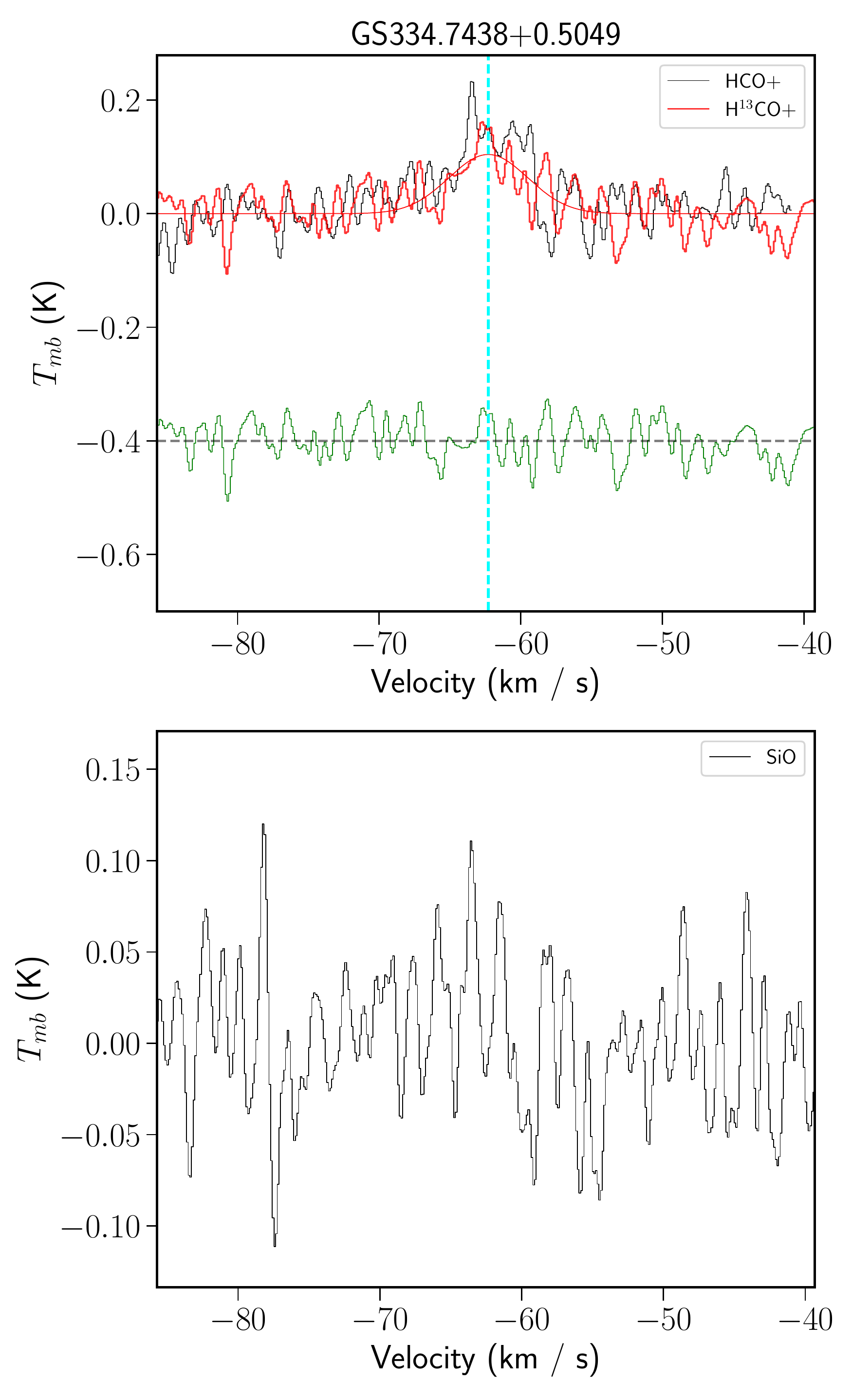}&
\hspace{-0.3cm}\includegraphics[width=6.cm]{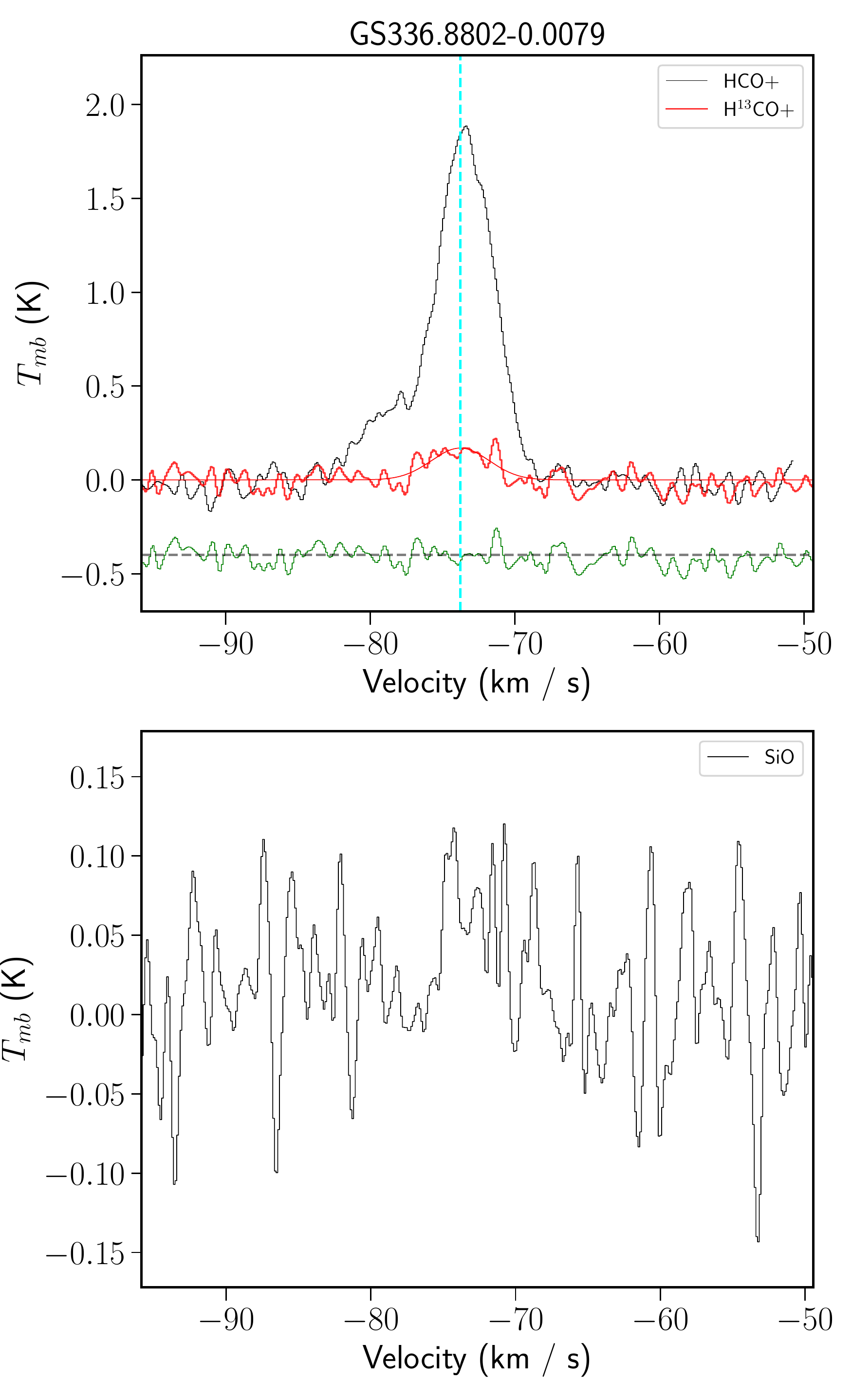}\\
\end{tabular}
\caption{Continued.}

\end{figure*}
\clearpage
\begin{figure*}
\hspace{-0.8cm}
\vspace{-0.2cm}
\begin{tabular}{p{6.2cm}p{6.2cm}p{6.2cm}}
\hspace{0.55cm}\includegraphics[width=6.cm]{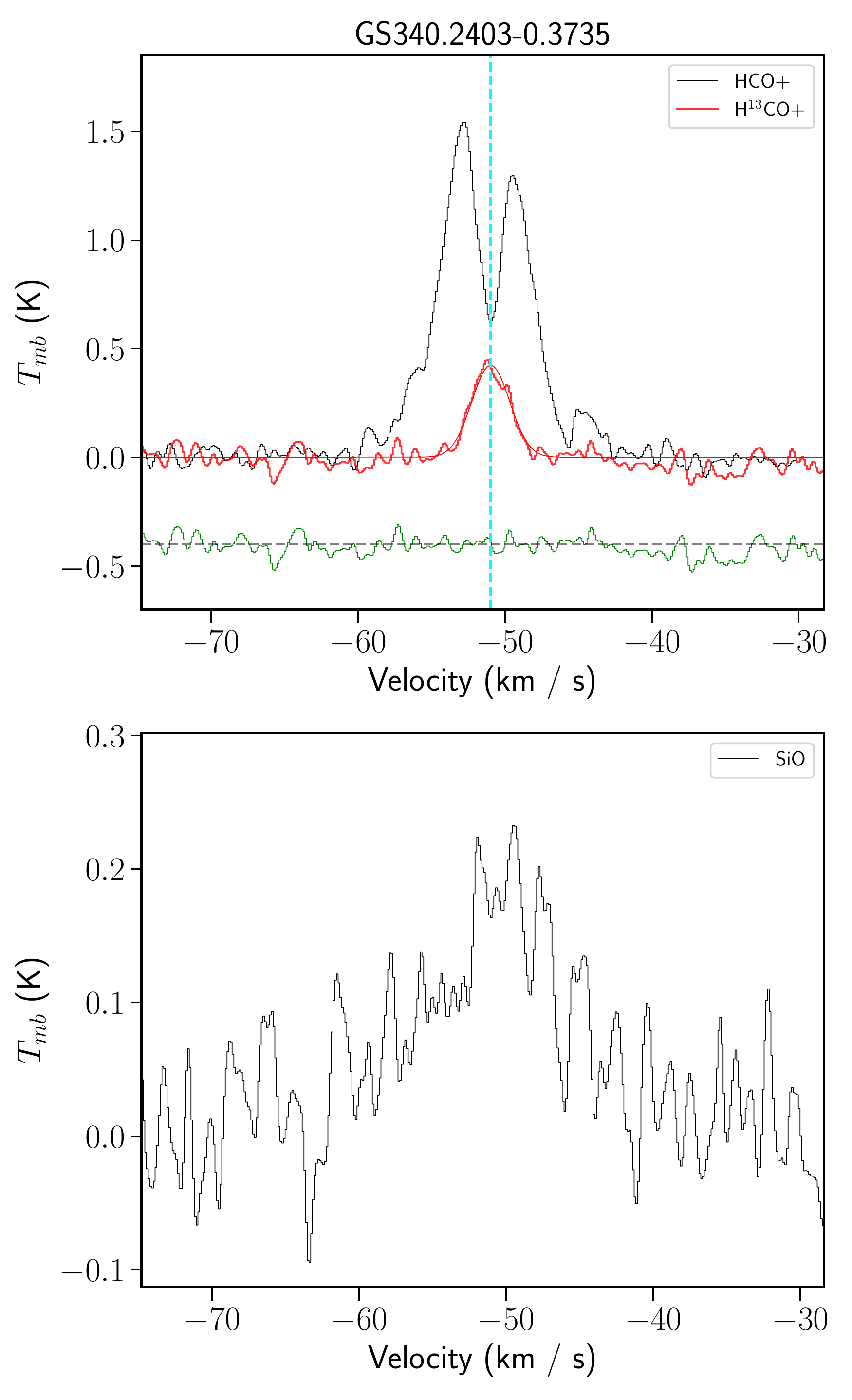}&
\hspace{-0.0cm}\includegraphics[width=6.cm]{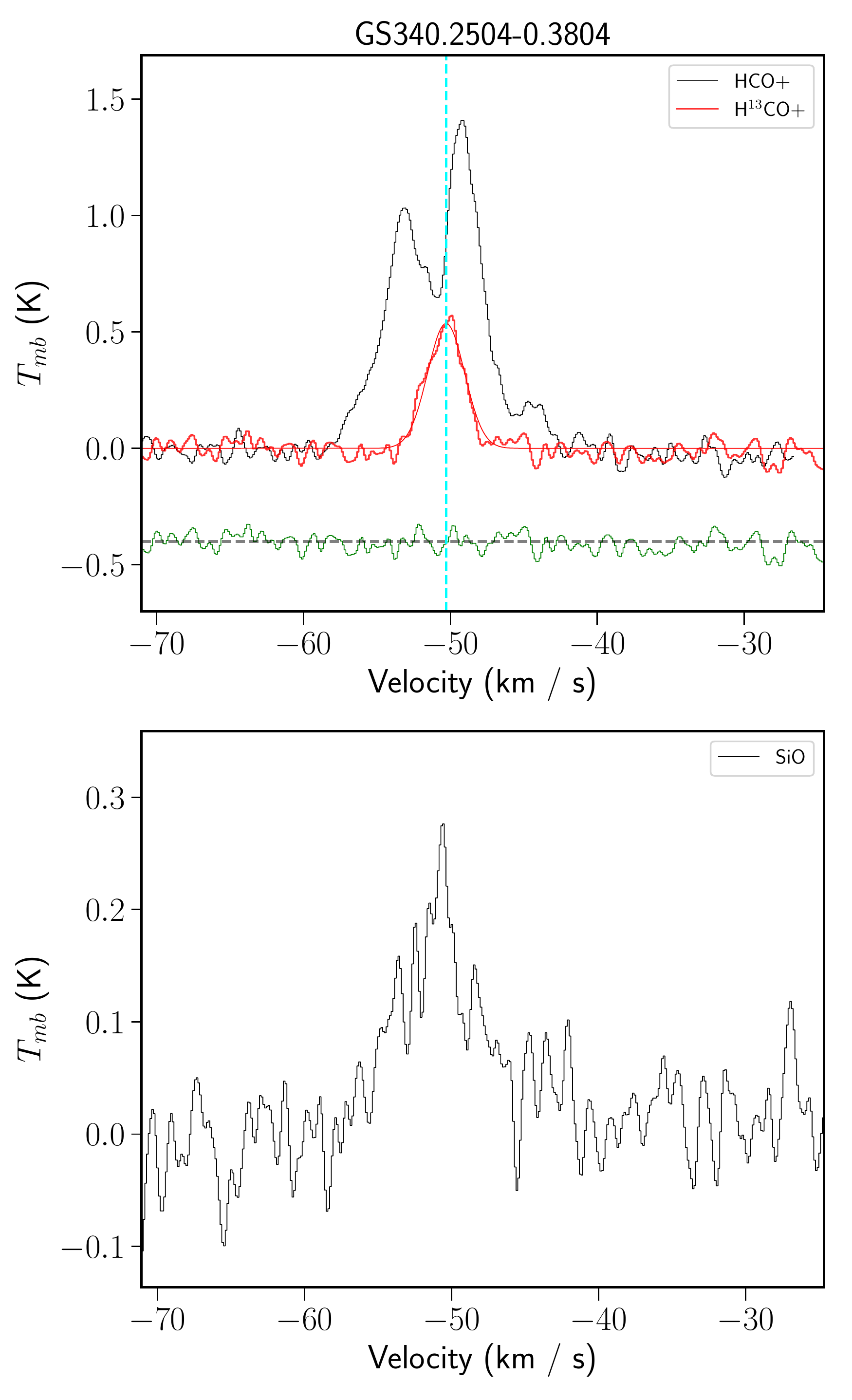}&
\hspace{-0.3cm}\includegraphics[width=6.cm]{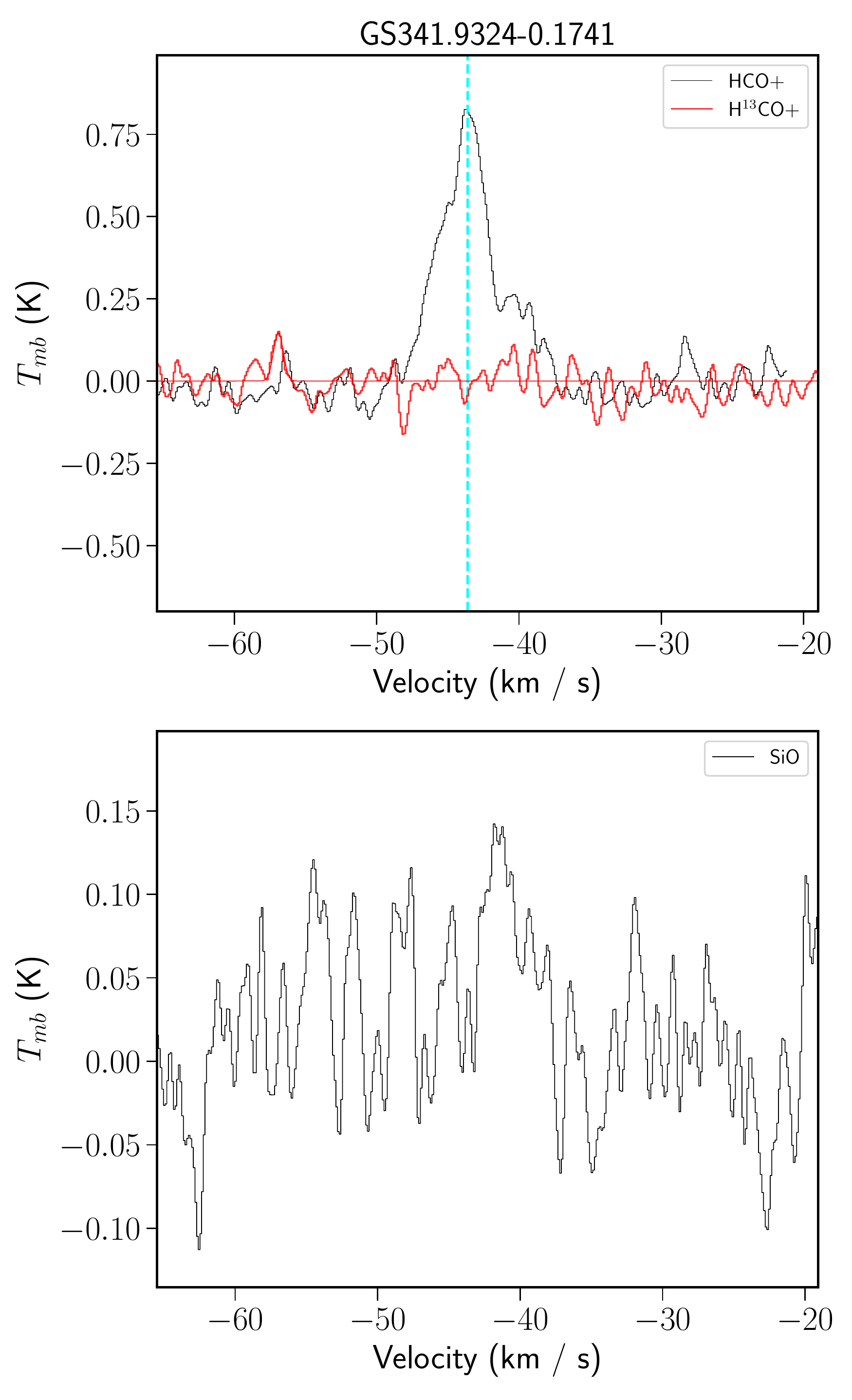}\\
\end{tabular}
\end{figure*}
\begin{figure*}
\hspace{-0.8cm}
\vspace{-0.2cm}
\begin{tabular}{p{6.2cm}p{6.2cm}p{6.2cm}}
\hspace{0.55cm}\includegraphics[width=6.cm]{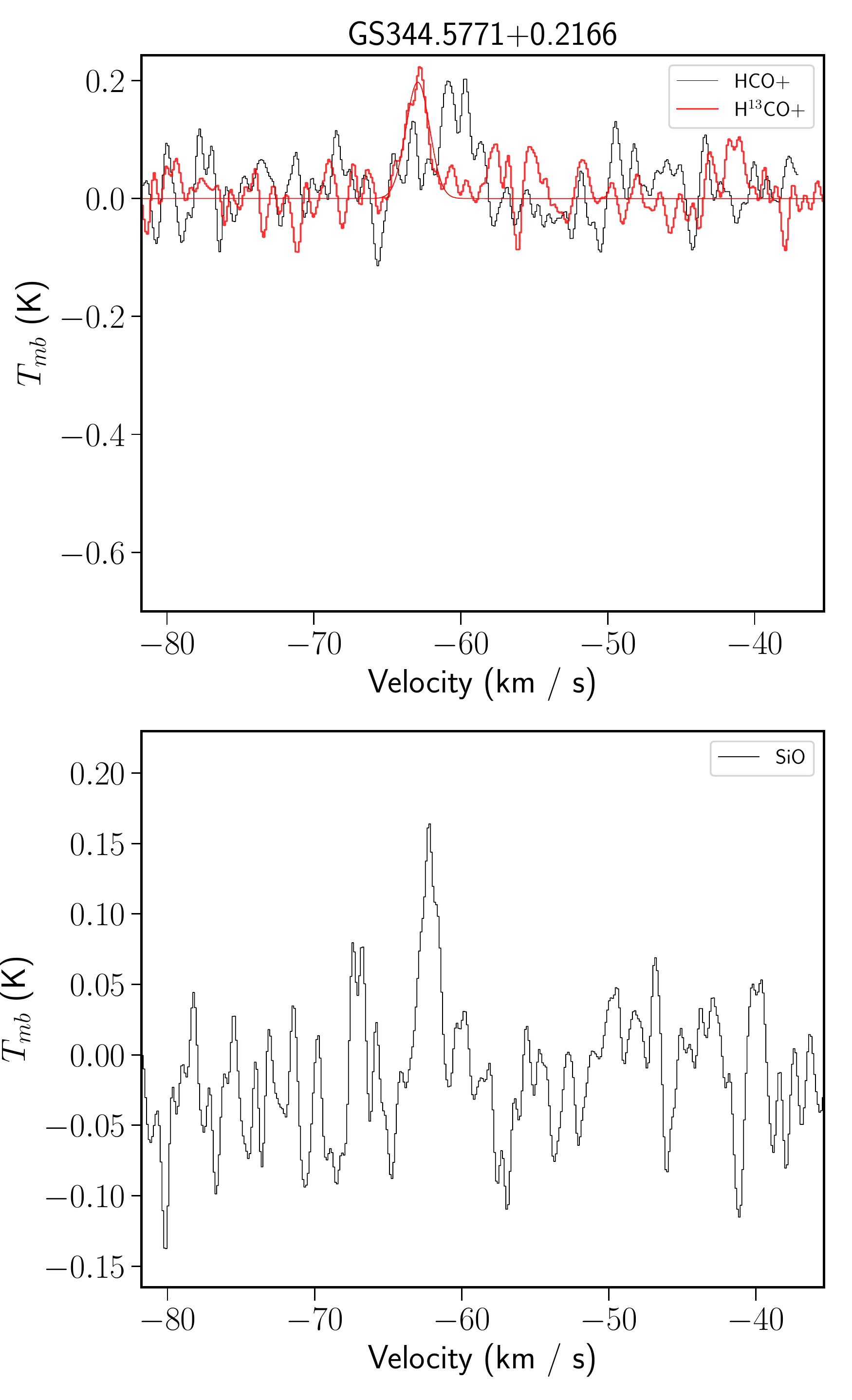}&
\hspace{-0.0cm}\includegraphics[width=6.cm]{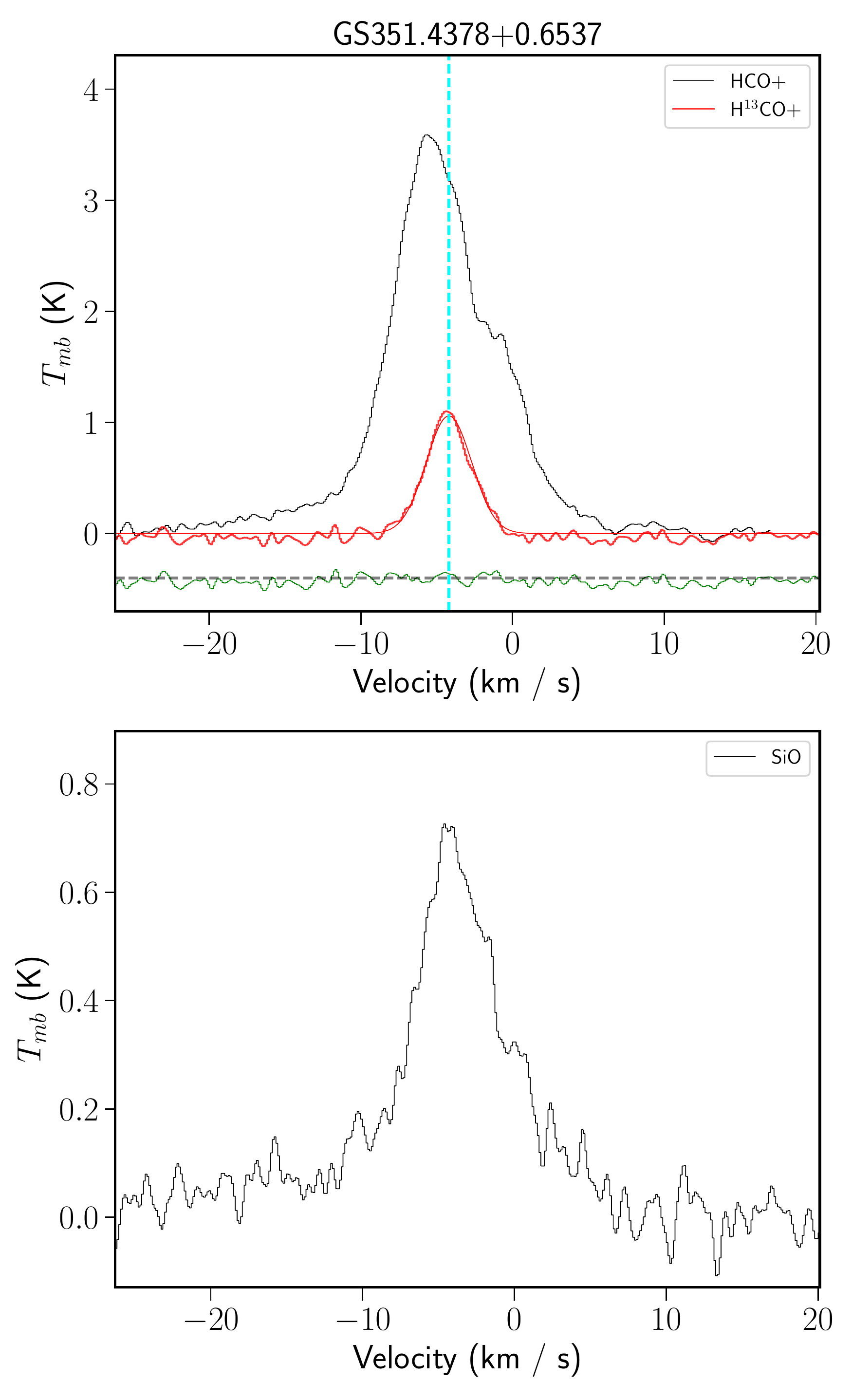}&
\hspace{-0.3cm}\includegraphics[width=6.cm]{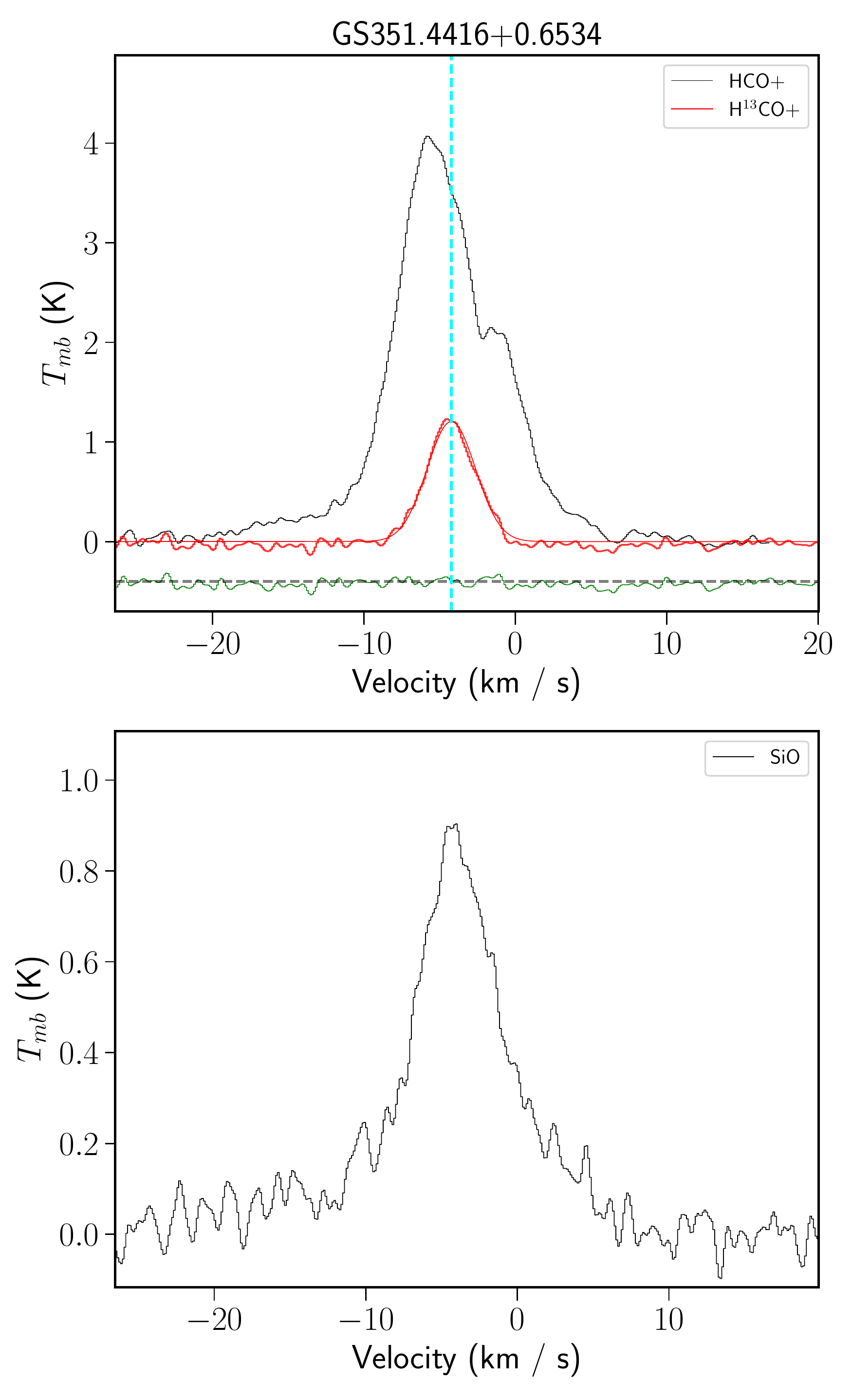}\\
\end{tabular}
\caption{Continued.}

\end{figure*}
\begin{figure*}
\hspace{-0.8cm}
\vspace{-0.2cm}
\begin{tabular}{p{6.2cm}p{6.2cm}p{6.2cm}}
\hspace{0.55cm}\includegraphics[width=6.cm]{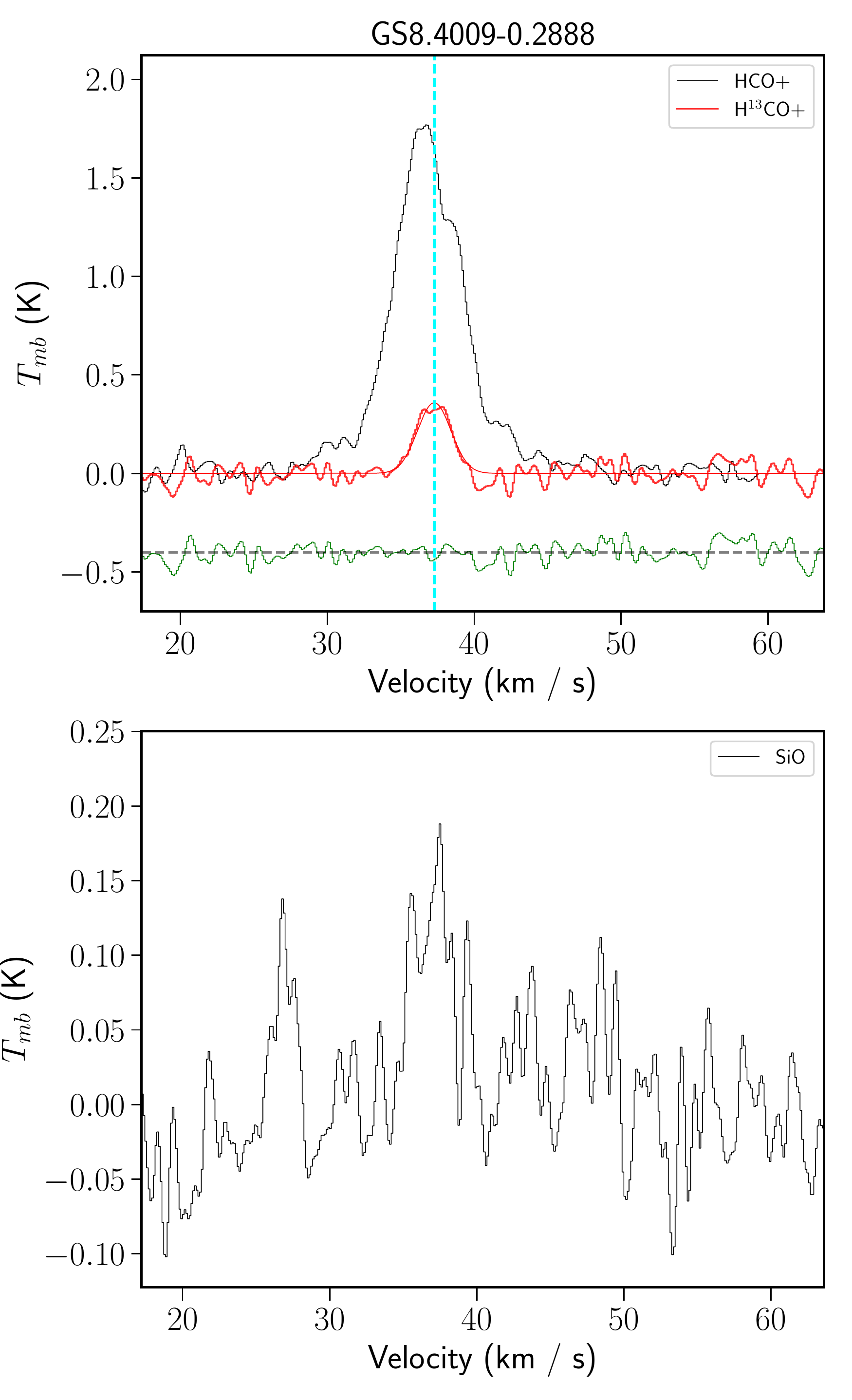}&
\hspace{-0.0cm}\includegraphics[width=6.cm]{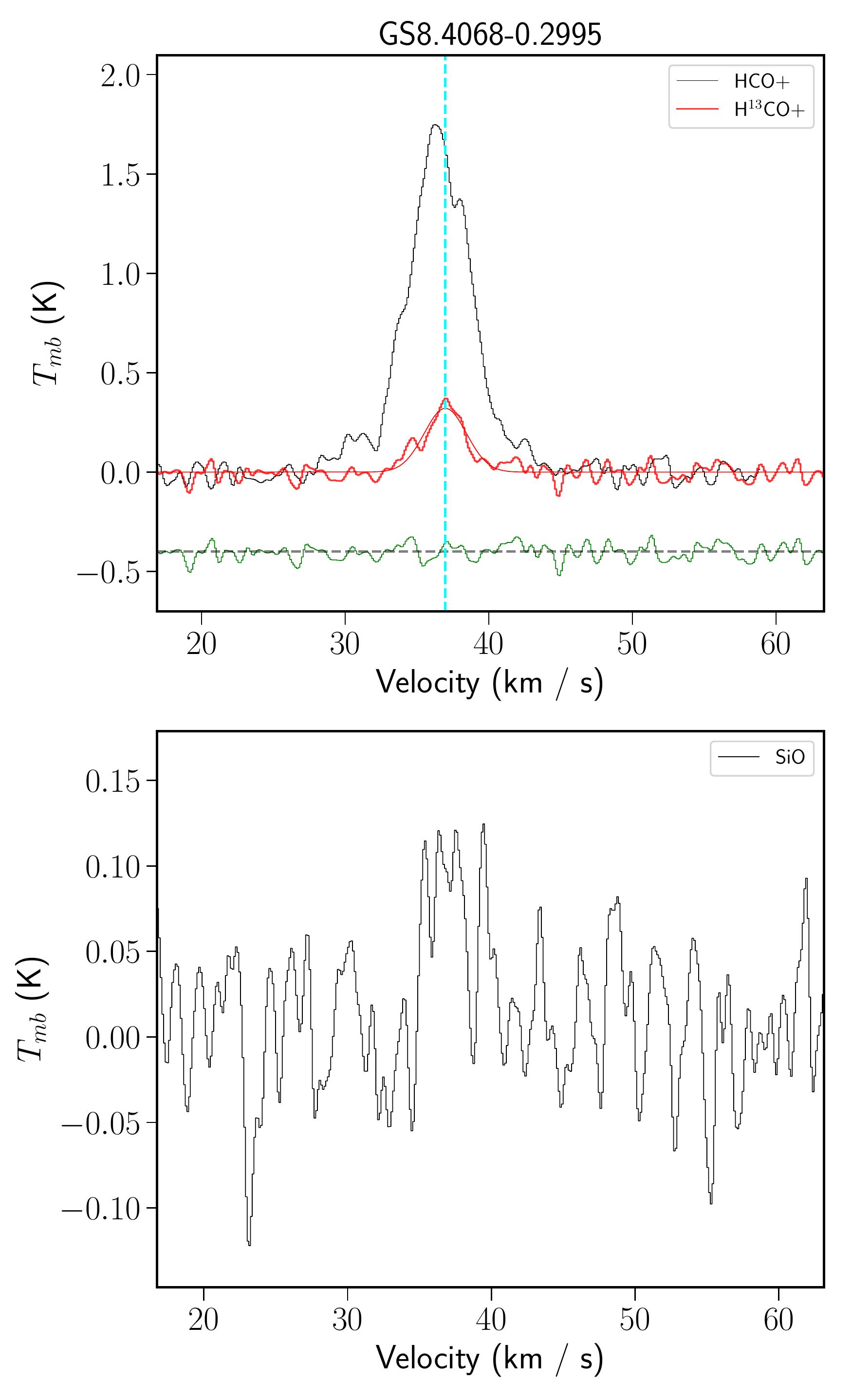}&
\hspace{-0.3cm}\includegraphics[width=6.cm]{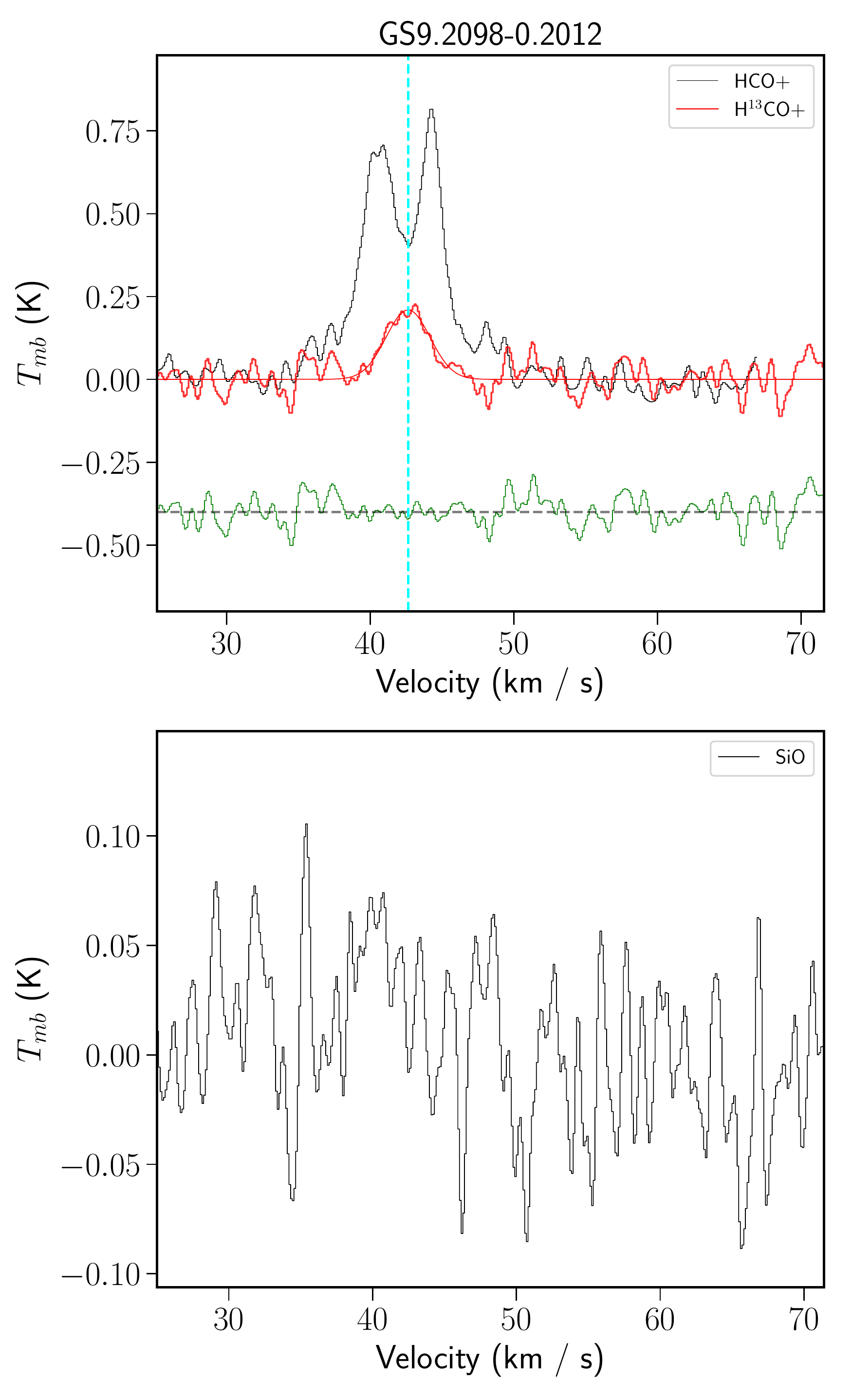}\\
\end{tabular}
\caption{Continued.}
\end{figure*}

\end{appendix}

\end{document}